%% file: article.tex
\documentclass[a4paper, 10pt]{article}
\usepackage{srcltx,graphicx,epstopdf}
\usepackage{amsmath, amssymb, amsthm}
\usepackage{color}
\usepackage{subfig} 

\usepackage{hyperref}
\usepackage[margin=1in]{geometry}
\usepackage{cleveref}
\crefname{section}{Section}{Sections}
\crefname{subsection}{Subsection}{Subsections}
\crefname{appendix}{Appendix}{Appendix}
\crefname{figure}{Figure}{Figures}
\crefname{table}{Table}{Tables}
\crefname{property}{Property}{Properties}
\crefname{theorem}{Theorem}{Theorem}
\usepackage{booktabs}
\usepackage{threeparttable}

\graphicspath{{images/}}

{\theoremstyle{remark} \newtheorem{remark}{Remark}}

\newcommand\pd[2]{\dfrac{\partial {#1}}{\partial {#2}}}

\newcommand\bxi{\boldsymbol{\xi}}
\newcommand\bA{\bf A}
\newcommand\bI{\bf I}
\newcommand\bB{\bf B}

\newcommand\bx{\boldsymbol{x}}

\newcommand\bu{\boldsymbol{u}}
\newcommand\bv{\boldsymbol{v}}

\newcommand\bM{{\bf M}}
\newcommand\bD{{\bf D}}

\newcommand\bomega{\boldsymbol{\omega}}

\newcommand\bg{\boldsymbol{g}}
\newcommand\bR{\boldsymbol{R}}

\newcommand\bbR{\mathbb{R}}
\newcommand\bbN{\mathbb{N}}

\newcommand\bq{\boldsymbol{q}}

\newcommand\dd{\,\mathrm{d}}
\newcommand\Kn{\mathit{Kn}}
\newcommand\mQ{\mathcal{Q}}
\newcommand\mM{\mathcal{M}}
\newcommand\mH{\mathcal{H}}

\newcommand\bw{\boldsymbol{w}}
\newcommand\He{\mathit{He}}
\def\bd{\boldsymbol{d}}
\def\bx{\boldsymbol{x}}
\def\bbR{\mathbb{R}}
\def\bu{\boldsymbol{u}}
\def\bq{\boldsymbol{q}}
\def\bsigma{\boldsymbol{\sigma}}
\def\bslambda{\boldsymbol{\lambda}}

\def\bF{\boldsymbol{F}}

\def\bg{\boldsymbol{g}}

\def\bm{\boldsymbol{m}}

\numberwithin{equation}{section}

\title {A hybrid moment method for multi-scale kinetic equations based
  on maximum entropy principle}

\author{Weiming Li\thanks{Laboratory of Computational Physics, Institute of 
  Applied Physics and Computational Mathematics, Beijing, China, 100088, email:
\tt{liweiming@pku.edu.cn}},~~Peng Song\thanks{Laboratory of Computational Physics, Institute of 
  Applied Physics and Computational Mathematics, Beijing, China, 100088, email:
\tt{song\_peng@iapcm.ac.cn}},~~ Yanli Wang\thanks{Beijing Computational
    Science Research Center, Beijing, China, 100193, email: {\tt
      ylwang@csrc.ac.cn}.} }

\begin{document}
\maketitle
\input{introduction}

\input{preliminary}
\input{method}
\input{numerical_results}
\input{article_conclusion}
\input{article_appendix}
\bibliographystyle{plain}
\bibliography{../article}

\end{document}

%% file: introduction.tex
\begin{abstract}
  We propose a hybrid moment method for the multi-scale kinetic
  equations in the framework of the regularized moment method
  \cite{NRxx}. In this method, the fourth order moment system is
  chosen as the governing equations in the fluid region. When
  transiting from the fluid regime to the kinetic regime, the maximum
  entropy principle is adopted to reconstruct the kinetic distribution
  function, so that the information in the fluid region can be
  utilized thoroughly.  Moreover, only one uniform set of numerical
  scheme is needed for both the fluid and kinetic regions.  Numerical
  tests show the high efficiency of this new hybrid method.

  \vspace{3pt}


\end{abstract}

\section{Introduction}

Traditional fluid models, such as Euler and the Navier-Stokes
equations, give accurate descriptions of gas flows when the system is
close to an equilibrium state.  However, many problems, such as
simulations of hypersonic flows, involve non-equilibrium processes,
where traditional fluid models break down. In this case, the kinetic
equations, such as the Boltzmann equation, are introduced to describe
the evolution of the particles. However, due to its high
dimensionality, the computational cost to simulate the kinetic
equation is quite expensive in comparison to using fluid models such
as Euler or Navier-Stokes equations.

Several parameters are introduced to describe how far the particles
are from the equilibrium, for example the Knudsen number. The Knudsen
number is defined as the ratio of the mean free path of the particles
to a typical macroscopic length. When the Knudsen number is large, the
flow is in the kinetic regime, and the simulations are studied by the
kinetic equations, for example the Boltzmann equation. As to the
Boltzmann equation, several kinds of numerical methods have been
proposed. The DSMC (Direct Simulation Monte Carlo) method \cite{Bird}
is a stochastic numerical method, which is quite efficient when the
Knudsen number is large, but may be expensive for the fluid regime.
There are also several classical deterministic methods, for example
discrete velocity method \cite{goldstein1989, Panferov2002}, and the
spectral method such as the Fourier spectral method
\cite{pareschi1996Spectral, Mouhot, Hu2016}. Moreover, Hermite
spectral method is also introduced to solve the Boltzmann equation
\cite{Gamba2018, QuadraticCol}, which can be traced back to Grad's
classical paper \cite{Grad}.  Despite the several kinds of numerical
methods that have been proposed, the numerical cost of Boltzmann
equations is still quite prohibitive due to the high dimensionality.

For the multi-scale problems, where the fluid and kinetic regions are
both included, the coupling of the kinetic and fluid dynamic equations
has become an important research area \cite{Levermore1998,
  FILBET2018841, Filbet2015}. In \cite{Levermore1998}, a moment
realizability matrix is introduced to differentiate the fluid region
and the kinetic region. Then the moment realizability matrix is
utilized in \cite{Filbet2015, FILBET2018841, xiongqiu} as a criterion
for solving the multi-scale kinetic problems. There are several other
hybrid methods, which are based on the local Knudsen number
\cite{KOLOBOV2007} or based on the viscosity and heat flux of the
Navier-Stokes equations \cite{TIWARI1998, ALAIA20125217}. Moreover,
the hybridization of the Monte Carlo method and the finite volume
method is proposed in \cite{Caflisch2016, Degond2011, Dimarco2007}.

Recently a globally hyperbolic moment method is proposed in
\cite{NRxx, Fan_new} to solve the Boltzmann equation. This method is
based on a Hermite spectral method where the expansion center is
chosen as the local velocity and temperature, where it is expected
that this expansion can describe the distribution function well even
when it is far from the equilibrium. A highly efficient numerical
scheme has been designed under the framework of this method, which
also has been successfully used to solve Vlasov and Wigner equations
\cite{Hu2014Simulation}. It is also proved that the moment system
derived by the hyperbolic moment method contains the fluid dynamic
equations such as Euler and Navier-Stokes equations.  This method is
verified to successfully simulate the movement of the particles with
any Knudsen number. But the expansion order may increase quickly with
the increase of the Knudsen number. Therefore, when there is a
coupling of kinetic and fluid regimes, for example in the simulations
of the hypersonic flows around a space vehicle during the re-entry
phase, there may exist large variation of the Knudsen number, in which
case, the expansion order is decided by the largest Knudsen number,
and the computational cost may be expensive.

In this paper, we propose a uniform hybrid method in the framework of
the regularized moment method \cite{NRxx}. Instead of Euler and the
Navier-Stokes equations, the moment equations with expansion order
four are utilized as the governing equations in the fluid region.  The
reason that we choose the fourth order moment model as our fluid model
is that the maximum entropy ansatz for reconstructing the kinetic
distribution function is utilized when we change from fluid to kinetic
description, and the fourth order system is the smallest moment system
in the maximum entropy theory where the distribution ansatz could
characterize non-equilibrium behavior. Another important point of the
hybrid method is the domain decomposition indicator. In our hybrid
method, the domain decomposition indicator is based on the moment
realizability matrices proposed in \cite{Levermore1998} and later used
in \cite{Filbet2015, FILBET2018841}. It has quite a simple form, where
instead of the derivatives of the macroscopic variables, only the
expansion coefficients of the distribution function with orders
smaller than four are used.  The criteria from/to hydrodynamic to/from
kinetic are proposed based on the indicator. Numerical tests show that
our method could describe the evolution of the particles in the fluid
region with quite few degrees of freedom and is therefore more
efficient than the traditional regularized moment method using a
uniform number of moments.  Another advantage of our method is that we
could use same numerical scheme for both the kinetic and fluid
regions. We do not need to strictly discriminate the numerical flux in
different domains, especially those at the kinetic and fluid
interfaces, since they all have the same form, which simplifies the
coupling of different domains. Numerical simulations are done for
several benchmark problems. In the numerical test, the BGK and Shakhov
collision term are applied, with the dimension reduction method in the
microscopic velocity space in \cite{Qiao} adopted to reduce
computation complexity. 

The rest of this paper is organized as follows: Section \ref{sec:pre}
introduces the Boltzmann equation, and two moment methods including
the maximum entropy method \cite{Levermore} and the regularized moment
methods \cite{NRxx} briefly. The detailed hybrid method is given in
Section \ref{sec:hybrid}, and the numerical algorithm is also proposed
in this section. Several numerical examples are exhibited in Section
\ref{sec:num}. Conclusion and future work are made in Section
\ref{sec:conclusion}. Detailed description of the numerical scheme and
the method to solve the maximum entropy method is given in the
Appendix.

%% file: preliminary.tex

\section{Boltzmann equation and the moment method}
\label{sec:pre}
In the kinetic theory, the Boltzmann equation is always adopted to
describe the movement of the particles, especially in the rarefied gas
dynamics. However, when the gas becomes dense, in other words when the
system approaches the fluid regime, the Boltzmann equation may reduce
to Euler or Navier-Stokes equations. In this section, we will first
introduce the Boltzmann equation \cite{Struchtrup2005} and some
related properties.

\subsection{Boltzmann equation}
In the Boltzmann equation, the distribution function $f(t, \bx, \bv)$
is used to describe the fluid state, where $t$ is time, $\bx$ is the
spatial space, and $\bv$ stands for the velocity of gas molecules.
The dimensionless form of the governing equation is
\begin{equation}
  \label{eq:Boltzmann}
  \pd{f}{t} + \bv \cdot \nabla_{\bx} f =  \frac{1}{\epsilon}\mQ[f], \qquad t \in
  \bbR^+, \quad \bx \in \bbR^3, \quad \bv \in \bbR^3,
\end{equation}
where $\epsilon$ is a formal smallness parameter which plays the role
of Knudsen number.  $\mQ[f]$ denotes particle collision, which may
have quite complex form. The collision operator satisfies the
following properties:
\begin{enumerate}
  \item conservation of the mass, momentum and kinetic energy
\begin{equation}
  \label{eq:conservation}
  \int_{\bbR^3 } \left(
    \begin{array}{c}
      1 \\
      \bv \\
      |\bv|^2
    \end{array}
  \right) \mQ[f] \dd \bv = 0.
\end{equation}
\item the Boltzmann's H-theorem
\begin{equation}
  \label{eq:H-Theorem}
  \qquad \int_{\bbR^3} \mQ[f] \log f \dd \bv \leqslant 0, 
\end{equation}
where equality holds if and only if $f$ is the Maxwellian, which is
the steady state of the Boltzmann equation \eqref{eq:Boltzmann}
\begin{equation}
  \label{eq:Maxwellian}
  \mM = \frac{\rho}{(2 \pi \theta)^{3/2}}\exp\left(-\frac{|\bv -
    \bu|^2}{2\theta}\right).
\end{equation}
Here $\rho(t, \bx)$ is the density, $\bu(t, \bx)$ is the macroscopic
velocity and $\theta(t, \bx)$ is the temperature of the particles,
whose relationships with the distribution function $f(t, \bx, \bv)$ is
as below
\begin{equation}
  \label{eq:macro}
\left(
  \begin{array}{c}
    \rho(t, \bx) \\[1mm]
    \rho(t, \bx)\bu(t, \bx) \\[1mm]
    \frac{1}{2}\rho |\bu(t, \bx)|^2 + \frac{3}{2}\rho(t, \bx)\theta(t,
    \bx)
  \end{array}
\right) = \int_{\bbR^3} \left(
  \begin{array}{c}
    1 \\
    \bv \\
    \frac{1}{2} |\bv|^2
      \end{array}
\right)  f(t, \bx, \bv) \dd \bv.
\end{equation}
Moreover, it also holds that
\begin{equation}
  \mQ(\mM) = 0.
\end{equation}
  \item Galilean invariance: $\mQ[f]$ commutes with translational and
    rotational operators.
\end{enumerate}
In this paper, we are mainly concerned with the hybrid method to solve
the Boltzmann equation, therefore, the simplified collision operator
is used here, for example the BGK collision operator
\begin{equation}
  \label{eq:BGK}
  \mQ^{\rm BGK}[f] = \frac{1}{\tau}(\mM - f), 
\end{equation}
where $\tau$ is the relaxation time, which is usually obtained from
the first approximation of the Chapman-Enskog theory \cite{Bird}.
$\tau$ is a parameter that related to the Knudsen number $\Kn$ and
some macroscopic variables such as the density. When $\tau$ is large,
very few molecular collisions occur, and the entire flow is
rarefied. For the classical BGK operator, it gives the wrong Prandtl
number which equals $1$ whereas the correct Prandtl number for a
monoatomic gas is $2/3$. Another simplified collision operator, the
Shakhov collisional operator preserves the correct Prandtl number
\cite{Struchtrup2005}, which has the form
\begin{equation}
  \label{eq:Q_shakhov}
  Q_{\rm Shakhov}(f) = \frac{1}{\tau}\left(f_s - f\right), \qquad
  f_s = P_3(t, \bx, \bv) \mM(t, \bx, \bv), 
\end{equation}
with
\begin{equation}
  P_3 = 1 + \dfrac{(1 - {\rm Pr})(\bv - \bu) \cdot \bq}{(D+2)\rho(t, \bx)
  [\theta(t, \bx)]^2}\left(\frac{|\bv - \bu|^2}{\theta(t, \bx)} -
  (D+2)\right), \qquad D = 3,
\end{equation}
where $D$ is the dimension number of the microscopic velocity space
and ${\rm Pr}$ is the Prandtl number.
$\bq$ is heat flux, which is defined as
\begin{equation}
  \label{eq:heatflux}
  q_i = \frac{1}{2}\int_{\bbR^3} |\bv - \bu|^2(v_i - u_i) f \dd
  \bv, \qquad i = 1, 2, 3. 
\end{equation}
There are other physical variables people are interested in, such as
the stress tensor $\sigma_{ij}$, which is defined as
\begin{equation}
  \label{eq:sigma}
  \sigma_{ij} = \int_{\bbR^3} \left((v_i - u_i)(v_j - u_j)  -
    \frac{1}{3}\delta_{ij}|\bv - \bu|^2\right) f \dd
  \bv, \qquad i,j = 1, 2, 3.
\end{equation}
Besides, the pressure tenser $p_{ij}$ is defined as
\begin{equation}
  \label{eq:p}
  p_{ij} = \int_{\bbR^3} (v_i - u_i)(v_j - u_j) f \dd \bv, \qquad
  \sigma_{ij} =  p_{ij} - \frac{1}{3}\delta_{ij} \sum\limits_{k=1}^3
  p_{kk}.   
\end{equation}

\begin{remark}
  In our numerical test, we choose the VHS model for $\tau$ as
  \begin{equation}
    \label{eq:tau0}
    \tau = \sqrt{\frac{\pi}{2}}\frac{15 \Kn}{(5 - 2\omega)(7 -
      2\omega))}  \frac{\theta^{\omega - 1}}{\rho},
  \end{equation}
  where $\omega$ is the viscous index dependent on the type of the
  particles, and $\Kn$ is the Knudsen number. Moreover, $\tau$ is also
  the same parameter to indicate the regime of the particles in some
  literature \cite{Filbet2015, FILBET2018841}.
\end{remark}

There has been much active research on algorithms for solving the
Boltzmann equation numerically, especially for problems in the
transitional regime, such as \cite{Jin2010, Gamba, Struchtrup2008.1,
  NRxx}. In this paper, we focus on developing a hybrid moment method
under the framework of the globally hyperbolic moment method
\cite{NRxx, Fan_new}, with a strategy that utilizes the closure method
based on the maximum entropy principle \cite{Levermore, Hauck,
  Groth}. In the next section, we will briefly introduce these two
methods.

\subsection{Maximum entropy moment method}
\label{sec:entropy}
For simplicity and without loss of generality, this section considers
the 1D case in microscopic velocity space. Define the $k$-th order
moment as
\begin{equation}
  \mu_k = \langle f \rangle_k = \int_{\bbR} f v^k \dd v.
\end{equation}
We could obtain the time evolution equation of $\mu_k$ by multiplying
\Cref{eq:Boltzmann} by $v^k$ and integrating against $v$ over $\bbR$:
\begin{equation}\label{eq:moment_eq}
  \pd{\mu_k}{t} + \pd{\mu_{k+1}}{x} = \langle
  \frac{1}{\epsilon}\mQ[f] \rangle_k,
  \quad k \in \bbN.
\end{equation}
Because in \Cref{eq:moment_eq}, the governing equation of $\mu_k$
depends on $\mu_{k+1}$, therefore the full system contains an infinite
number of equations. We do a truncation in order to obtain a reduced
model for \Cref{eq:Boltzmann}. Specifically, we choose a fixed integer
$M$ and discard all equations for $\mu_k$, $k > M$. Still, as the
truncated moment system depends on $\mu_{M+1}$, it is not a closed
system, so a {\it moment closure} is needed in the system. One way of
specifying a closure is to construct an ansatz for the distribution
function.  Specifically, given moments $\mu_k$, $k = 0, \cdots, M$,
one could reconstruct an ansatz of the distribution function,
$\hat{f}$, which satisfies
\begin{equation}
  \langle \hat{f}(v;\mu_0, \cdots, \mu_M) \rangle_k = \mu_k, \quad k =
  0, \cdots, M.
\end{equation}
Then the moment closure could be given by
\begin{equation}
  \mu_{M+1} = \langle \hat{f}(v;\mu_0, \cdots, \mu_M) \rangle_{M+1}.
\end{equation}
The maximum entropy moment method closes the system
\eqref{eq:moment_eq} by specifying an ansatz of $f$ based on the
entropy minimization principle \cite{Dreyer, Jaynes, Levermore,
  Muller1993}.  Specifically, the ansatz $\hat{f}$ finds the most
likely distribution function under constraints of the given moments by
solving the following functional minimization problem
\begin{equation}\label{eq:maximum_entropy}
  \hat{f} = \text{argmin} H(f), \quad s.t. \langle \hat{f}(v)
  \rangle_k = \mu_k, \quad k = 0, \cdots, M,
\end{equation}
with $H(f) = \langle f \log f - f\rangle$.  Direct computation shows
the unique solution of \Cref{eq:maximum_entropy} has the form
\begin{equation}\label{eq:maximum_entropy_ansatz}
  \hat{f} = \exp\left(\sum\limits_{k=0}^M \lambda_k v^k \right),
\end{equation}
where $\lambda_k$, $k = 0,\cdots, M$ are called the Lagrange
multipliers.  Since \Cref{eq:maximum_entropy_ansatz} is in the
integrable function space, the highest order of the polynomial
$\sum\limits_{k=0}^M \lambda_k v^k$ should be even.

The simplest model in the maximum entropy hierarchy that contains all
the conserved variables is the Euler equation. It is sufficient for
the description of processes in local thermodynamic equilibrium, but
even in moderately rarefied gas flows, stronger deviations from
equilibrium may be expected, making the closing relationship of Euler
equation unsound. The next model in the maximum entropy hierarchy is
the fourth order moment model, which extends the validity of the Euler
system by including some higher order moments such as the heat
flux. Levermore \cite{Levermore} proved that the maximum entropy
moment model has many nice properties, the most important of which are
that it is globally hyperbolic, dissipates the physical entropy, and
that it always corresponds to a non-negative ansatz.  Unfortunately,
for the maximum entropy moment hierarchy, if moments beyond the second
order are included, no closed-form expressions for the closing
relationships are available.  This entails solving an optimization
problem which is usually computationally expensive \cite{Abramov,
  Mcdonald2013, Schaerer}.

On the other hand, the regularized moment method is another approach
to obtain hyperbolic closures for a hierarchy of moment system.
However, for the regularized moment method, closed-form expressions
for closing fluxes are always available, making simulation with the
regularized moment method much more efficient for high order moment
systems. We will briefly review the regularized moment method in the
next section.

\subsection{The regularized moment method}
The globally hyperbolic regularized moment method was first proposed
in \cite{NRxx} to solve the Boltzmann equations, which is also
validated to be efficient to solve the Wigner, Vlasov equations
\cite{Hu2014Simulation}, and also applied to the radiative transfer
problems.  Under the framework of the regularized moment method, the
distribution function $f$ is approximated by series expansion of the
basis functions and a special regularization is adopted to the
globally hyperbolic moment systems.  Precisely, the distribution
function $f(t, \bx, \bv)$ is expanded as
\begin{equation}
  \label{eq:expansion}
  f(t, \bx, \bv) = \sum_{\alpha \in \bbN^3} f_{\alpha}(t, \bx)
  \mH_{\alpha}^{\bu, \theta}(\bv), 
\end{equation}
where $\alpha = (\alpha_1, \alpha_2, \alpha_3)$ is a three-dimensional
multi-index, and the basis functions $\mH_{\alpha}^{\bu,
  \theta}(\bv)$ are defined as 
\begin{equation}
  \label{eq:Hermite}
  \mH_{\alpha}^{\bu, \theta}(\bv) = \prod_{i = 1}^3
  \frac{1}{\sqrt{2\pi}}
  \theta^{-\frac{\alpha_i +1}{2}}\He_{\alpha_i}(\xi_i)\exp\left(
    -\frac{\xi_i^2}{2}\right), \qquad \xi_i = \frac{v_i -
    u_i}{\sqrt{\theta}},
\end{equation}
where $\He_{\alpha_i}$ is the Hermite polynomial. Under this
expansion, the relationship between the moment coefficients
$f_{\alpha}$ and the macroscopic variables are as below
\begin{equation}
  \label{eq:macro_f}
  \begin{gathered}
      f_{0} = \rho(t, \bx), \quad  f_{e_i} = 0, \quad \sum\limits_{j=1}^3
  f_{2e_j} = 0,  \\
  q_i = 2f_{3e_i} + \sum\limits_{d=1}^3 f_{2e_d+e_i}, \quad
  \sigma_{ij} = (1 + \delta_{ij})f_{e_i+e_j}, \qquad i,j=1,2,3.
  \end{gathered}
\end{equation}
With this expansion, the Maxwellian \eqref{eq:Maxwellian} can be
expressed by only the zeroth-order of expansion as 
\begin{equation}
  \label{eq:exp_max}
  \mM = f_0(t, \bx)
  \mH_{0}^{\bu, \theta}(\bv), \qquad  f_0 = \rho.  
\end{equation}

With a truncation of \eqref{eq:expansion}, we can get a finite
approximation of the distribution function. Precisely, we let
$M\geqslant 3$ be a positive number and only the coefficients in the
set $\mathcal{S} = \{f_{\alpha}\}_{|\alpha| \leqslant M}$ are
considered. Thus, the expansion is truncated as
\begin{equation}
  \label{eq:truncation}
  f(t, \bx, \bv) \approx \sum_{|\alpha| \leqslant M} f_{\alpha}(t, \bx)
  \mH_{\alpha}^{\bu, \theta}(\bv). 
\end{equation}
Following the procedure in \cite{NRxx}, we can get the globally
hyperbolic moment systems. Precisely, substituting the expansion
\eqref{eq:Hermite} into the Boltzmann equation \eqref{eq:Boltzmann}
and adopting the regularization in \cite{NRxx}, we can get the
globally hyperbolic moment equations (HME) for the Boltzmann equation,

  \begin{equation}\label{eq:ms}
    \begin{aligned}
     &  \pd{f_{\alpha}}{t} + 
      \sum_{j=1}^D \left( \theta \pd{f_{\alpha - e_j}}{x_j} 
        +u_j\pd{f_{\alpha}}{x_j}
        +(1-\delta_{M,|\alpha|})(\alpha_j + 1) \pd{f_{\alpha+e_j}}{x_j}
      \right) +\sum_{d=1}^D \pd{u_d}{t} f_{\alpha-e_d} \\   
      &+ 
      \sum_{j,d=1}^D \pd{u_d}{x_j} \left( 
        \theta f_{\alpha-e_d-e_j} 
        +u_jf_{\alpha-e_d}
        + (1-\delta_{M,|\alpha|})(\alpha_j + 1) f_{\alpha-e_d+e_j} 
      \right)    + \frac{1}{2} \pd{\theta}{t} \sum_{d=1}^D f_{\alpha-2e_d}  \\ 
   &+ 
      \sum_{j,d=1}^D
      \frac{1}{2}\theta \pd{\theta}{x_j} \left( \theta f_{\alpha-2e_d-e_j} 
        +u_jf_{\alpha-2e_d}
        + (1-\delta_{M,|\alpha|})(\alpha_j + 1) f_{\alpha-2e_d+e_j}
      \right)  = Q_{\alpha},
      \qquad |\alpha|\leqslant M,
    \end{aligned}
  \end{equation}
  where $\delta$ is the Kronecker's delta, and $f_{\alpha}$ is taken
  as zero if any component of $\alpha$ is negative. $Q_{\alpha}$ is
  the expansion of the collision term. If we take the BGK collision
  model \eqref{eq:BGK}, then
  \begin{equation}
    \label{eq:Q_BGK}
    Q^{\rm BGK}_{\alpha} = \left\{
        \begin{array}{cc}
          -\frac{1}{\tau} f_{\alpha}, & |\alpha| \geqslant 2, \\
          0, & \text{otherwise.}
        \end{array}
\right.
\end{equation}
For the Shakhov collision model, the collision coefficients
$Q_{\alpha}$ have the form as
\begin{equation}
  \label{eq:col_shakhov}
  Q^{\rm shakhov}_{\alpha} =  \left\{
        \begin{array}{cc}
          \frac{1}{\tau}\left( \frac{1-Pr}{5}q_j -  f_{\alpha}\right), & \alpha = 2e_i + e_j, \quad
                                        i,j = 1,2,3, \\[2mm]
          Q^{\rm BGK}_{\alpha}, & \text{otherwise.}
        \end{array}
\right.
\end{equation}
Collecting all the independent variables of $f_{\alpha}$, $\bu$ and
$\theta$ as a vector $\bw$, then the system \eqref{eq:ms} can be
written in a quasi-linear form
\begin{equation}\label{eq:ms_quasi}
  {\bD }(\bw)\pd{\bw}{t}+\sum_{j=1}^D\bM_j(\bw)\bD(\bw)\pd{\bw}{x_j}=\bg(\bw), 
\end{equation}
where $\bD\pd{\bw}{t}$ corresponds to the time derivative in
\eqref{eq:ms} while $\bM_j\bD\pd{\bw}{x_j}$ describes the convection
term on the $x_j$ direction, and $\bg$ denotes the right hand side of
\eqref{eq:ms}. The detailed expression can be found in \cite{Qiao},
and the form is put here only for the sake of convenience.  In
\cite{Fan_new, Torrilhon2006}, the moment system \eqref{eq:ms_quasi}
is proved to be globally hyperbolic and a highly efficient numerical
scheme is designed in the framework of the regularized moment method,
where the general finite volume method is used in the spatial
space. The detailed numerical scheme is introduced in the Appendix
\ref{app:numerical}.

%

For the rarefied gas, especially when the Knudsen number is large,
quite a large expansion number $M$ is needed to approximate the
distribution function. In the framework of the regularized moment
method, the expansion number of the distribution function $M$ is
decided according to the largest Knudsen number, which may be quite
expensive for the problems with large variation of the Knudsen
number. It is still a large waste for the area where the Knudsen
number is small. Therefore, the hybrid kinetic/fluid scheme in the
framework of the regularized moment method is proposed in this paper
to reduce the computational cost, with an indicator designed to adjust
the expansion number according to the Knudsen number.


%% file: method.tex
\section{Hybrid regularized moment method}
\label{sec:hybrid}
Many engineering problems involve both the fluid regime and the
kinetic region. For the fluid regimes, the Euler or Navier-Stokes like
fluid description is enough to solve the problem.  However, for the
kinetic region, we should use the kinetic model to accurately
describe the system. The kinetic model is always quite expensive to
solve, even in the framework of the regularized moment
method. Therefore, for the sake of computational efficiency, several
so-called hybrid kinetic/fluid schemes are proposed
\cite{FILBET2018841, Filbet2015} with an automatic
domain-decomposition criterion. In this section, we will also
introduce a hybrid method in detail.

\subsection{Regime indicator and criterion}
In this section, the indicator matrix for the hybrid moment method and
the specific criterion is proposed. We will start from the analysis of
the moment system \eqref{eq:ms} derived in the regularized moment
method. Following the method in \cite{NRxx}, we can deduce the
equations of density, velocity and temperature when setting
$\alpha = 0$, $e_i$ and $2e_i$ respectively as
\begin{equation}
  \label{eq:moment_Euler}
  \begin{gathered}
    \pd{\rho}{t} + \sum\limits_{j=1}^{3}\left(u_j\pd{\rho}{x_j} + \rho
      \pd{u_j}{x_j}\right) = 0, \\
    \rho\left(\pd{u_d}{t} + \sum_{j=1}^3u_j \pd{u_d}{x_j} \right) +
    \sum_{j=1}^3 \pd{p_{jd}}{x_j} = 0,\qquad  d= 1, 2,3,\\
    \rho\left(\pd{\theta}{t} + \sum_{j=1}^3u_j \pd{\theta}{x_j}
    \right) + \frac{2}{3}\sum_{j=1}^3\left(\pd{q_j}{x_j} +
      \sum_{d=1}^3 p_{jd} \pd{u_d}{x_j}\right) = 0.
  \end{gathered}
\end{equation}
When the distribution function $f$ is Maxwellian
\eqref{eq:Maxwellian}, the stress tensor and heat flux are reduced
into
\begin{equation}
  \label{eq:p_euler}
  p_{ij} = \delta_{ij} \rho \theta, \qquad q_i = 0, \qquad i,j = 1,2,3.
\end{equation} 
Then with some rearrangement, we can find that the moment system
\eqref{eq:moment_Euler} is reduced into the Euler equations
\begin{equation}
  \label{eq:Euler}
  \begin{array}{c}
    \left\{\begin{array}{l}
             \pd{\rho}{t} + \nabla_{\bx} \cdot(\rho \bu) = 0, \\[2mm]
             \pd{(\rho \bu)}{t} + \nabla_{\bx} \cdot(\rho \bu \otimes \bu + \rho
             \theta \bI) = \boldsymbol{0}, \\[2mm]
             \pd{E}{t} + \nabla_{\bx} \cdot(\bu(E + \rho \theta)) = 0,
  \end{array}\right.
  \end{array}
\end{equation}
where $E = \frac{1}{2}\rho |\bu|^2 + \frac{3}{2}\rho \theta$ is the
kinetic energy.  Similarly, we can also get the Navier-Stokes equation
from \eqref{eq:moment_Euler}
\begin{equation}
  \label{eq:NS}
  \left\{\begin{array}{l}
           \pd{\rho}{t} + \nabla_{\bx} \cdot(\rho \bu) = 0, \\[2mm]
           \pd{(\rho \bu)}{t} + \nabla_{\bx} \cdot \big(\rho \bu \otimes \bu + \rho
           \theta \bI  \big)=  \nabla_{\bx} \cdot \bsigma, \\[2mm]
           \pd{E}{t} + \nabla_{\bx} \cdot (\bu(E + \rho \theta)) =  
           \nabla_{\bx} \cdot (\bsigma \cdot \bu + \bq),
         \end{array}\right.
\end{equation}
where the relationship between $\bsigma, \bq$ and $\bu, \theta$ will
be given by some kind of Fourier law \cite{Struchtrup2005}, which will
lead to a closed equation system. From the analysis above, we can see
that Euler and Navier-Stokes equations are included in the moment
equations \eqref{eq:ms}. With the same procedure, we can find that the
high order equations similar to Burnett and Super-Burnett equations
can also be deduced.  With the increase of the extension number, the
moment equations are also validated to describe the movements of the
particles in the rarefied gas problems.

Those macroscopic equations can also be deduced from the
Chapman-Enskog expansion of the distribution functions, where the
distribution function is expanded as
 \begin{equation}
  \label{eq:ce-expansion}
  f = \mM \left[1 + \epsilon g^{(1)} + \epsilon^2 g^{(2)} +
    \cdots \right],
\end{equation}
with $g^{(i)}$ the fluctuations which depend on $\rho$, $\bu$,
$\theta$ and their finite derivatives. The Euler equation could be
derived by assuming a zeroth order expansion with respect to
$\epsilon$, in other words, $\hat{f} = \mM$. The first order expansion
with respect to $\epsilon$ would yield Navier-Stokes equations, while
higher order expansions corresponds to Burnett and super-Burnett
equations.

Then we will introduce the indicator matrix for the moment equations,
which could distinguish the fluid region and the kinetic region. We
will first introduce the two variables
\begin{equation}
  \label{eq:AB}
  {\bf A}(\bv) = \int_{\bbR^3}\left(\bxi \otimes \bxi - \frac{|\bxi|^2}{3}{\bf
      I} \right)f(t. \bx, \bv) \dd \bv, \qquad
  {\bf B}(\bxi)
  = \int_{\bbR^3}\frac{1}{2}\big(|\bxi|^2 - 5\big)\bxi f(t,\bx,\bv)
  \dd \bv, \quad \bxi = \frac{\bv - 
    \bu}{\sqrt{\theta}}.
\end{equation}
If the distribution function $f$ is reduced into Maxwellian, we can
get the Euler equations. In this case, it holds that
\begin{equation}
  \label{eq:Euler_result}
  \int_{\bbR^3} {\bf A} f \dd \bv  = \int_{\bbR^3}{\bf B} f \dd \bv =
  0. 
\end{equation}
If $f \approx \mM (1 + \epsilon g^{(1)})$, then we can deduce
Navier-Stokes equations, in which case,
\begin{equation}
  \label{eq:NS_result}
  \int_{\bbR^3} {\bf A} f \dd \bv  \neq 0, \qquad  \int_{\bbR^3}{\bf B}
  f \dd \bv \neq   0. 
\end{equation}
If the distribution function $f$ has the form \eqref{eq:ce-expansion},
\eqref{eq:NS_result} still holds.  Thus, it is natural to use $\bf A$
and $\bf B$ to get the regime indicators in order to discriminate the
particle regimes. Actually, it has already been successfully used in
several works to detect the fluid regions, such as \cite{Filbet2015,
  FILBET2018841}, where a moment realizability matrix is proposed
based on $\bA$ and $\bB$ to describe the domain decomposition criteria
as

\begin{equation}
  \label{eq:real_m}
  {\bf M} := \bm \otimes  \bm   \qquad \bm = \left(1, \bxi,
    \sqrt{\frac{2}{3}}\left(\frac{|\bxi|^2}{2} -  \frac{3}{2}\right)\right),
  \qquad \bxi = \frac{\bv - \bu}{\sqrt{\theta}}.
\end{equation}
In the framework of the regularized moment method, we can find that
the corresponding coefficient matrix of the moment realizability
matrix \eqref{eq:real_m} is
\begin{equation}
  \label{eq:indicator_matrix}
  {\bf I} =\frac{1}{\rho} \int_{\bbR^3} {\bf M} f(t, \bx , \bv)\dd \bv = \left(
    \begin{array}{ccccc}
      1& 0& 0& 0 & 0 \\
      0 & \frac{p_{11}}{\rho \theta} & 0 & 0
                 & \frac{ \sqrt{2}q_1}{ \sqrt{3}\rho
                   \theta^{3/2}} \\
      0 & 0 &  \frac{p_{22}}{\rho \theta} & 0
                 &   \frac{\sqrt{2}q_2}{\sqrt{3}\rho
                   \theta^{3/2}} \\
      0 & 0 & 0 & \frac{p_{33}}{\rho \theta}
                 &\frac{\sqrt{2}q_3}{\sqrt{3}\rho
                   \theta^{3/2} }\\
      0 &\frac{\sqrt{2}q_1}{\sqrt{3}\rho
          \theta^{3/2}} & \frac{\sqrt{2}q_2}{\sqrt{3}\rho
                         \theta^{3/2} } &  \frac{\sqrt{2}q_3}{\sqrt{3}\rho
                                         \theta^{3/2} } & C
    \end{array}
\right),
\end{equation}
where
\begin{equation}
  \label{eq:C}
  C = \frac{1}{\rho} \int_{\bbR^3} \frac{2}{3} \left(\frac{|\bxi|^2}{2}
    - \frac{3}{2}\right)^2 f(t, \bx, \bv) \dd \bv  =
  \frac{4}{\rho\theta^2}\sum_{j=1}^3f_{4e_k} +
  \frac{2}{3\rho\theta^{2}} 
  \sum_{i,j=1}^3(1-\delta_{ij})f_{2e_i+2e_j} 
  + 1.  
\end{equation}

For the moment, we have introduced the moment realizability matrix
made of the expansion coefficients of the distribution function. Then
we will introduce the criteria of the hybrid method.  Before that, the
governing equation in the fluid regime is proposed first.

From the moment systems \eqref{eq:ms}, we can find that the moment
systems contain Euler and Navier-Stokes equations. However, instead of
the Euler and Navier-Stokes equations, the fourth order moment system
is utilized in the fluid region. Our reason for choosing a
fourth-order moment system on the macroscopic subdomains is as
follows: for the fluid regime, we hope our method could cover the
Navier-Stokes limit as it has wider applications than the Euler
equation. At the same time, noting that if we need to switch from the
fluid regime to the kinetic regime, we would need to specify an ansatz
of the distribution function from the few moments which are all the
information of the solution on the fluid subdomain to determine the
data for the kinetic subdomain. We wish to use the maximum entropy
distribution function to recover the kinetic distribution, as it is
the most likely distribution function corresponding to the limited
given information. The lowest order system in the maximum entropy
model hierarchy which includes the Navier-Stokes limit is the
fourth-order moment model, which has a similar hyperbolic form as the
fourth-order moment model in the regularized moment hierarchy, with
only the closure being different. Therefore, choosing the fourth-order
moment model as the model used in the fluid regime allows us to not
only retain the wide applicability of the Navier-Stokes equation, but
also maintain the accuracy of the distribution function provided by
the maximum entropy principle when the regime is changing from fluid
to kinetic.

As a result, we propose a coupling procedure between a fourth-order
moment system and an arbitrarily high order moment system under the
framework of the regularized moment method. For the Euler equations,
we can find that the matrix \eqref{eq:indicator_matrix} is reduced
into the Identity matrix.  When the moment number equals four, we can
get the same number of the moment coefficients least demanded by the
maximum entropy method in \ref{sec:entropy}.  Moreover, the indicator
matrix will remain the same for any expansion number larger than
four. Therefore, the specific criteria and the closure method from the
fluid region to kinetic region are proposed based on the moment
realizability matrix \eqref{eq:real_m} and the maximum entropy method.

\paragraph{From kinetic to fluid}
Supposing the eigenvalues of \eqref{eq:indicator_matrix} other than
$1$ are $\lambda_i, i =1, 2, 3$. We will first propose the criterion
from the kinetic region to the fluid region. There exist a large
number of methods \cite{FILBET2018841, Filbet2015, Levermore1998,
  KOLOBOV2007, Caflisch2016} to decide how far the gas is from the
equilibrium. Here, we adopt the simple comparison between the
eigenvalues $\lambda_i$ and $1$. The detailed criterion is then as
follows: The kinetic description could be reduced into the fluid
description if the two conditions below are satisfied.
\begin{enumerate}
\item the eigenvalues are all close to the eigenvalues of the Euler
  equations which all equal $1$.
  \begin{equation}
  \label{eq:criterion_kf1}
 \max_{i= 1,2,3}|\lambda_i -1| \leqslant \epsilon_1.
\end{equation}
\item the $l_1$ norm of the expansion coefficients of the distribution
  function $f_{\alpha}$ when $|\alpha| > 4$ is small.
  \begin{equation}
  \label{eq:criterion_kf2}
  \sum_{4 <|\alpha|\leqslant M} |f_{\alpha}| \leqslant \epsilon_2. 
\end{equation}
\end{enumerate}

If both criteria are satisfied, the expansion number of the
distribution function is set as $M = 4$. Then the region is changed
from kinetic to fluid regime.

\begin{remark}
  The two parameters $\epsilon_1$ and $\epsilon_2$ are
  problem-dependent. In our examples, they are set as
  $\epsilon_1 = 10^{-3}$ and $\epsilon_2 = 10^{-4}$.
\end{remark}

\paragraph{From fluid to kinetic}
The criterion from fluid region to kinetic region is similar, where
the eigenvalues $\lambda_i$ are also utilized. The detailed criterion
is then as follows: The fluid description should be recovered to
kinetic description if the condition below is satisfied.
\begin{enumerate}
\item at least one of the eigenvalues is far from the eigenvalues of the
  Euler equations.
  \begin{equation}
  \label{eq:criterion_fk1}
  \max_{i=1,2,3}|\lambda_i -1| > \epsilon_1.
\end{equation}
\end{enumerate}

If the criterion is satisfied, the regime should be changed from the
fluid region to the kinetic region.  The expansion number of the
distribution function is then reset to the maximum expansion number.

The only point left here is how to set the expansion coefficients of
the distribution function when $\alpha > 4$. The easiest way is just
to set them to zero. However, the fact that we need to reset the
expansion number implies that the low order expansion and also setting
the high order coefficients to zero may not be accurate enough to
characterize the distribution function. Therefore, in this hybrid
method, the maximum entropy method is applied here to derive the high
order coefficients. This is also one of the reasons that we choose the
fourth-order moment system instead of the Euler or Navier-Stokes
equations in the fluid region.

In the maximum entropy method, we need to first recover a kinetic
distribution function from the low order moments, which is all the
information we have in the fluid regime. We do this by numerically
solving the optimization problem \Cref{eq:maximum_entropy} where the
known moments in the fluid regime (the first four order moments) are
the given constraints. Then the recovered kinetic distribution
function can be computed easily, based on which we can get the high
order expansion coefficients. The algorithm to solve the specific
optimization problem \Cref{eq:maximum_entropy} where the constraints
are exactly the first four order moments is part of the subject of
another paper in preparation, and some details are listed in Appendix
\ref{app:maximum_entropy_solver} for the completeness.  Due to the
extreme complexity of solving the maximum entropy optimization
problem, our current fast algorithm for solving the maximum entropy
distribution function is constrained to 1D problems.  In our numerical
examples, the recovering of the maximum entropy distribution function
from known moments is done for the dimension reduced distribution
function.

For now, we have introduced the criterion to distinguish different
regimes. In the next section, the detailed numerical scheme will be
proposed, which is applied under the framework of the regularized
moment with some adjustment in the interface between the fluid and
kinetic regime.

\subsection{Numerical Schemes}
In the regularized moment method, the standard finite volume
discretization is adopted in the spatial space. Firstly, the standard
splitting method is applied, where the moment equations are split into
the convection part and the collision part. The collision part is the
same as that in the regularized moment method and we refer \cite{NRxx}
for more details. For the convection part, the difference is the case
that the expansion numbers are different in the adjacent cells.  Here,
we take the 1D spatial space as an example. Suppose $\Gamma_h$ to be a
uniform mesh in $\bbR$, and each cell is identified by an index
$j$. For a fixed $x_0\in \bbR$ and $\Delta x > 0$, and the numerical
solution to approximate the distribution function $f$ at $t = t_n$ is
denoted as
\begin{equation}
  \label{eq:f_tn}
  f_h(x, \bv)=  f_j^n(\bv) = \sum_{|\alpha| \leqslant M_j^n} f_{\alpha,
    j}^n\mH^{u_{1,j}^n, \theta_j^n}_{\alpha}(\bv), \qquad x \in \Gamma_j, 
\end{equation}
where $M_j^n$ is the expansion number at cell $j$ and time $t=t_n$
with $u_{1,j}^n$ the first entry of $\bu_j^n$.  Assuming the expansion
number on the cell $j+1$ are $M_{j+1}^n$. If $M_{j+1}^n$ equals
$M_{j}^n$, then the same HLL flux in \cite{NRxx} is utilized
here. However, if they are different, we just take the algorithm below
\begin{enumerate}
  \label{alg:hybrid}
\item Let $M_j= \max(M_{j+1}^n, M_j^n)$;
\item Change the expansion number of $f_{j+1}^n$ and $f_{j}^n$ to
  $M_j$, where the increased expansion coefficients are set as $0$.
  The reset distribution functions are labeled as $f_{j}^{n,\ast}$ and
  $f_{j+1}^{n,\ast}$;
\item Calculate the numerical flux $F_{j+1/2}^{n, \ast}$ using HLL
  scheme with $f_{j}^{n,\ast}$ and $f_{j+1}^{n,\ast}$;
\item For the $j-$th cell, derive the numerical flux $F_{j+1/2}^{n}$
  by setting the expansion number of $F_{j+1/2}^{n,\ast}$ as
  $M_{j}^n$, where the expansion coefficients whose orders are higher
  than $M_j^n$ are simply cut off.
  \end{enumerate}

  For the completeness of this paper, the detailed numerical method
  for the regularized method is listed in Appendix
  \ref{app:numerical}.


%% file: numerical_results.tex

\section{Numerical Experiments}
\label{sec:num}
By now, we are ready to carry out our numerical tests to see the
performance of the hybrid method. In order to reduce the numerical
computation, the dimension reduction method in the microscopic
velocity space is utilized here, which we will introduce in the next
section.

\subsection{Dimension reduction method}

In some benchmark problems, for example the shock wave problem, we
only focus on the spatial 1D and microscopic 3D problems. In this
case, the distribution function has some symmetric properties in the
direction $v_2$ and $v_3$. Therefore, the dimension reduction method
is always adopted to reduce the computational complexity. In this
paper, the dimension reduction method in \cite{Qiao} is utilized for
the regularized moment method, where the BGK and Shakhov collision
terms are discussed. In this paper, the total freedom of the
distribution function in the 3D microscopic velocity space is
described by two distribution functions in 1D microscopic velocity
space. The reduced distribution functions are defined
\begin{equation}
  \label{eq:reduced_f}
  g(t, x, v_1) = \int_{\bbR^2} f(t, x, \bv) \dd v_2 \dd v_3, \qquad h(t, x, v_1)
  = \int_{\bbR^2} \frac{v_2^2 + v_3^2}{2} f(t, x, \bv) \dd v_2 \dd v_3. 
\end{equation}
Assuming the expansion number of the initial distribution function
$f(t, x, \bv)$ is $M$, then $g(t, x, v_1)$ and $h(t, x, v_1)$ can be
approximated as
\begin{equation}
  \label{eq:expansion_reduce}
  g(t, x, v_1) \approx \sum_{\alpha_1 \leqslant M_g} g_{\alpha_1}(t,
  x)\mH_{\alpha_1}^{u_1, \theta}(\xi_1),  \qquad 
  h(t, x, v_1) \approx \sum_{\alpha_1 \leqslant M_h} h_{\alpha_1}(t,
  x)\mH_{\alpha_1}^{u_1, \theta}(\xi_1),  \qquad \xi_1 = \frac{v_1 -
    u_1}{\sqrt{\theta}},
  \end{equation}
  where $M_g = M$ and $M_h = M - 2$ and $\theta$ is the temperature of
  the distribution function $f(t, x, \bv)$. Substituting the expansion
  \eqref{eq:expansion_reduce} into the Boltzmann equation, we can
  derive similar moment equations for $g(t, x, v_1)$ and
  $h(t, x, v_1)$. For the reduced model, the criterion is applied to
  both distribution functions. Precisely, for the distribution
  function $g(t, x, v_1)$, the same criterion in the last section is
  adopted.  For the distribution function $h(t, x, v_1)$, when the
  distribution function is Maxwellian, it equals zero. Then the
  criterion is changed as below: supposing the eigenvalues of the
  indicator matrix get by $g(t, x, v_1)$ is $\lambda_i, i=1,2,3$, the
  criterion in both regions is changed into
  \paragraph{From kinetic to fluid}
  \begin{enumerate}
  \item the eigenvalues are all close to the eigenvalues of the Euler
    equations.
    \begin{equation}
      \label{eq:criterion_kfh1}
      \max_{i= 1,2,3}|\lambda_i -1| \leqslant \epsilon_1.
    \end{equation}
  \item the $l_1$ norm of the expansion coefficients of the
    distribution function $g_{\alpha}$ and $h_{\alpha}$ are small.
  \begin{equation}
  \label{eq:criterion_kfh2}
  \sum_{4 <|\alpha|\leqslant M} |g_{\alpha}| \leqslant \epsilon_2,
  \qquad   \sum_{2 <|\alpha|\leqslant M-2} |h_{\alpha}| \leqslant \epsilon_2. 
\end{equation}
\end{enumerate}

\paragraph{From fluid to kinetic}
\begin{enumerate}
\item at least one of the eigenvalues is far to the eigenvalues of the
  Euler equations.
  \begin{equation}
    \label{eq:criterion_fkh1}
    \max_{i=1,2,3}|\lambda_i -1| > \epsilon_1.
  \end{equation}
\end{enumerate}

Moreover, the maximum entropy method is applied on both distribution
functions to complete the transition from the fluid region to the
kinetic region.

\subsection{Numerical examples}
In this section, several numerical examples are tested to validate the
efficiency of our hybrid algorithm. In all test cases, 1D in spatial
space and 3D in microscopic velocity space problems are studied. The
BKG or Shakhov collision operators are adopted with the relaxation
time corresponding to the Maxwell molecules model with $\omega$. Thus
\eqref{eq:tau0} is reduced into
\begin{equation}
  \label{eq:tau}
  \tau = \frac{K_n}{\rho}.
\end{equation}
The reference solution is computed by the discrete velocity method
with mesh fine enough.  When simulating with HME and our hybrid
method, the linear reconstruction in the spatial space \cite{Cai2018} is
applied to improve the efficiency of the numerical scheme and the time
step is determined by the CFL condition \eqref{eq:CFL} for all the
examples. Computations for the Euler solutions, HME and hybrid method
are performed in sequential code on a computer with Intel i7-7500U CPU
and $16$ GB memory.

\subsubsection{Shock tube problem}
\label{sec:test1}
This section considers the shock tube problem for the BGK collision
term.  The computation domain is infinite but taken to be $[-1.5,1.5]$
in the simulations. As in \cite{Torrilhon2006}, the initial condition
is $f(0,x,\bv) = f_M(\rho, u_1,\theta)$, with
  \begin{equation}
    \rho(0, x) = \left\{
    \begin{array}{l}
      7.0, \quad \text{if}~x < 0, \\
      1.0, \quad \text{if}~x > 0, 
    \end{array}\right.
    \quad
    p(0, x) = \rho \theta = \left\{
      \begin{array}{l}
        7.0, \quad \text{if}~x < 0, \\
        1.0, \quad \text{if}~x > 0, 
      \end{array}\right.
    \quad
    u_x(0,x) = 0.
  \end{equation}
  We consider different Knudsen numbers ranging from the kinetic
  regime to the fluid limit. The numerical solutions get by the hybrid
  method and HME are studied, where the numerical solutions to Euler
  equations and that to Boltzmann equation get by DVM is also computed
  as reference. In all these computations, we take $1000$ evenly
  spaced grids in spatial space for both the hybrid method and HME,
  while the Euler solution is computed on a sufficiently fine mesh to
  ensure convergence. For the full HME system and the hybrid model,
  the maximum expansion order is set as $M = 40$. The ending time is
  $t =0.3$. In our test, the Knudsen number is set as $\Kn = 0.0001$,
  $0.05$, $0.5$ and $5$, where the computational region is changing
  from the fluid regime to the kinetic regime.

\begin{figure}[htbp]
  \centering
  \subfloat[$\rho$ ]{
    \label{fig:rho_test1_0p0001}
  \includegraphics[width=0.48\textwidth]{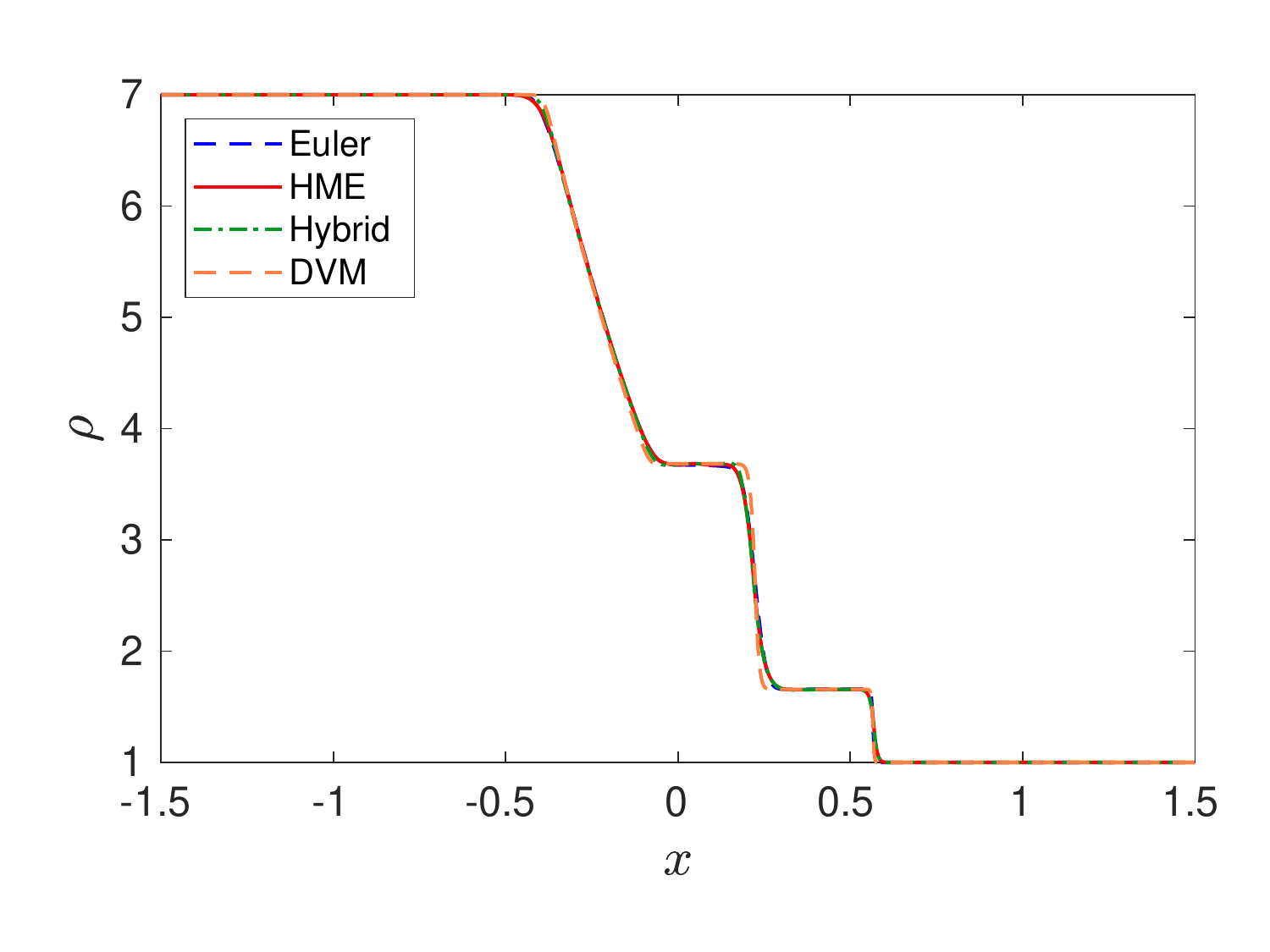}}
\hfill
\subfloat[$u_x$]{\label{fig:u_test1_0p0001}
  \includegraphics[width=0.48\textwidth]{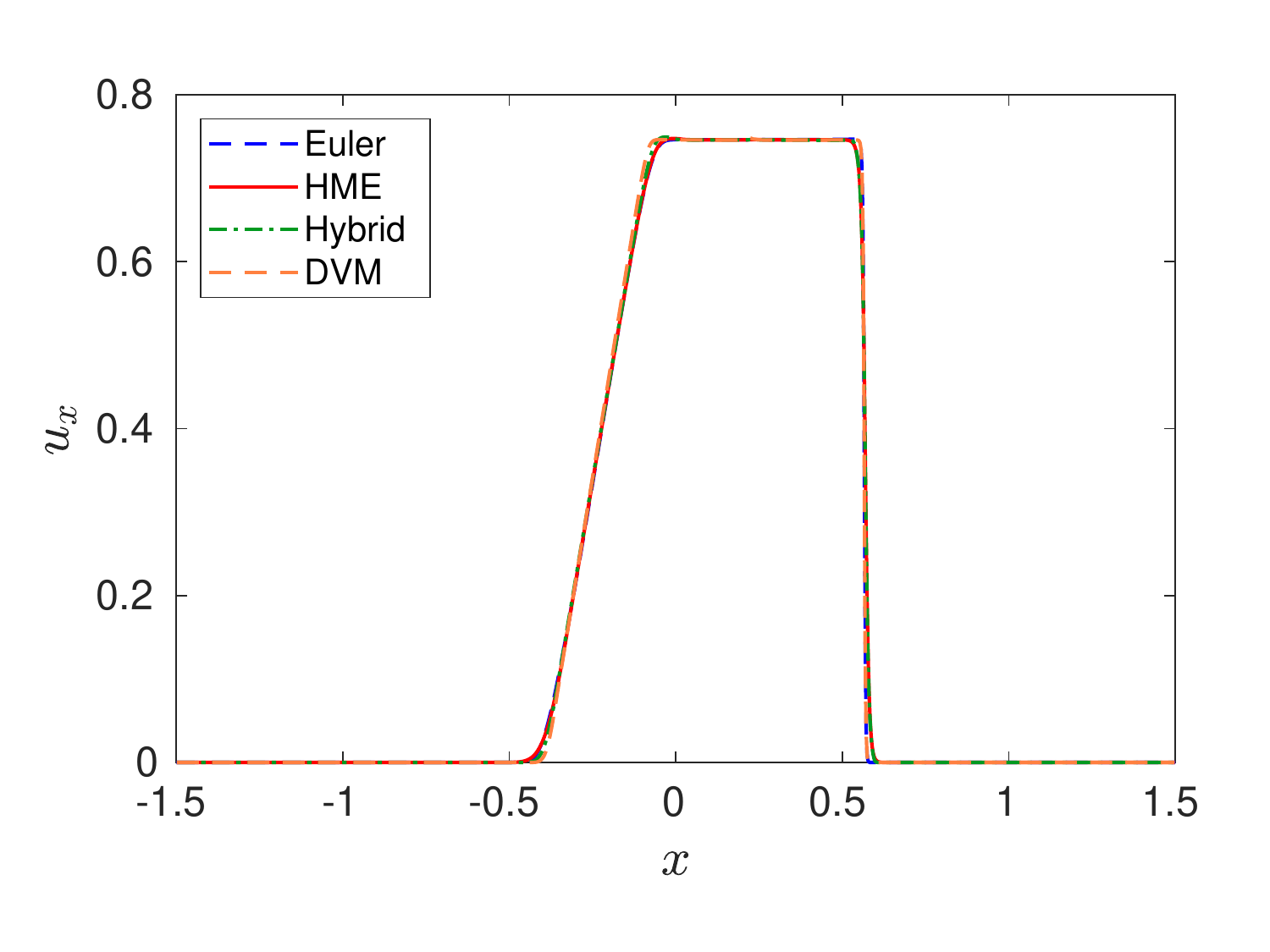}}
  \\
   \subfloat[$\theta$]{\label{fig:theta_test1_0p0001}
  \includegraphics[width=0.48\textwidth]{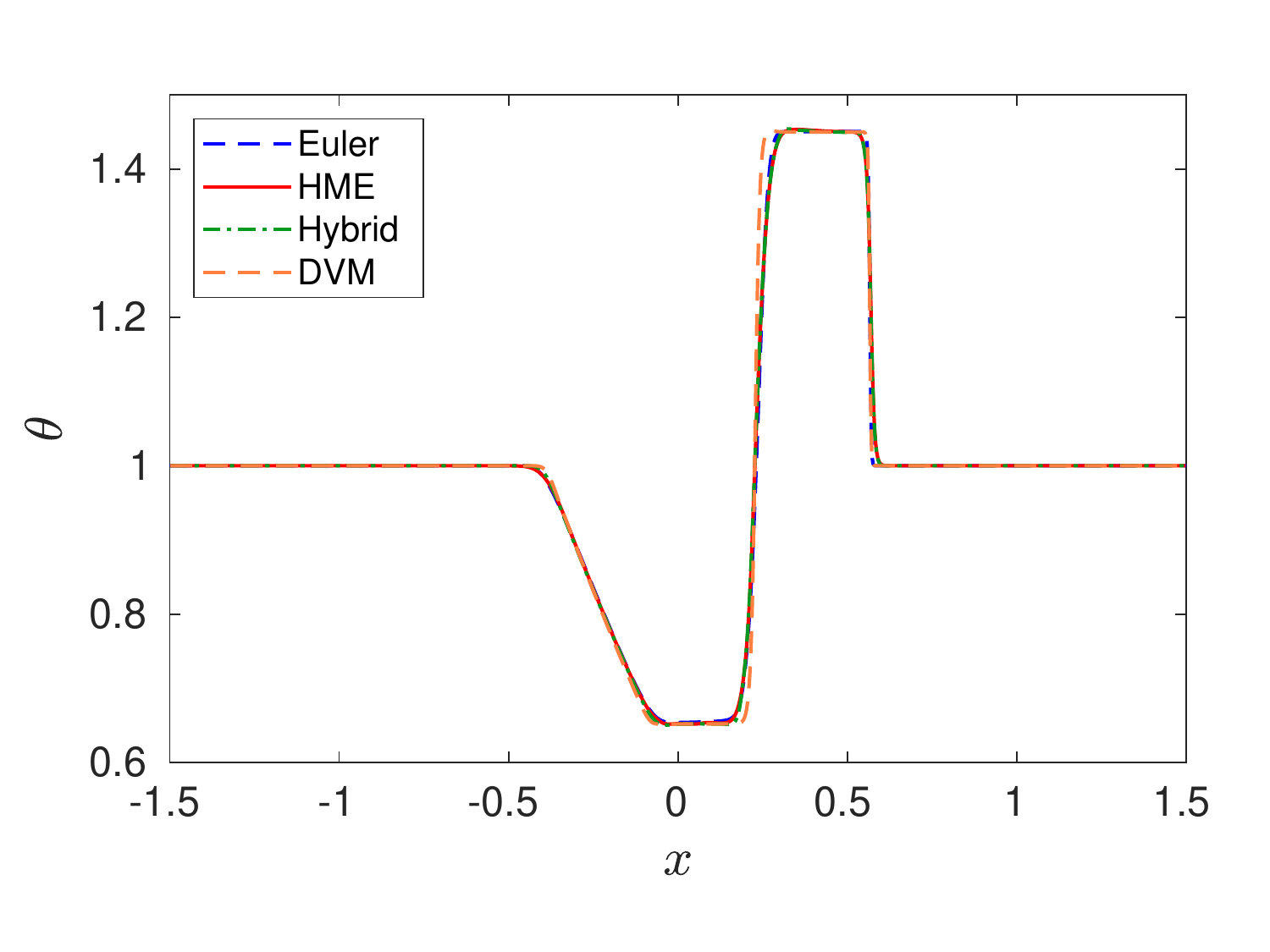}}
  \hfill
  \subfloat[$q_x$]{\label{fig:q_test1_0p0001}
  \includegraphics[width=0.48\textwidth]{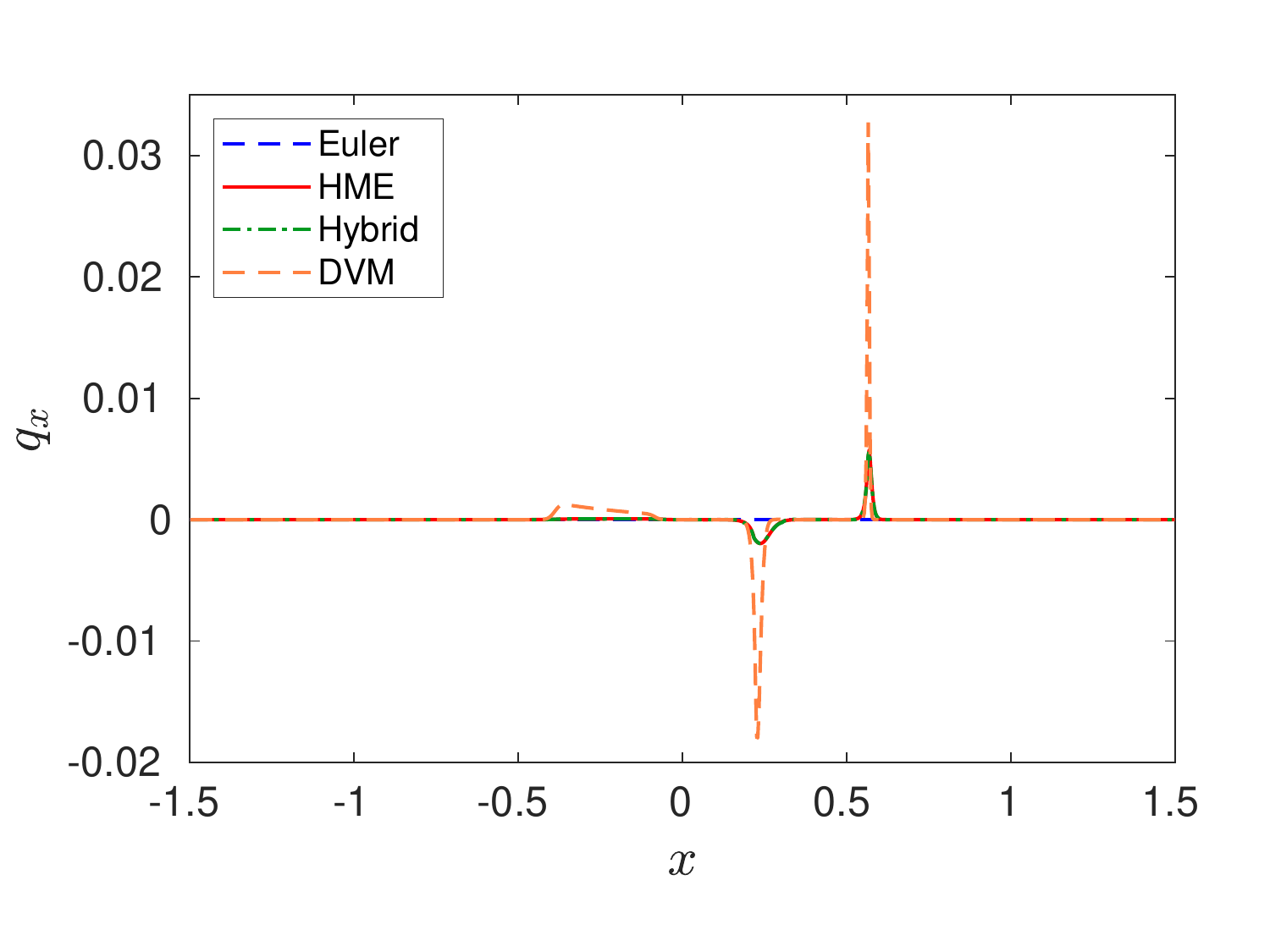}}
\caption{Comparisons of solutions in Sec \ref{sec:test1} for $\rho$,
  $u_x$, $\theta$ and $q_x$ as functions of $x$ for $\Kn = 10^{-4}$.}
  \label{fig:shocktube_test1_0p0001}
\end{figure}

\begin{figure}[htbp]
  \centering
  \subfloat[$\rho$ ]{
    \label{fig:rho_test1_0p05}
  \includegraphics[width=0.48\textwidth]{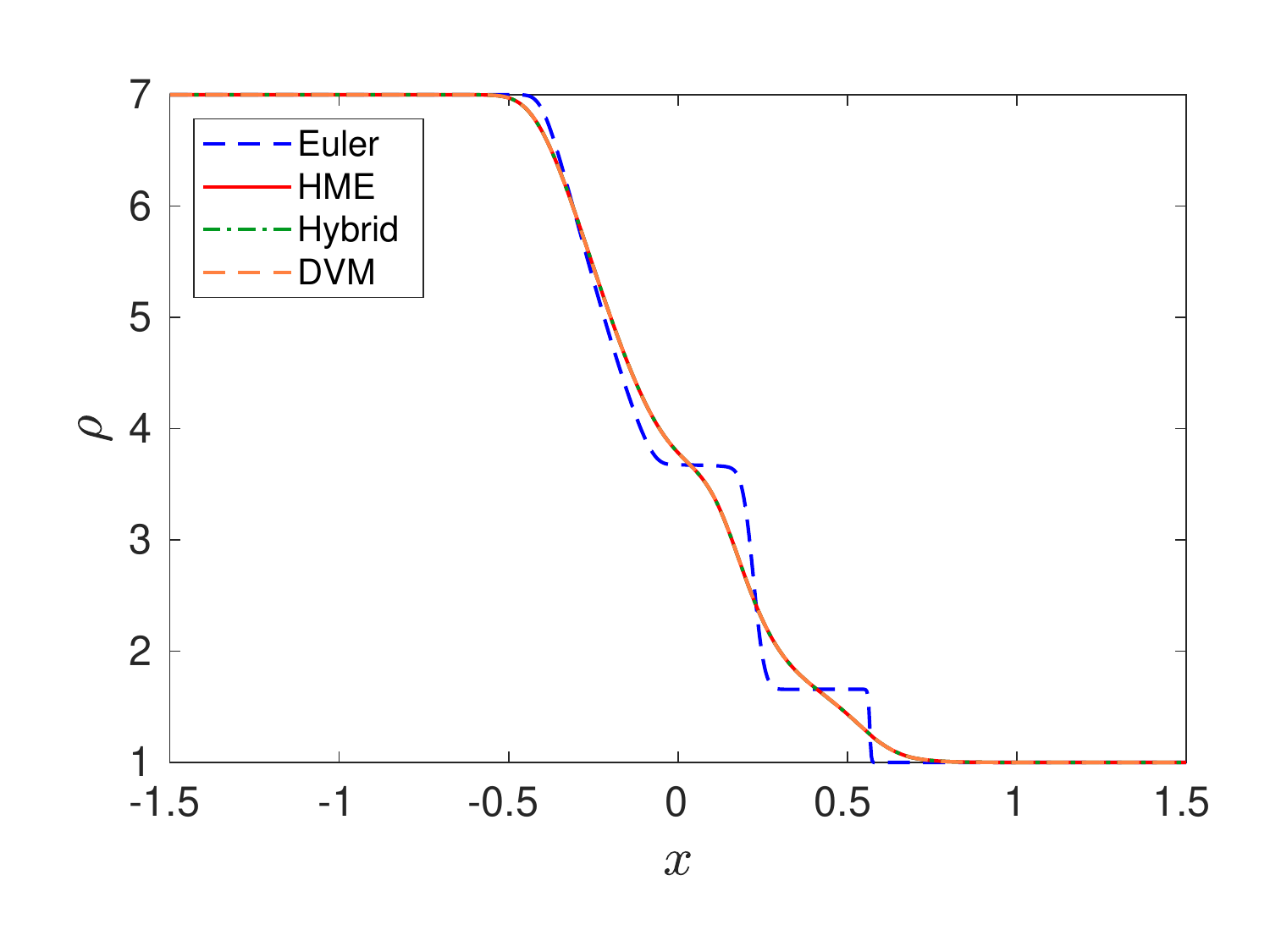}}
\hfill
\subfloat[$u_x$]{\label{fig:u_test1_0p05}
  \includegraphics[width=0.48\textwidth]{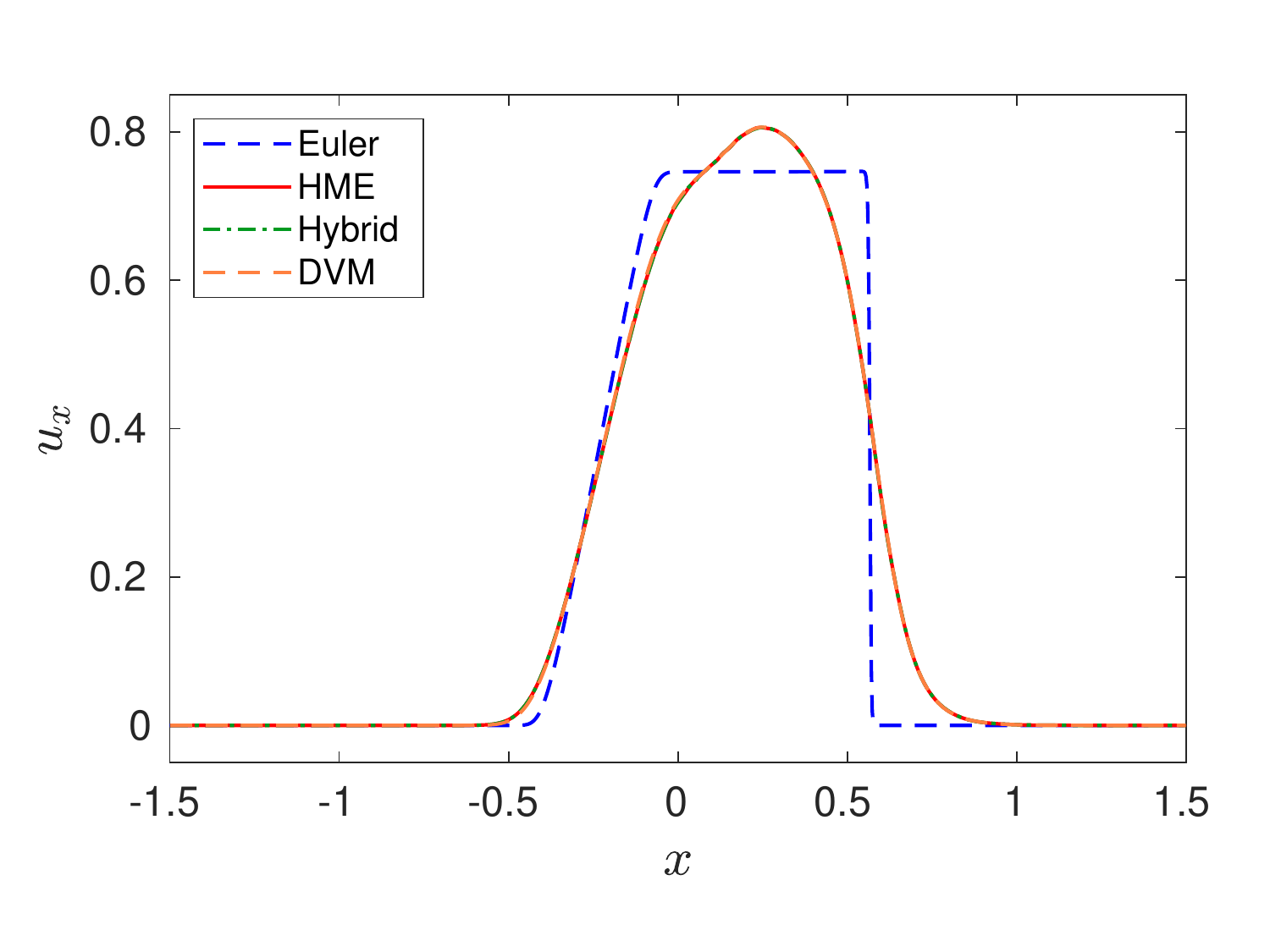}}
  \\
   \subfloat[$\theta$]{\label{fig:theta_test1_0p05}
  \includegraphics[width=0.48\textwidth]{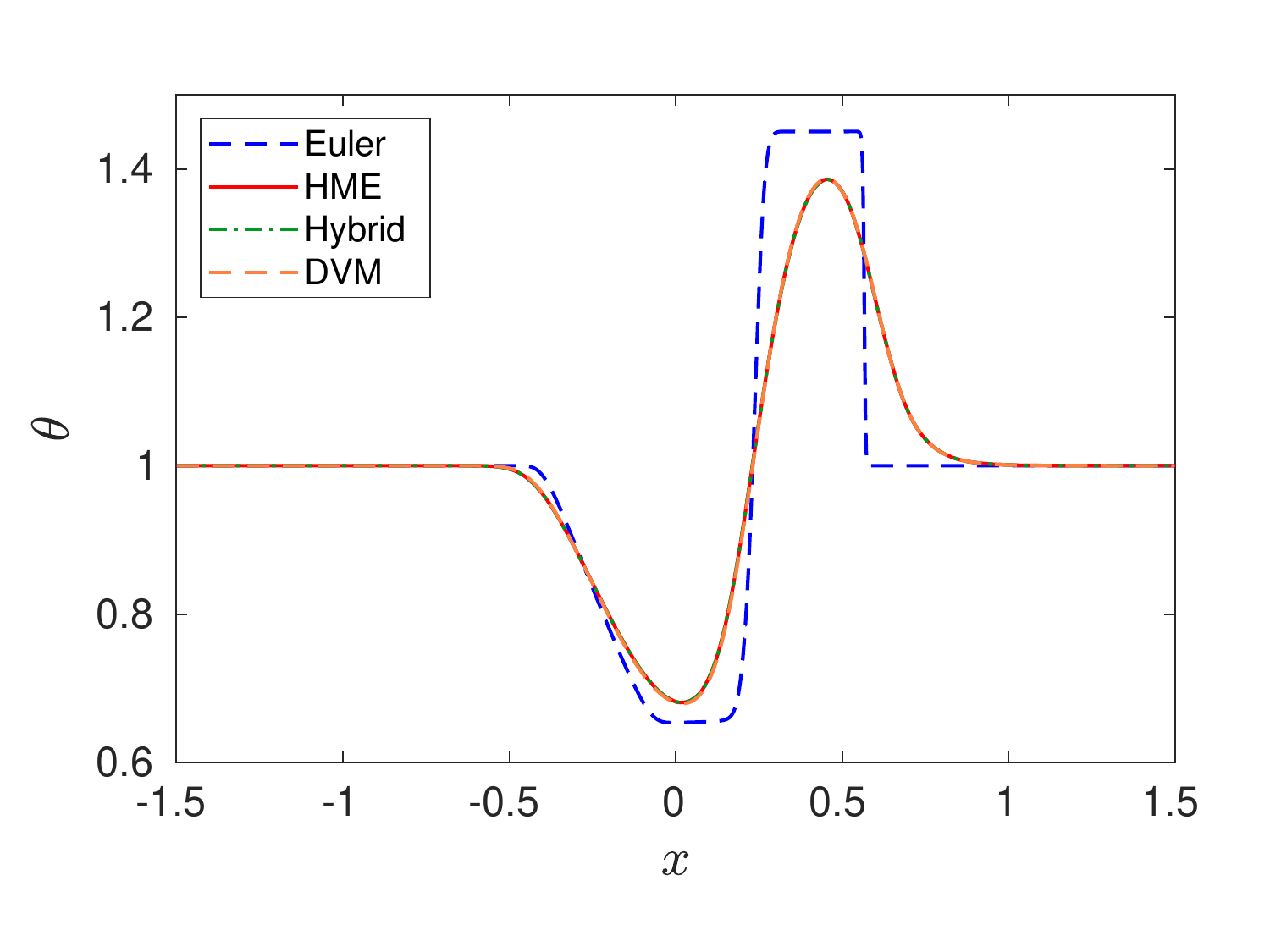}}
  \hfill
  \subfloat[$q_x$]{\label{fig:q_test1_0p05}
  \includegraphics[width=0.48\textwidth]{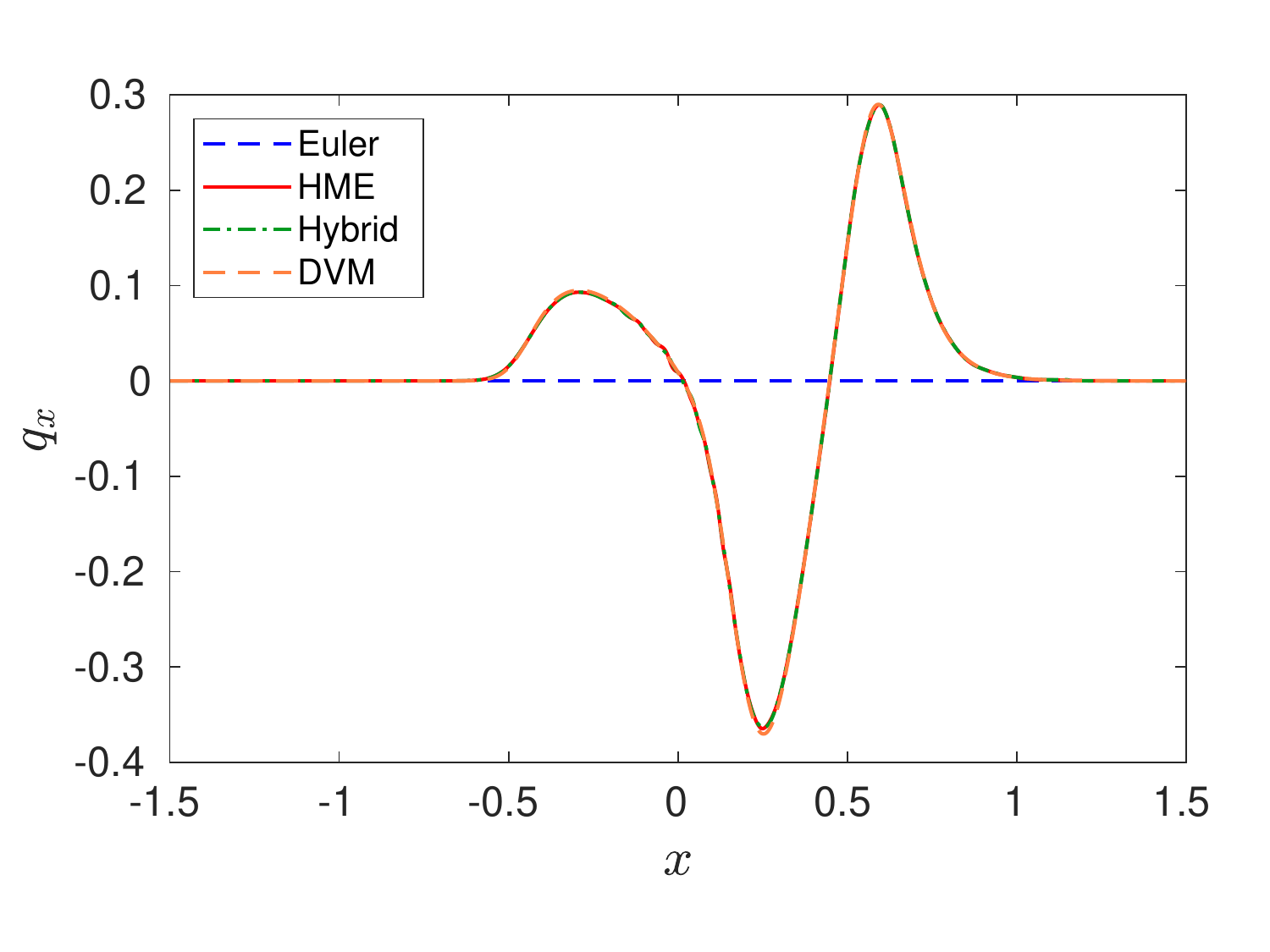}}
\caption{Comparisons of solutions in Sec \ref{sec:test1} for $\rho$,
  $u_x$, $\theta$ and $q_x$ as functions of $x$ for $\Kn = 0.05$.}
  \label{fig:shocktube_test1_0p05}
\end{figure}

\begin{figure}[htbp]
  \centering
  \subfloat[$\rho$ ]{
    \label{fig:rho_test1_0p5}
  \includegraphics[width=0.48\textwidth]{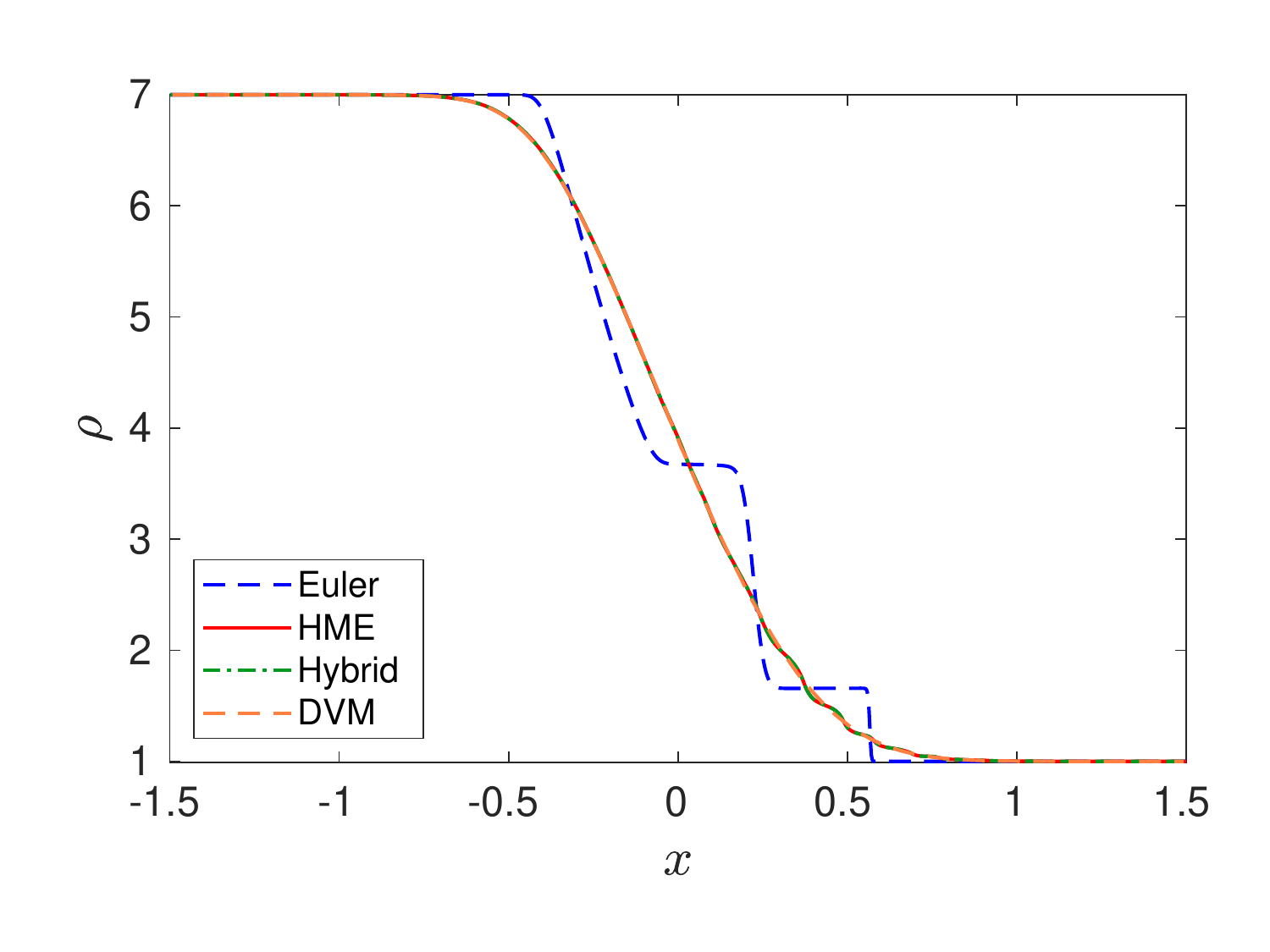}}
\hfill
\subfloat[$u_x$]{\label{fig:u_test1_0p5}
  \includegraphics[width=0.48\textwidth]{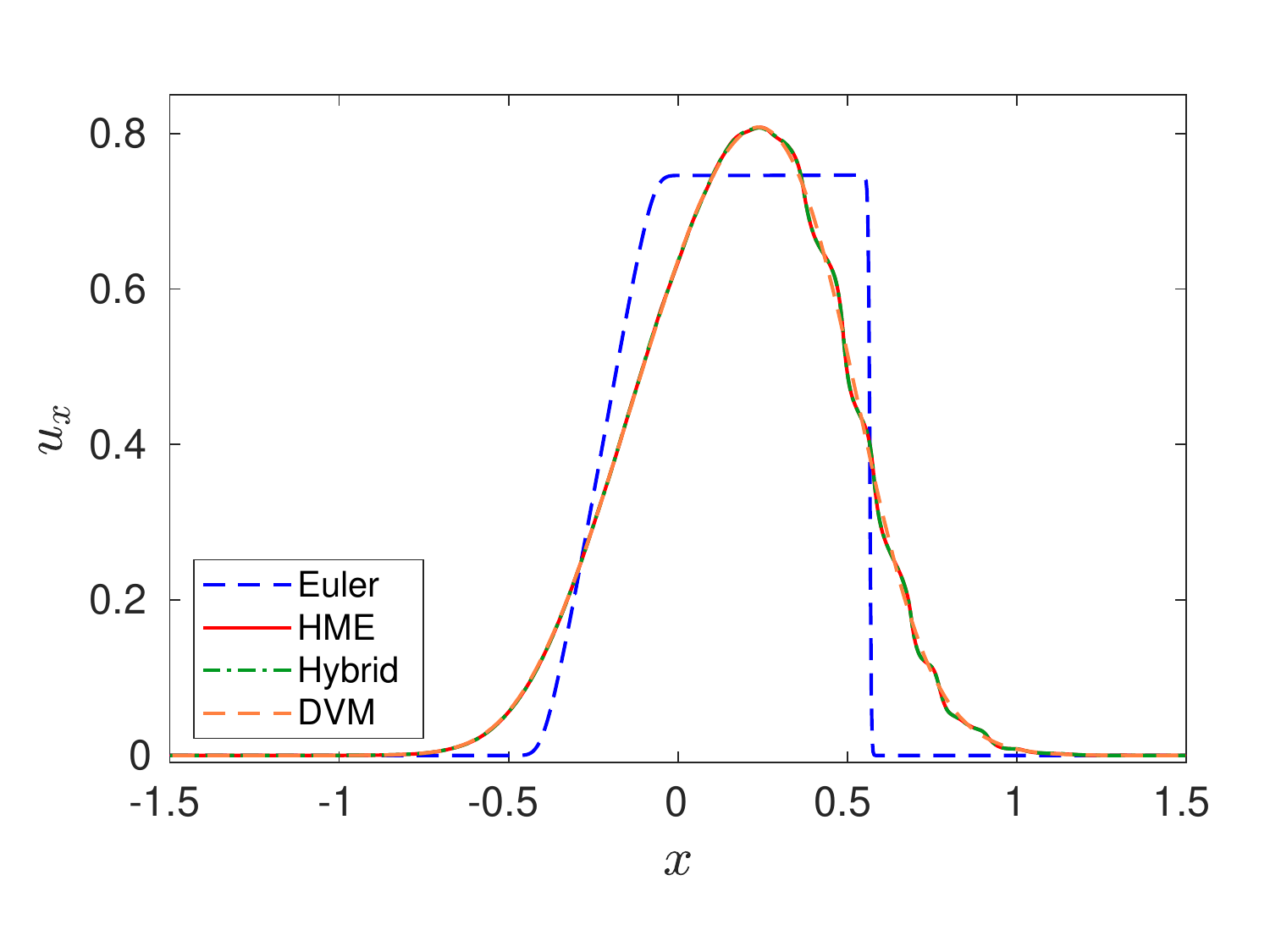}}
  \\
   \subfloat[$\theta$]{\label{fig:theta_test1_0p5}
  \includegraphics[width=0.48\textwidth]{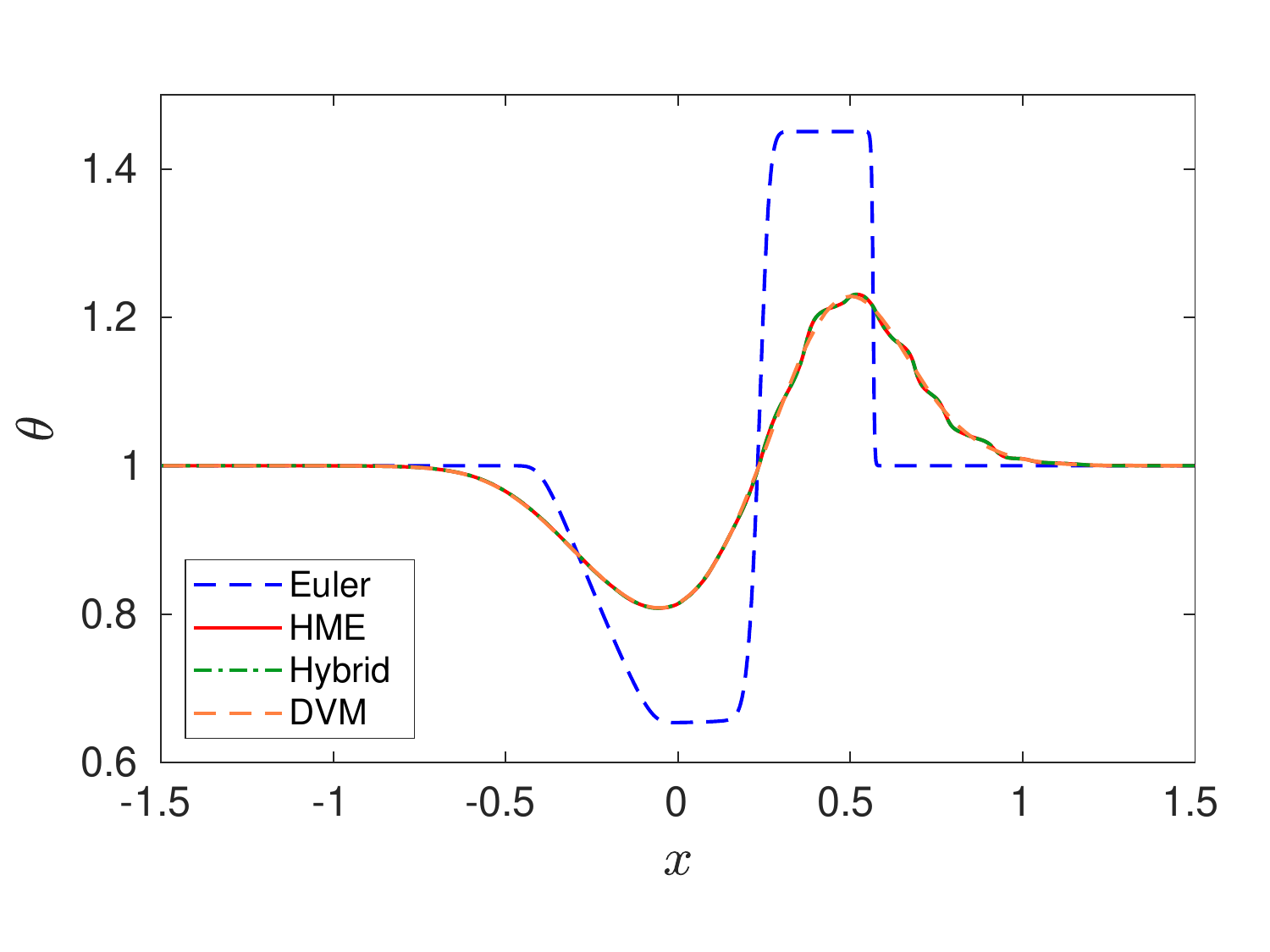}}
  \hfill
  \subfloat[$q_x$]{\label{fig:q_test1_0p5}
  \includegraphics[width=0.48\textwidth]{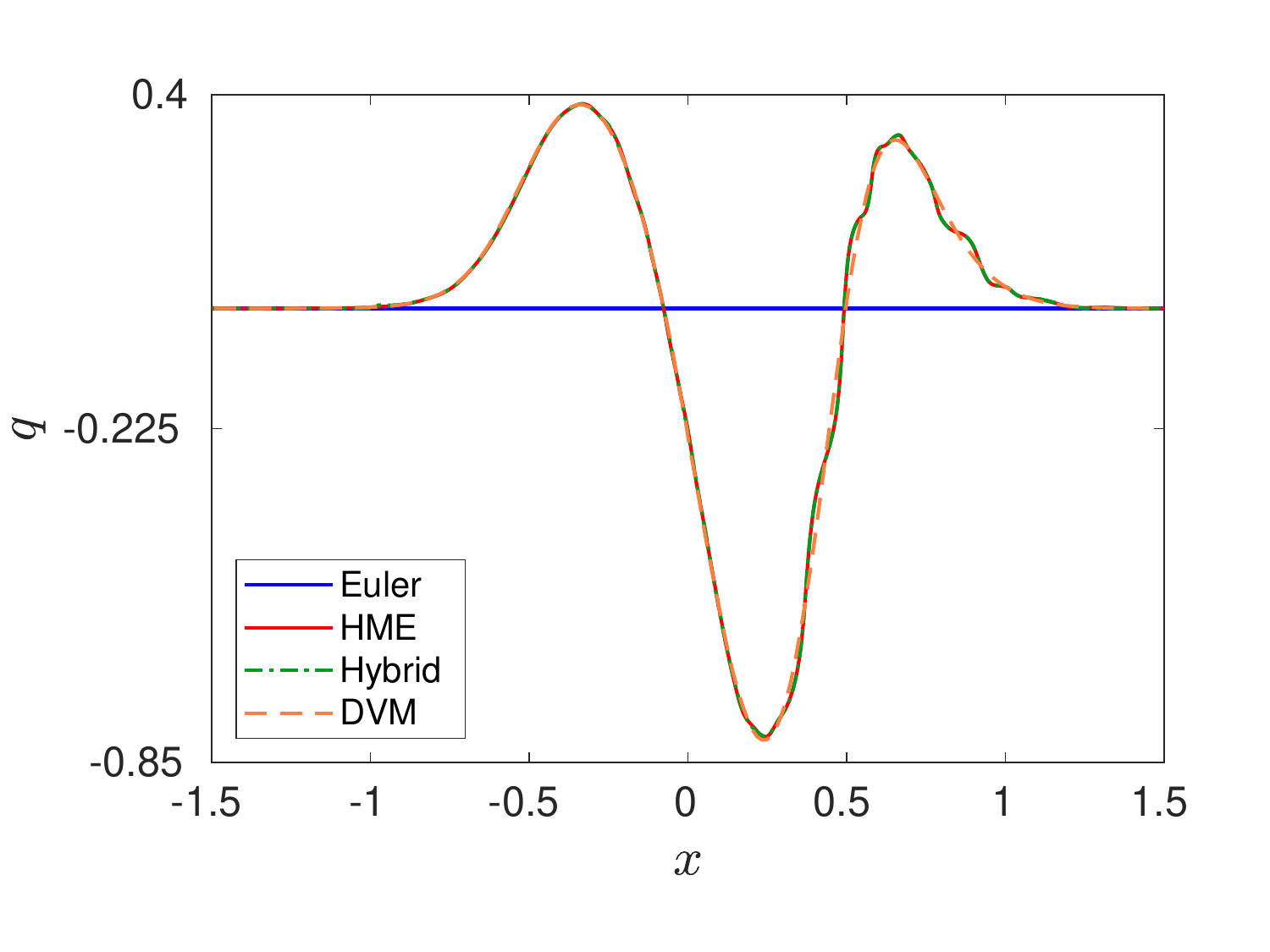}}
\caption{Comparisons of solutions in Sec \ref{sec:test1} for $\rho$,
  $u_x$, $\theta$ and $q_x$ as functions of $x$ for $\Kn = 0.5$.}
  \label{fig:shocktube_test1_0p5}
\end{figure}

\begin{figure}[htbp]
  \centering
  \subfloat[$\rho$ ]{
    \label{fig:rho_test1_5}
    \includegraphics[width=0.48\textwidth]{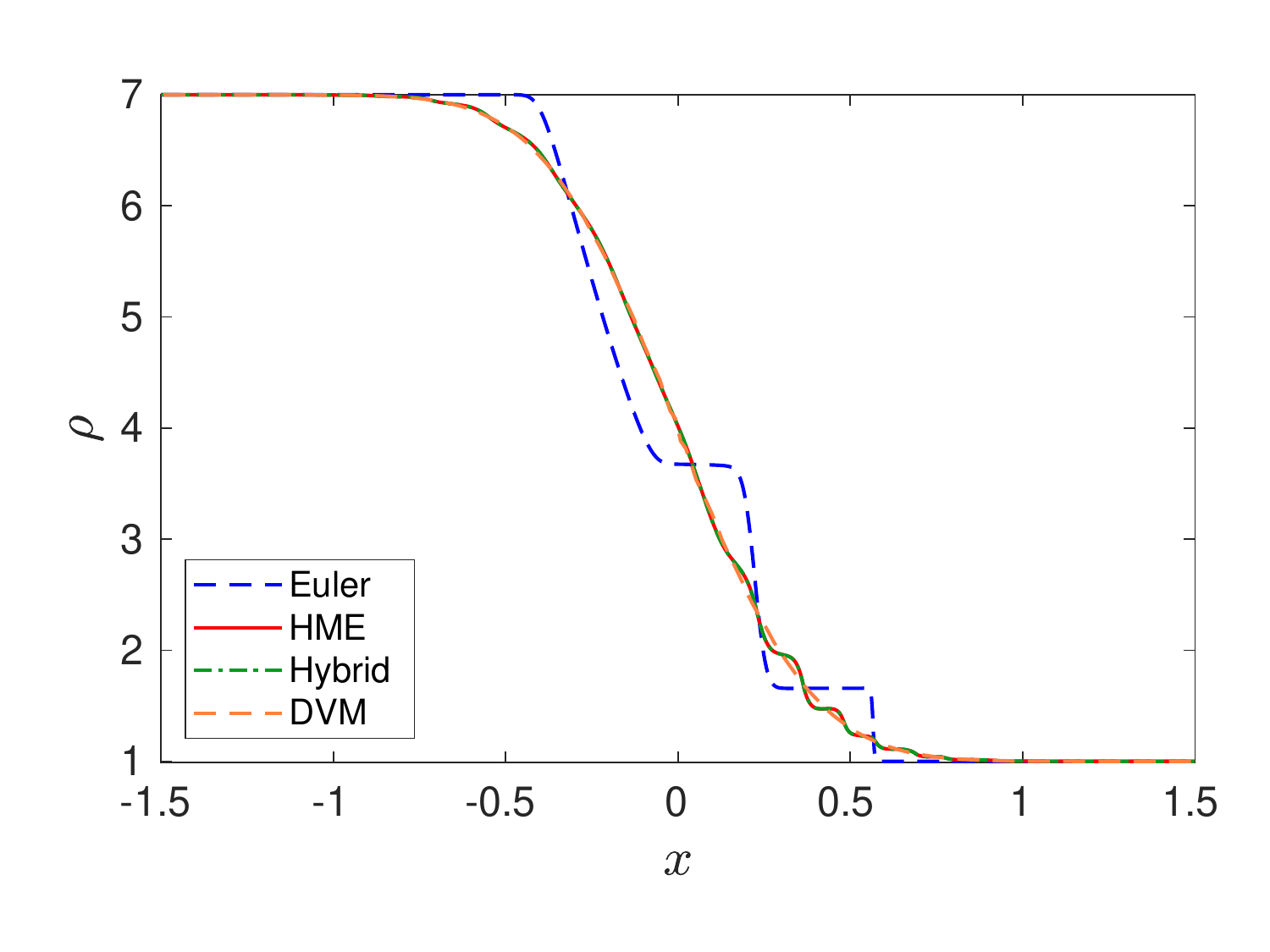}}
  \hfill \subfloat[$u_x$]{\label{fig:u_test1_5}
    \includegraphics[width=0.48\textwidth]{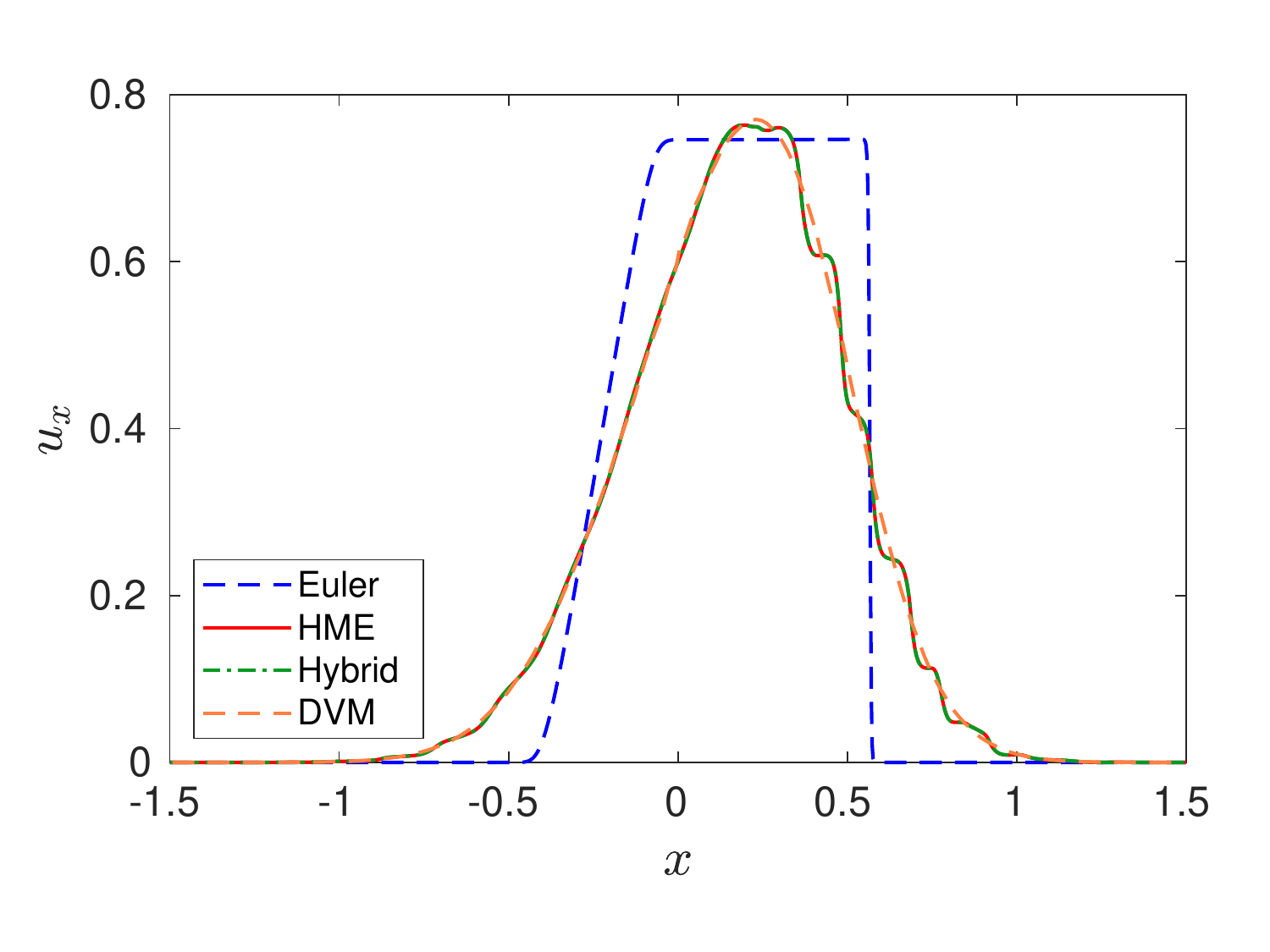}}
  \\
  \subfloat[$\theta$]{\label{fig:theta_test1_5}
    \includegraphics[width=0.48\textwidth]{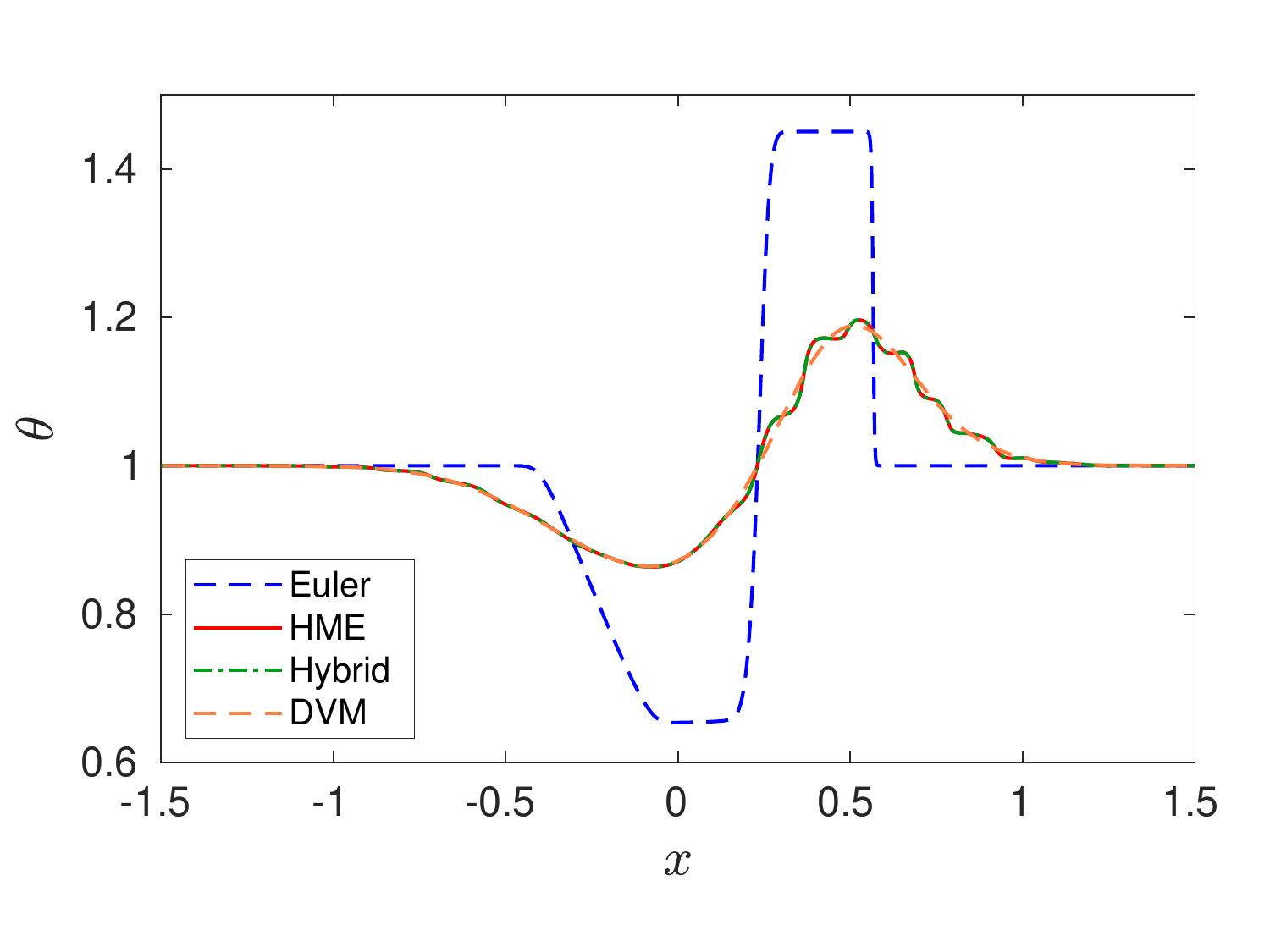}}
  \hfill \subfloat[$q_x$]{\label{fig:q_test1_5}
    \includegraphics[width=0.48\textwidth]{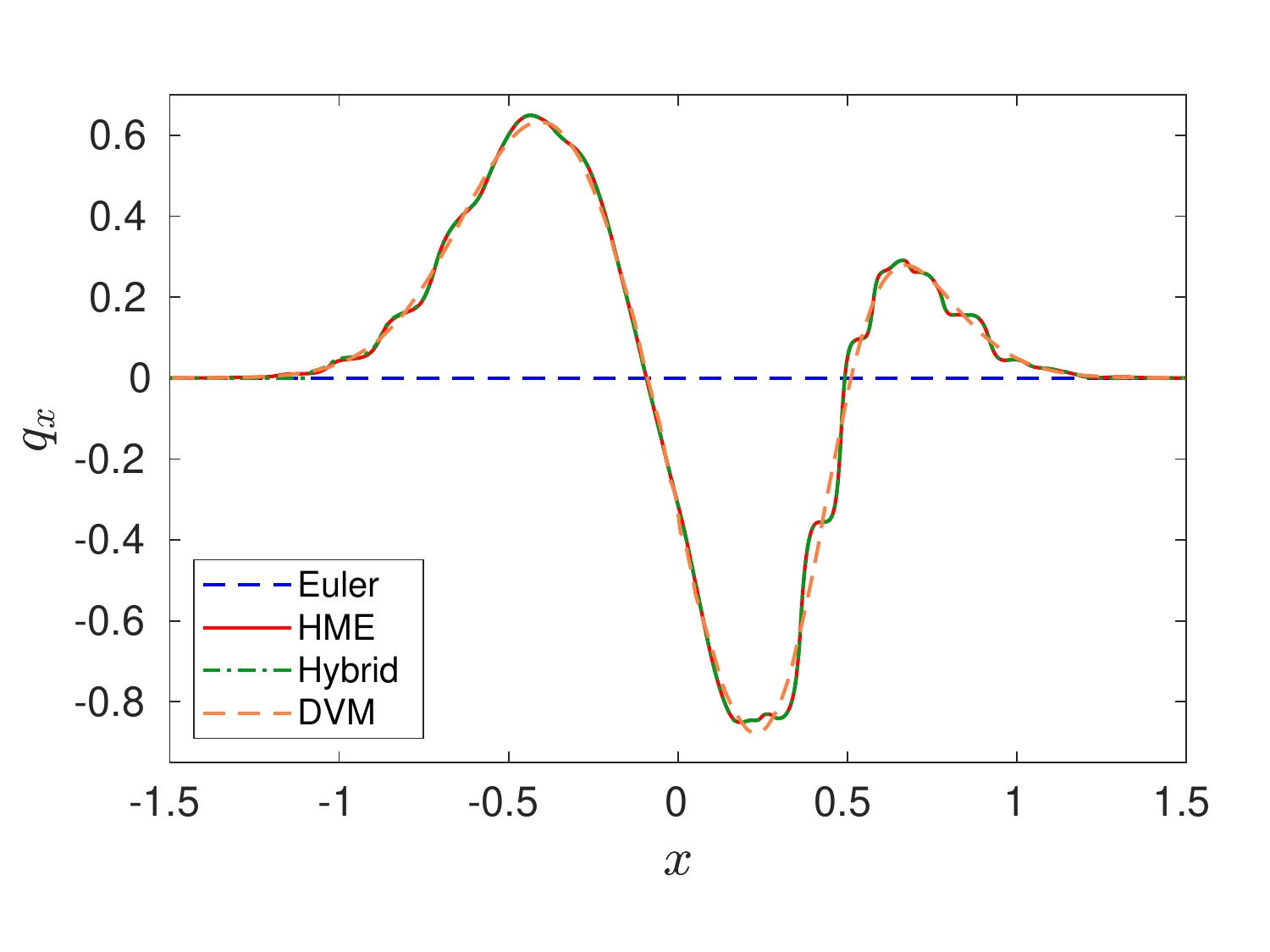}}
  \caption{Comparisons of solutions in Sec \ref{sec:test1} for $\rho$,
    $u_x$, $\theta$ and $q_x$ as functions of $x$ for $\Kn = 5$.}
  \label{fig:shocktube_test1_5}
\end{figure}

  In \Cref{fig:shocktube_test1_0p0001},
  \ref{fig:shocktube_test1_0p05}, \ref{fig:shocktube_test1_0p5} and
  \ref{fig:shocktube_test1_5}, the numerical solutions of density
  $\rho$, macroscopic velocity in the $x-$ direction $u_x$,
  temperature $\theta$ and heat flux in the $x-$ direction $q_x$ for
  the hybrid method and HME with different Knudsen numbers are
  plotted.  From these figures, we can see that for all these Knudsen
  numbers, the numerical solutions get by the hybrid method and HME
  are all on top of each other. Moreover, they are all matching well
  with the reference solutions get by DVM. It means that the hybrid
  method could capture the movement of the particles for all the
  regimes.  Especially, when $\Kn = 0.0001$ where it belongs to the
  fluid regime, it is expected that Euler equations could describe the
  movement of the particles. From \Cref{fig:shocktube_test1_0p0001},
  we can see that the numerical solutions of the hybrid method and HME
  are all consistent with that of the Euler equation. With the
  increasing of the Knudsen number, the numerical solutions are moving
  further and further away from that of the Euler equations. When
  $\Kn = 5$, the particles are in the transitional area, and can only
  be described by the kinetic theory. From
  \Cref{fig:shocktube_test1_5}, we can find that numerical solutions
  of the hybrid method and HME are still almost the same, but all
  these solutions are quite different from that of Euler equations.

\begin{figure}[htbp]
  \centering
  \subfloat[$\Kn = 0.0001$ ]{
    \label{fig:order_test1_0p0001}
    \includegraphics[trim=100 225 100
    250,clip,width=0.48\textwidth]{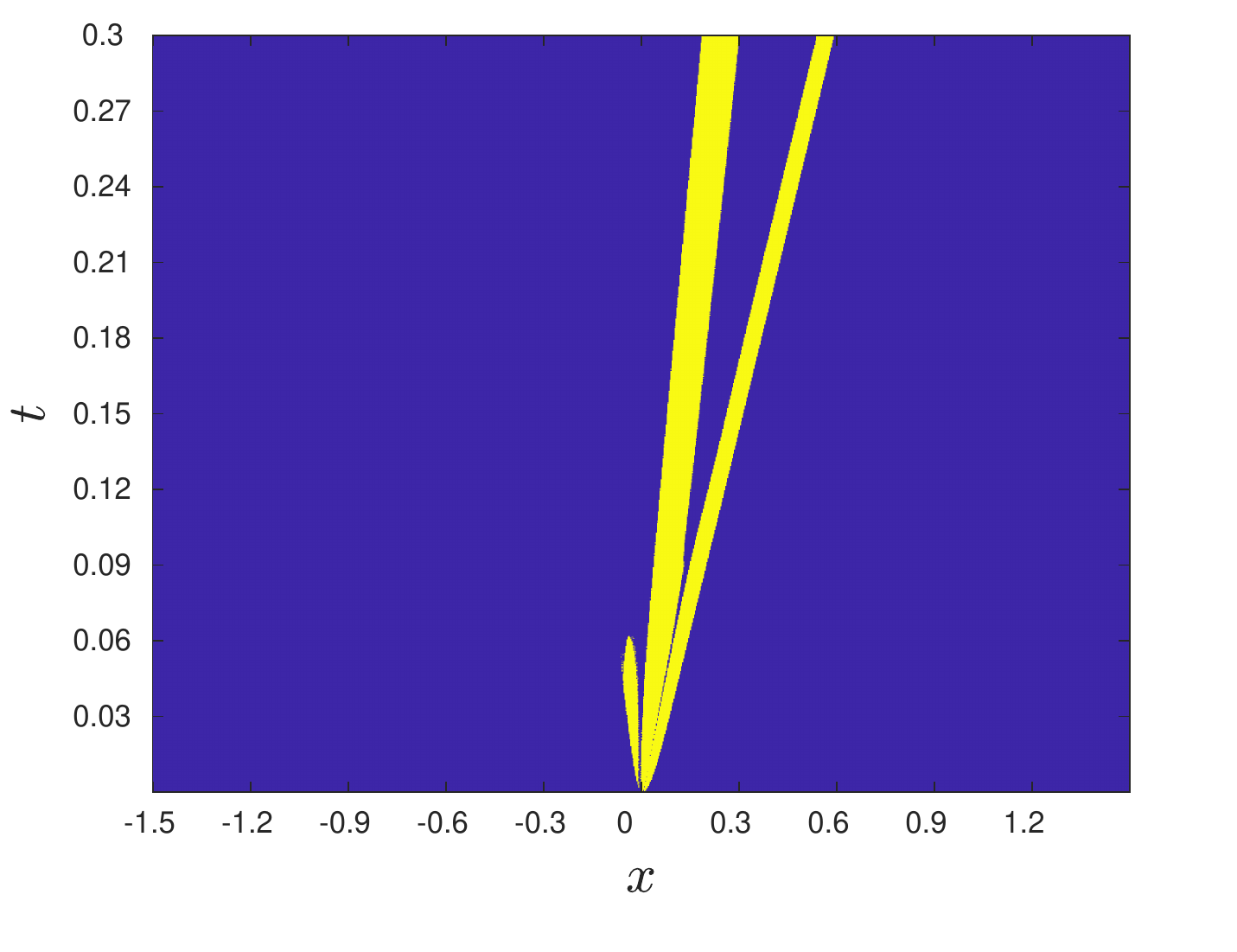}}
  \hfill \subfloat[$\Kn = 0.05$]{\label{fig:order_test1_0p05}
    \includegraphics[trim=100 225 100
    250,clip,width=0.48\textwidth]{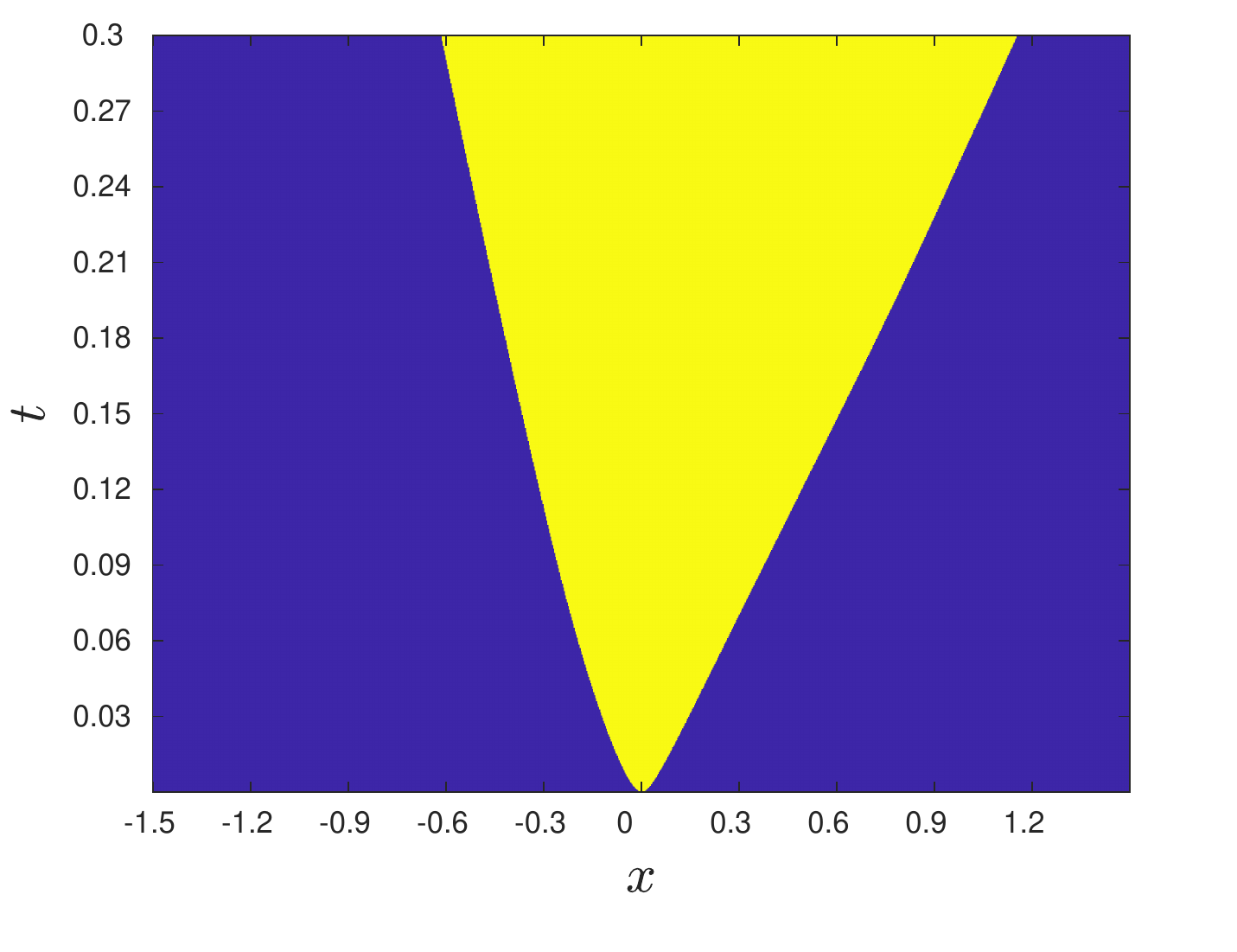}}
  \\
  \subfloat[$\Kn = 0.5$]{\label{fig:order_test1_0p5}
    \includegraphics[trim=100 225 100
    250,clip,width=0.48\textwidth]{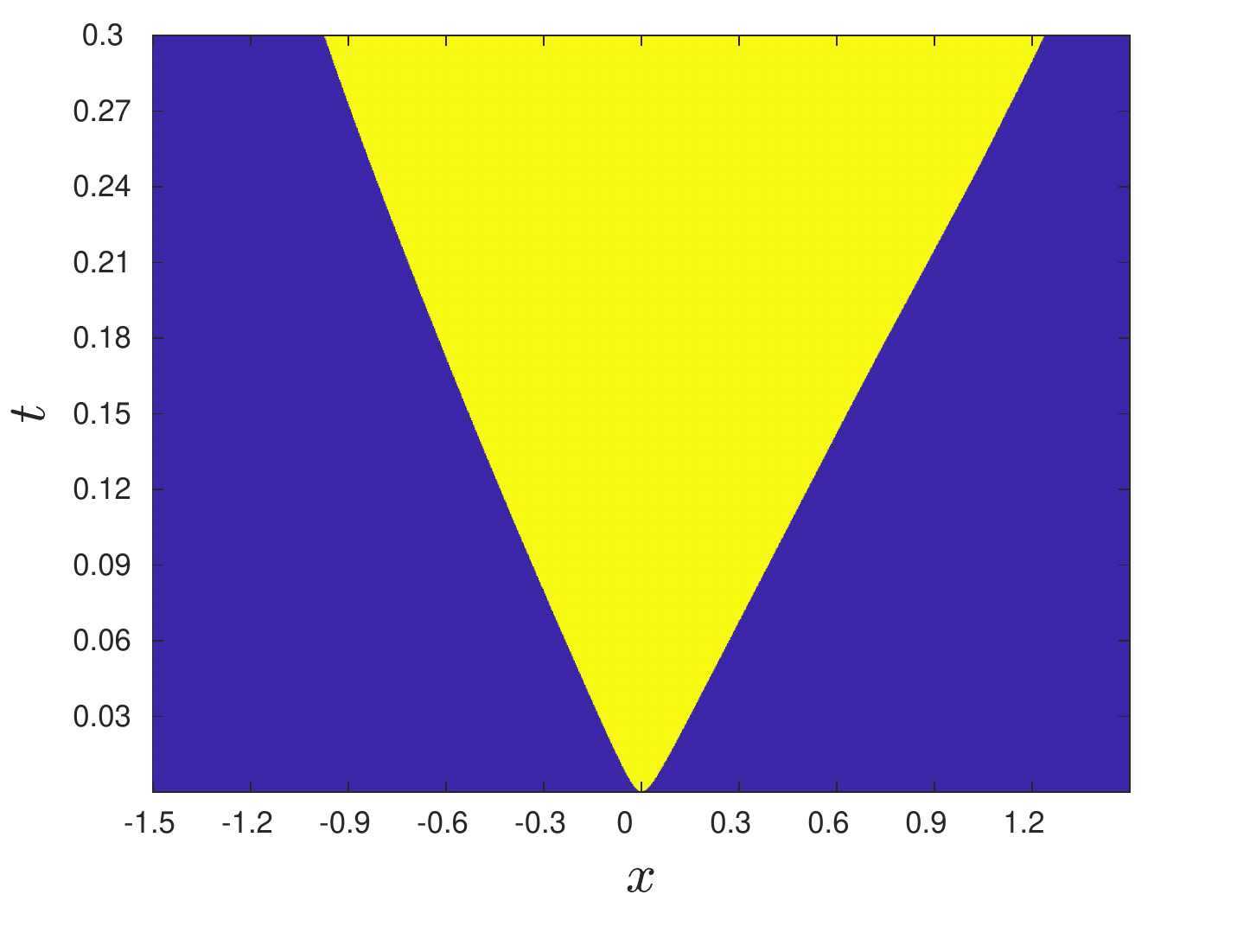}}
  \hfill \subfloat[$\Kn = 5$]{\label{fig:order_test1_5}
    \includegraphics[trim=100 225 100
    250,clip,width=0.48\textwidth]{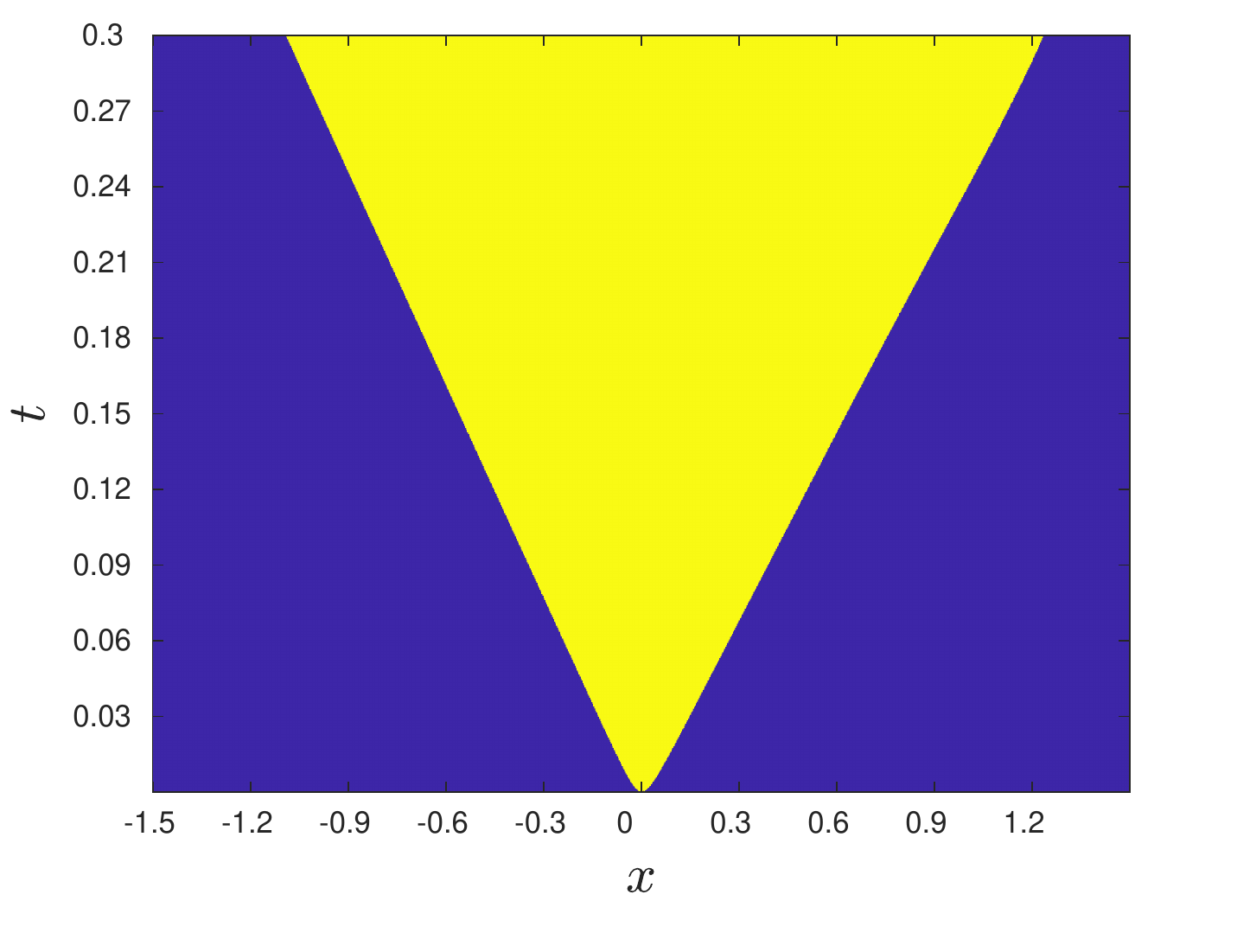}}
  \caption{The maximum order used in each region at each time step for
    different Knudsen numbers in Sec \ref{sec:test1}.}
  \label{fig:shocktube_test1_order}
\end{figure}

In order to show the efficiency of the hybrid method, the time
variation of the expansion order for each grid used by the hybrid
method at each time step is tested. \Cref{fig:shocktube_test1_order}
shows the change of the expansion order, where the blue region is
those with order $4$ and the yellow region is those with order
$40$. From it, we can find that when $\Kn = 0.0001$, most of the
regions is blue, which means that the particles are in the fluid
regime. With the increasing of $\Kn$, the yellow region is becoming
bigger and bigger. From the numerical results in
\Cref{fig:shocktube_test1_0p0001} to \ref{fig:shocktube_test1_5}, we
can find that for all the Knudsen number tested, the middle part of
the spatial area belongs to the kinetic regime, which is also
consistent with the expansion order showed in
\Cref{fig:shocktube_test1_order}, which indicates that this hybrid
method could exactly detect different regimes. The computation time
for the hybrid method and HME for the different Knudsen numbers is
compared in \Cref{tab:timecompare-test1}. We could see that the hybrid
method could save the computational time for all the Knudsen numbers,
which means that the hybrid method works with a high efficiency in all
the regimes. For the fluid regime, the hybrid method saves up to
almost half of the computation cost. Even for $\Kn = 5$ when we are in
a rarefied regime, the hybrid solver reduces the computation cost by
around $30$ percent.

\begin{table}[htbp]
  \centering
    \def\arraystretch{1.5}
    \footnotesize
    \setlength{\tabcolsep}{5mm}{
    \begin{tabular}{ccc}
      $\Kn$  & Full HME  & Hybrid \\
      \toprule
      $0.0001$ & $223$s  & $112$s\\
      \hline
      $0.05$ & $220$s & $143$s \\
      \hline
      $0.5$ & $221$s & $152$s \\
      \hline
      $5$ &  $222$s &  $153$\\
    \end{tabular}}
  \caption{Comparison of computational time of the hybrid method and
    HME with different Knudsen numbers in Sec \ref{sec:test1}.}
  \label{tab:timecompare-test1}
\end{table}

\subsubsection{Shock tube problem with two rarefaction waves}
\label{sec:test2}
The shock tube problem with two rarefaction waves is tested in this
section, a similar set-up could be found in \cite{ALAIA20125217}.
The computation domain is infinite but taken to be $[-1.5, 1.5]$ in
the simulation. The initial condition is
$f(0, x, \bv) = f_M(\rho, u_x, \theta)$, with
\begin{equation}
  \rho(0, x) = 1.0, \quad
  u_x(0, x) = \left\{
    \begin{array}{l}
      -1.0, \quad \text{if}~x<0, \\
      1.0, \quad \text{if}~x > 0,
    \end{array}\right.
    \quad
    \theta = 1.0.
  \end{equation}
  We take $\Kn = 0.01$ and $\Kn = 1$ respectively and compare the
  different solutions for the Shakhov collision operator. In the test,
  the grid size is set as $N = 1000$ and the maximum expansion order
  is $M = 40$.

\begin{figure}[htbp]
  \centering
  \subfloat[$\rho$ ]{
    \label{fig:rho_test2_0p01}
  \includegraphics[width=0.48\textwidth]{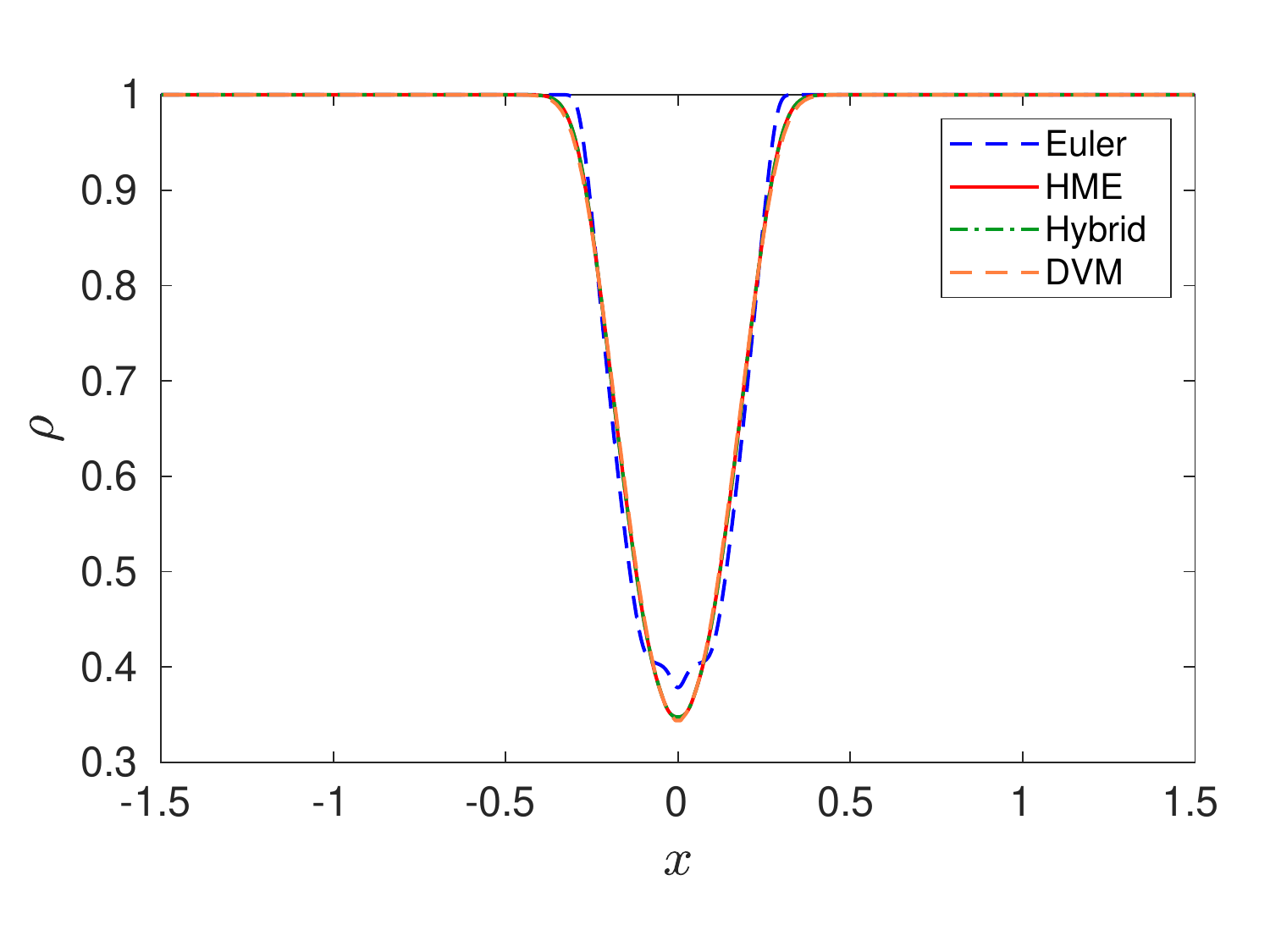}}
\hfill
\subfloat[$u_x$]{\label{fig:u_test2_0p01}
  \includegraphics[width=0.48\textwidth]{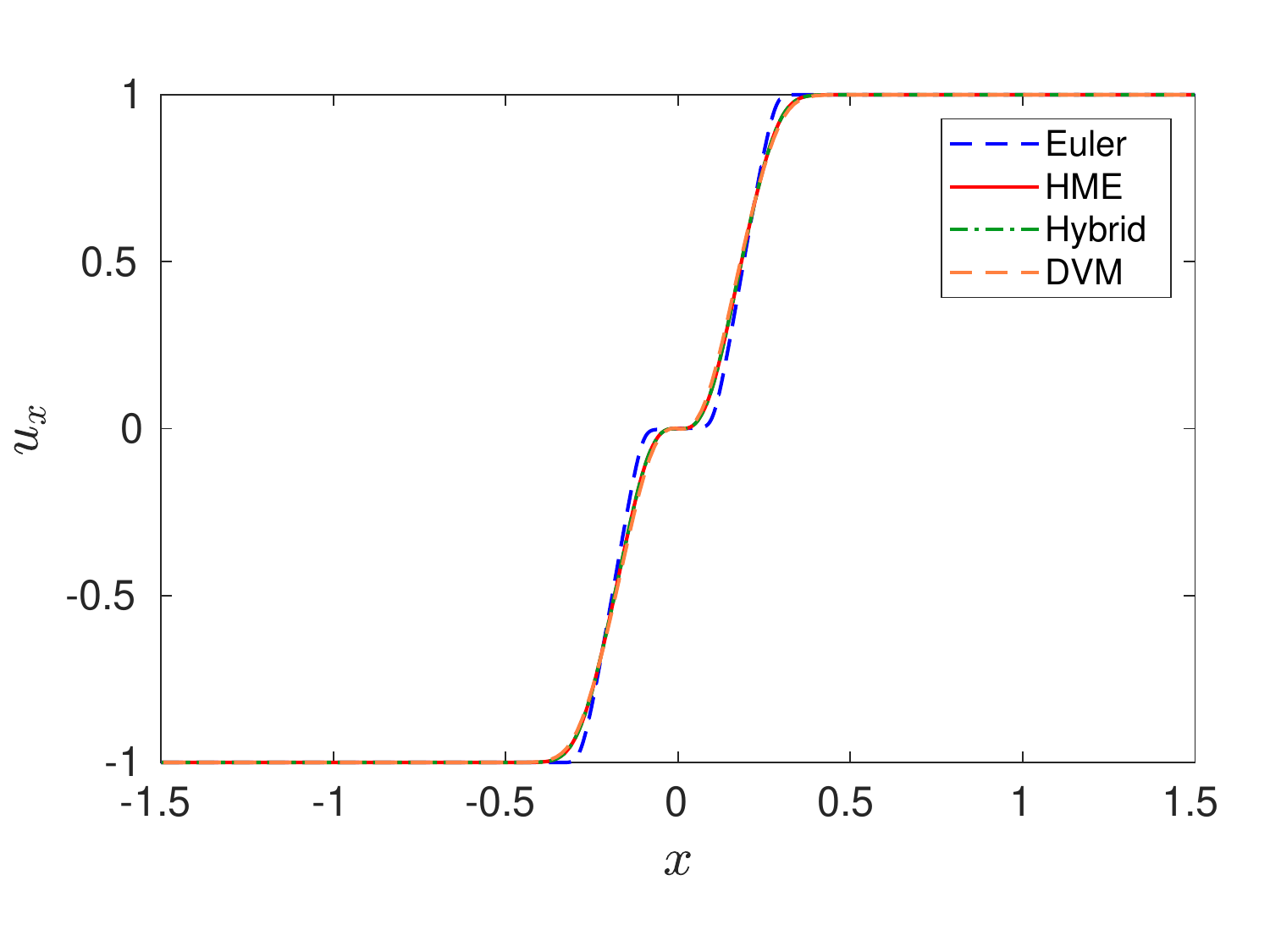}}
  \\
   \subfloat[$\theta$]{\label{fig:theta_test2_0p01}
  \includegraphics[width=0.48\textwidth]{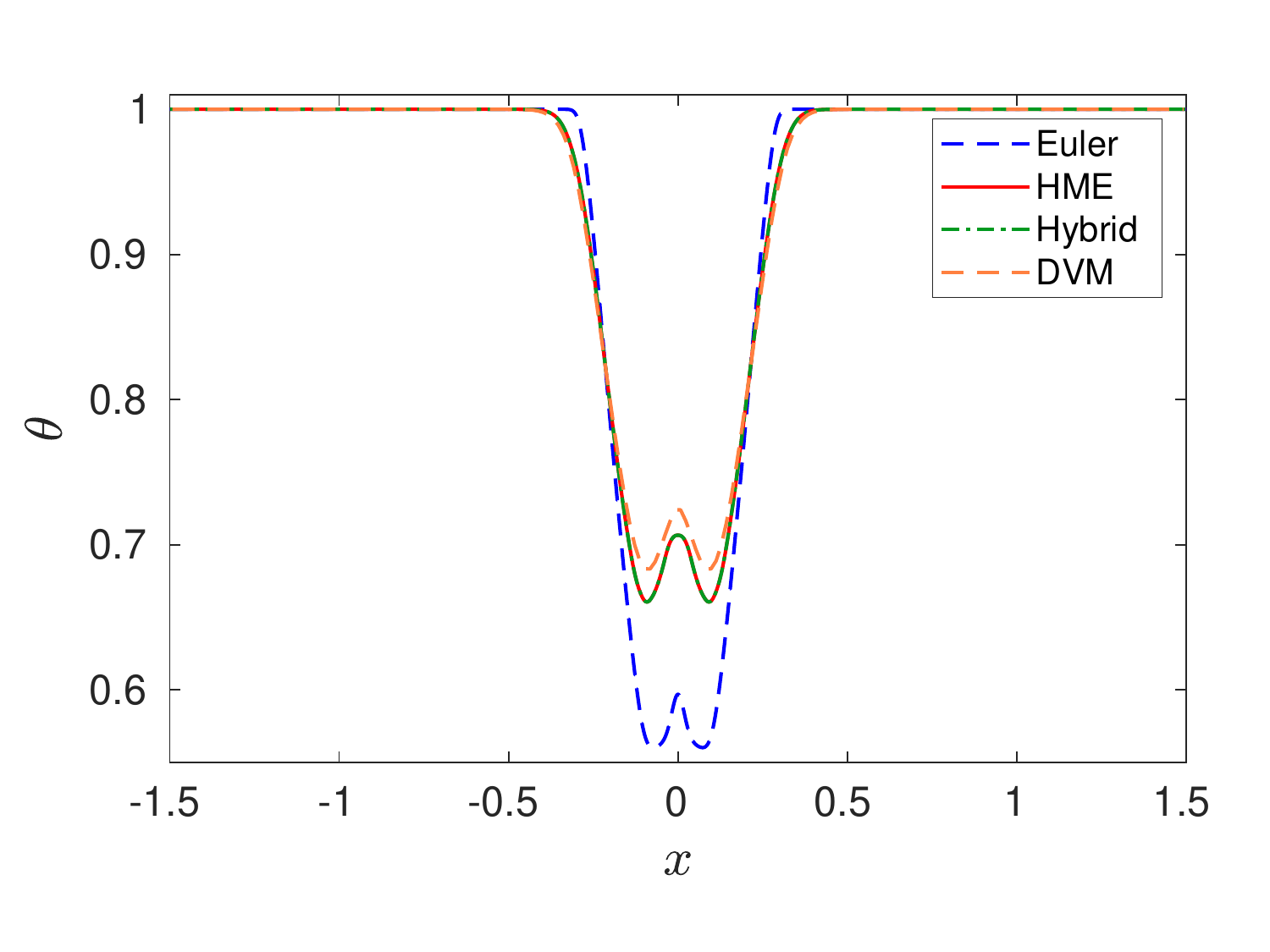}}
  \hfill
  \subfloat[$q_x$]{\label{fig:q_test2_0p01}
  \includegraphics[width=0.48\textwidth]{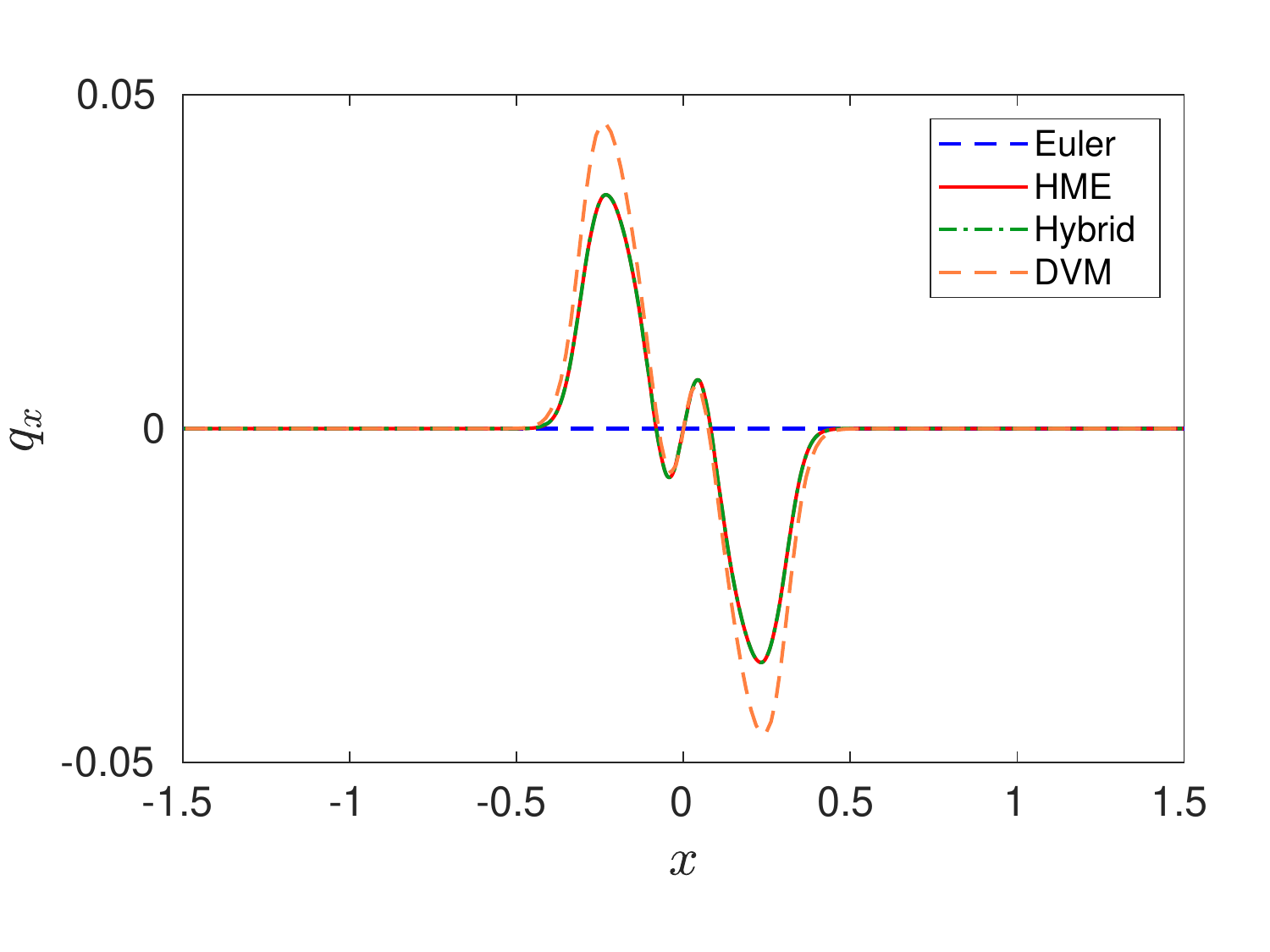}}
\caption{Comparisons of solutions for $\rho$, $u_x$, $\theta$ and
  $q_x$ as functions of $x$ for $\Kn = 0.01$ in Sec \ref{sec:test2}.}
  \label{fig:test2_0p01}
\end{figure}

\begin{figure}[htbp]
  \centering
  \subfloat[$\rho$ ]{
    \label{fig:rho_test2_1}
    \includegraphics[width=0.48\textwidth]{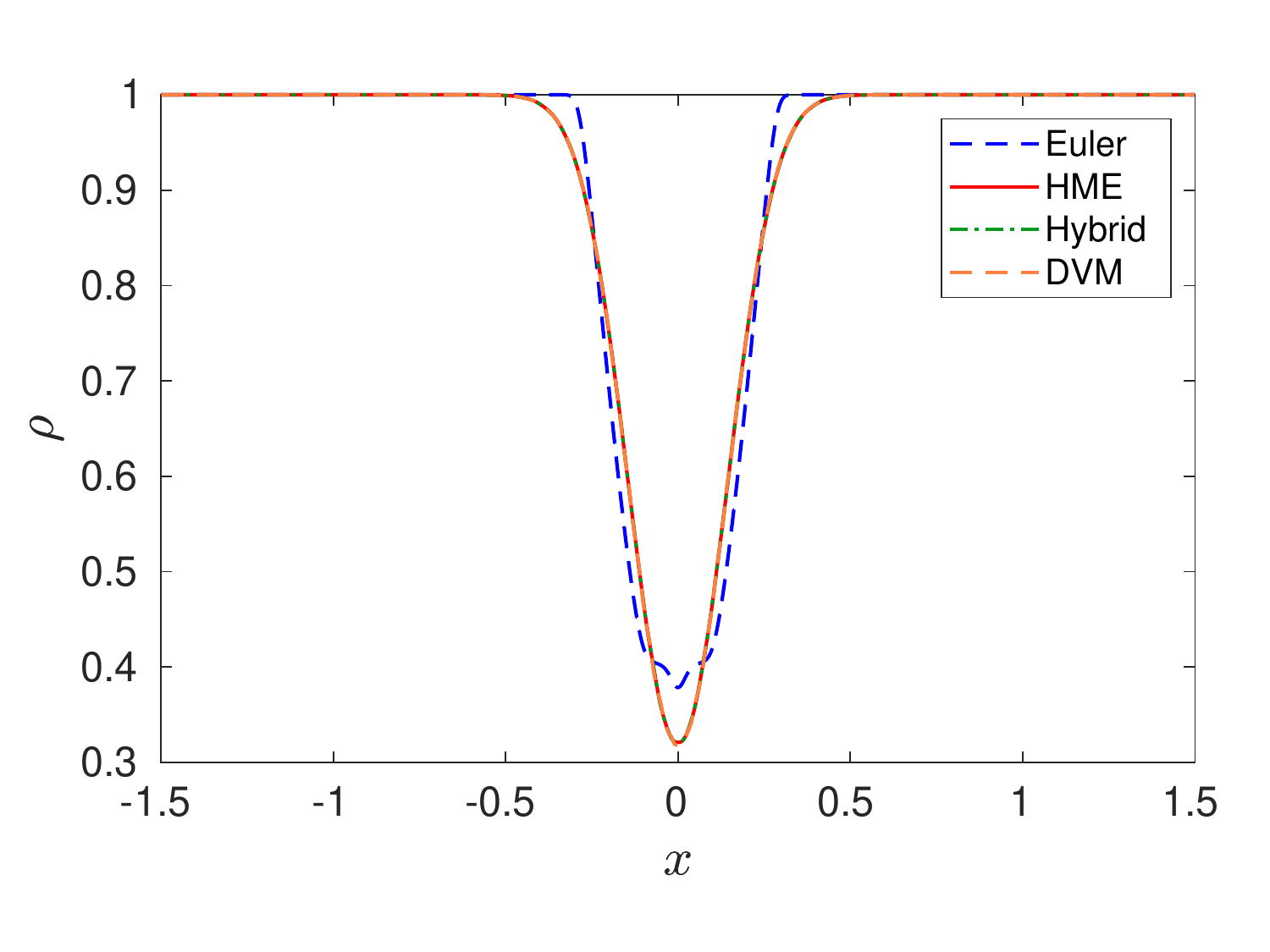}}
  \hfill \subfloat[$u_x$]{\label{fig:u_test2_1}
    \includegraphics[width=0.48\textwidth]{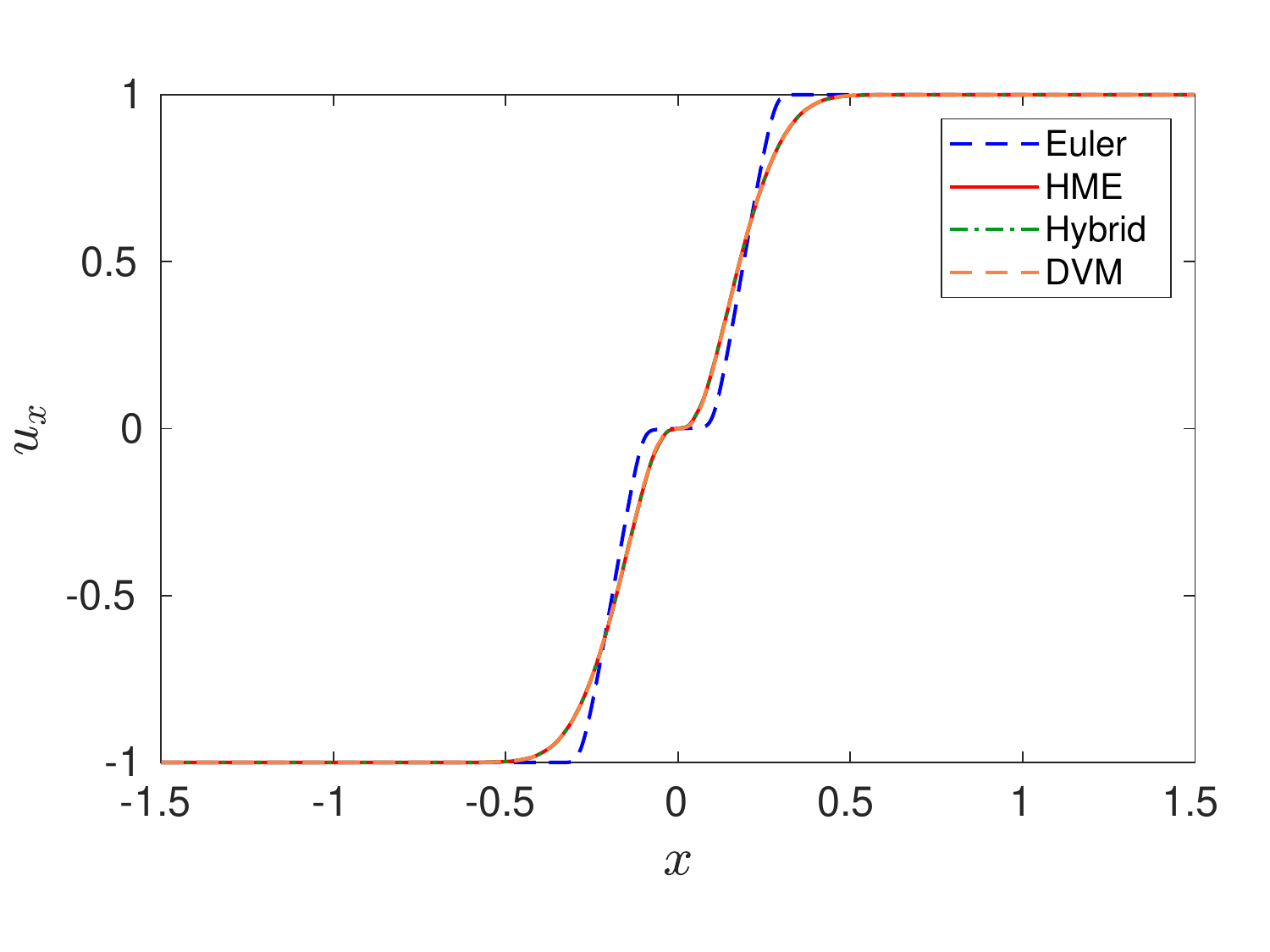}}
  \\
  \subfloat[$\theta$]{\label{fig:theta_test2_1}
    \includegraphics[width=0.48\textwidth]{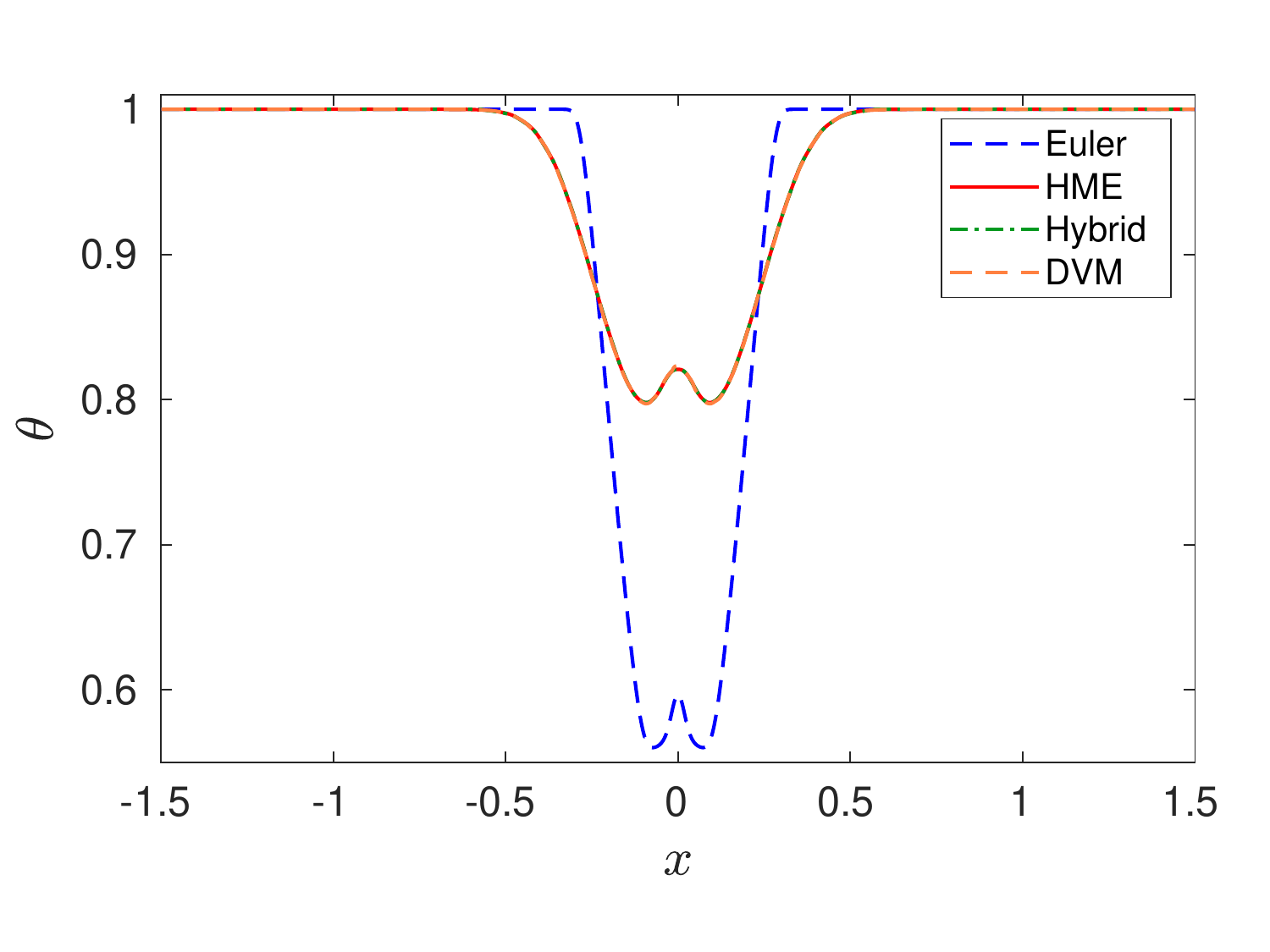}}
  \hfill \subfloat[$q_x$]{\label{fig:q_test2_1}
    \includegraphics[width=0.48\textwidth]{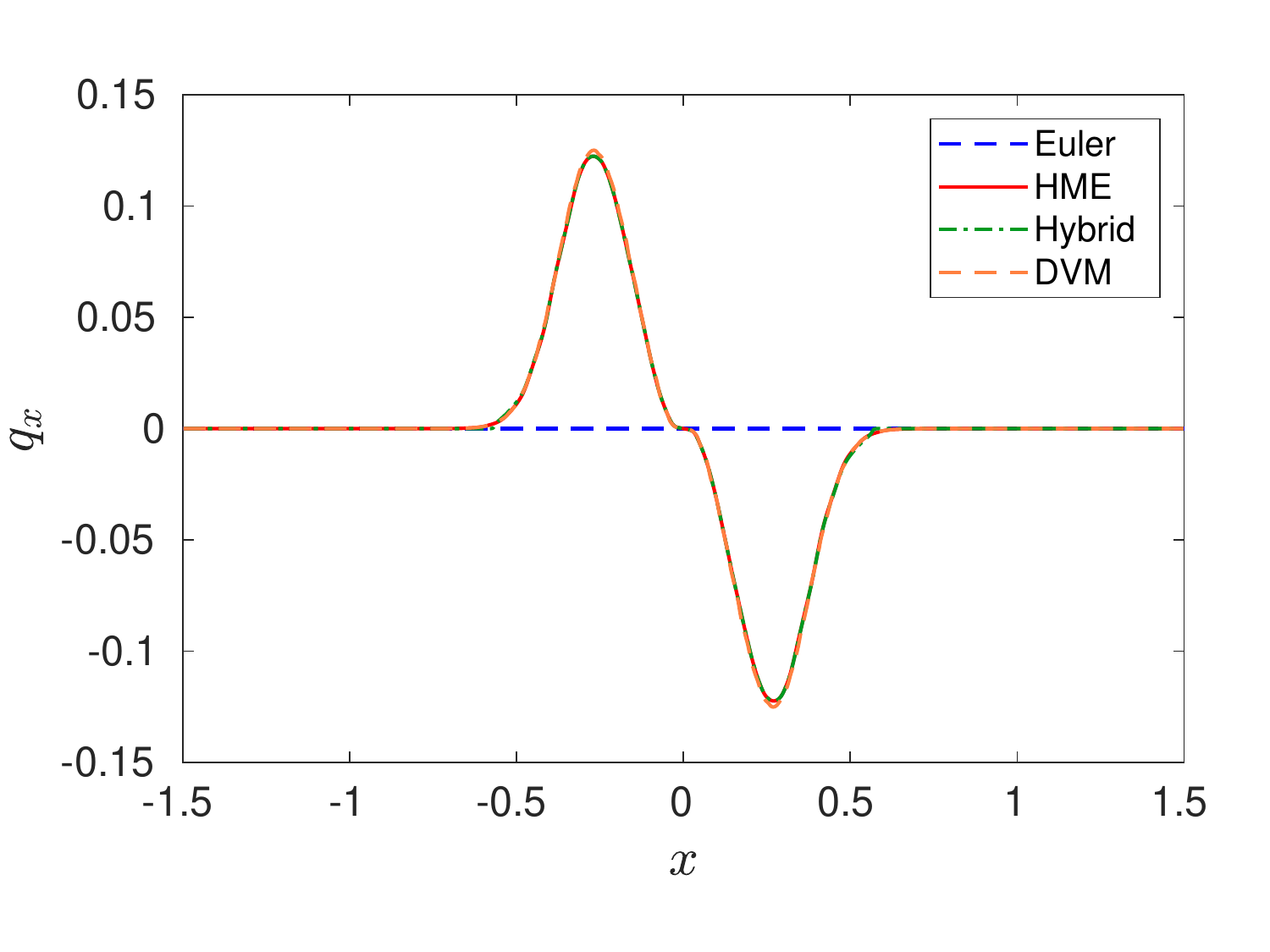}}
  \caption{Comparisons of solutions for $\rho$, $u_x$, $\theta$ and
    $q_x$ as functions of $x$ for $\Kn = 1$ in Sec \ref{sec:test2}.}
  \label{fig:test2_1}
\end{figure}

  \Cref{fig:test2_0p01} and \ref{fig:test2_1} show the macroscopic
  variables $\rho$, $u_x$, $\theta$ and $q_x$ at $t = 0.12$ with
  different Knudsen numbers. For both Knudsen number, we can find that
  the numerical solutions get by the hybrid method are almost the same
  as HME, which all agree with the reference get by DVM. This means
  that for problems with two rarefaction waves, the hybrid method can
  capture the movement of the particles for all the regimes.

\begin{figure}[htbp]
  \centering
  \subfloat[$\Kn = 0.01$ ]{
    \label{fig:order_test2_0p01}
    \includegraphics[trim=100 225 100
    250,clip,width=0.48\textwidth]{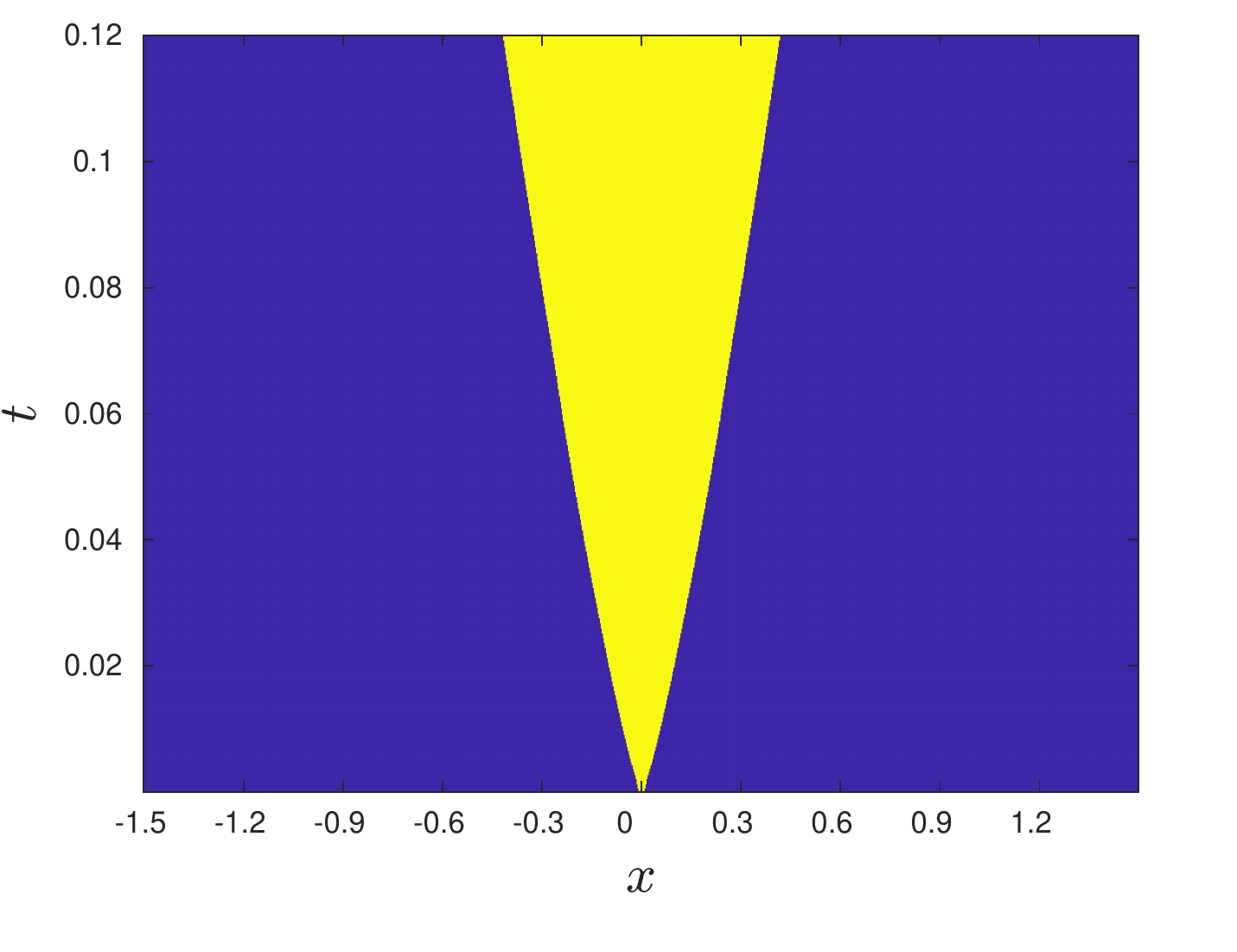}}
  \hfill \subfloat[$\Kn = 1$]{\label{fig:order_test2_1}
    \includegraphics[trim=100 225 100
    250,clip,width=0.48\textwidth]{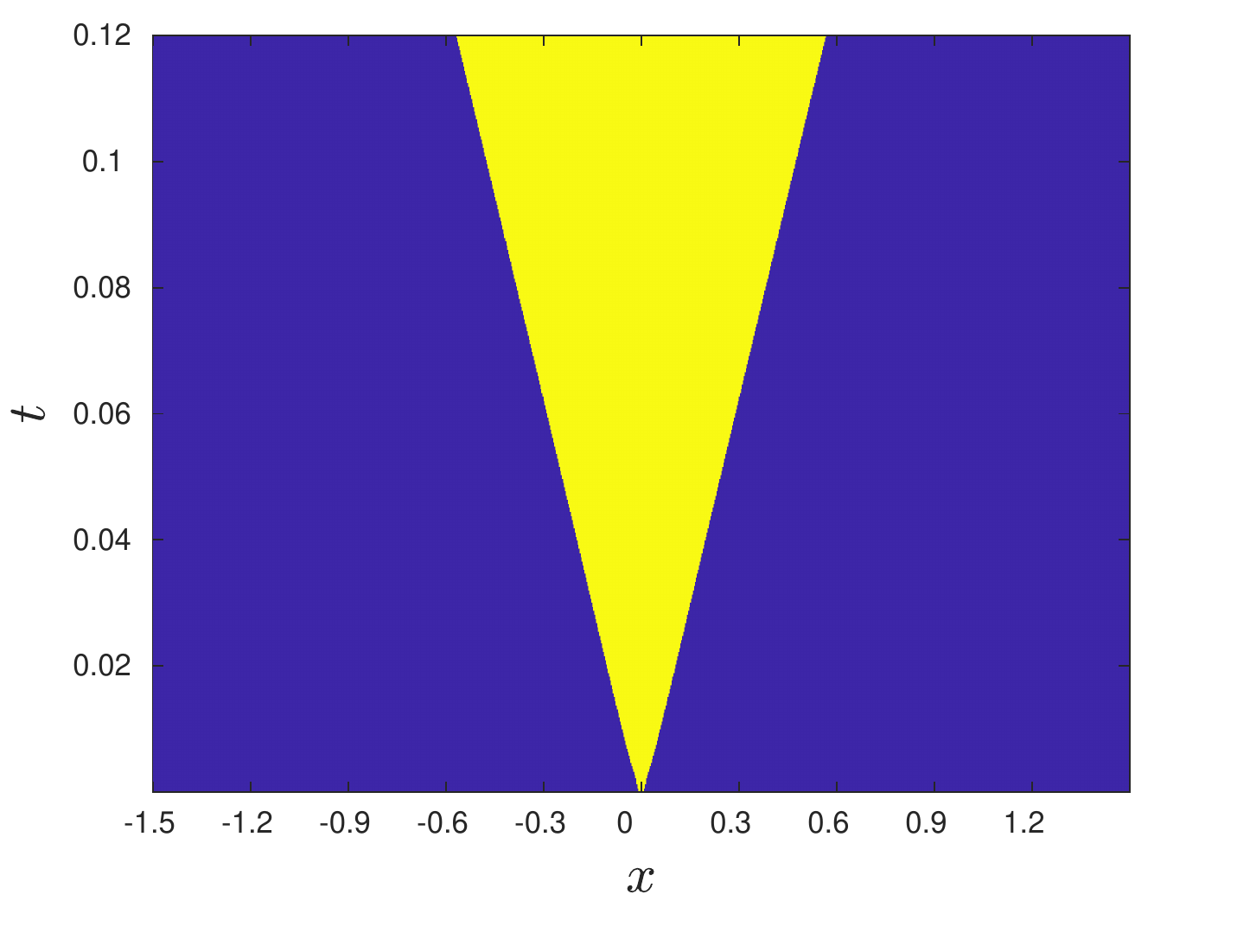}}
  \caption{The maximum order used in each spatial gird at each time
    step for different Knudsen numbers in Sec \ref{sec:test2}.}
\label{fig:test2_order}
\end{figure}

  The efficiency of the hybrid method is also tested.
  \Cref{fig:test2_order} shows the time evolution of the maximum order
  actually used by the hybrid method at each time step for each grid
  in this example. As in the previous example, the blue region
  indicates a maximum expansion order of $4$, while the yellow region
  means a maximum expansion order of $40$.  From it, we can find that
  in the middle of the spatial area, where it is the kinetic regime,
  the expansion order is $40$, while in the left and right sides of
  the spatial area, the expansion order four is kept. All this means
  that the hybrid scheme could detect the rarefied zones correctly for
  the rarefaction wave problems. In \Cref{tab:timecompare-test2} we
  compare the computation time for the hybrid method and HME for the
  two Knudsen numbers we tested. For both cases, we see that almost
  $50$ percent of reduction in computation cost, which shows the high
  efficiency of the hybrid method.

\begin{table}[htbp]
  \centering
  \def\arraystretch{1.5}
  \footnotesize
  \label{tab:timecompare-test2}
  \setlength{\tabcolsep}{5mm}{
    \begin{tabular}{ccc}
      $\Kn$  & Full HME  & Hybrid \\
          \toprule
      $0.01$ & $74$s  & $43$s\\
      \hline
      $1$ &  $75$s &  $44$s\\
    \end{tabular}}
  \caption{Comparison of computational time of the hybrid method and
    HME with different Knudsen numbers in Sec \ref{sec:test2}.}
\end{table}

\subsubsection{Blast wave}
\label{sec:test4}
In this section, a similar blast wave problem as in \cite{Filbet2015}
is tested. The computation domain is taken to be $[-1, 1]$ in the
simulations. The initial condition is
$f(0, x, \bv) = f_M(\rho, u_x, \theta)$, with
\begin{equation}
  \rho(0, x) = 1.0, \quad
  u_x(0, x) = \left\{
    \begin{array}{l}
      1.0, \quad \text{if}~x \leqslant -0.3, \\
      0, \quad \text{if}~ -0.3 < x < 0.3, \\
      -1.0, \quad \text{if}~x \geqslant 0.3,
    \end{array}\right.
    \quad
    \theta = \left\{
      \begin{array}{l}
        1.5, \quad \text{if}~x \leqslant -0.3, \\
        1.0, \quad \text{if}~-0.3 < x < 0.3, \\
        1.5, \quad \text{if}~x \geqslant 0.3.
      \end{array}\right.
\end{equation}
We take $\Kn = 0.001$ and $\Kn = 0.01$. In the test, the grid size is set as
$N = 1000$ and the maximum expansion order is $M = 40$.

\begin{figure}[htbp]
  \centering
  \subfloat[$\rho, t= 0.05$ ]{
    \label{fig:rho_test4_Kn0p001_t0p05}
  \includegraphics[width=0.3\textwidth]{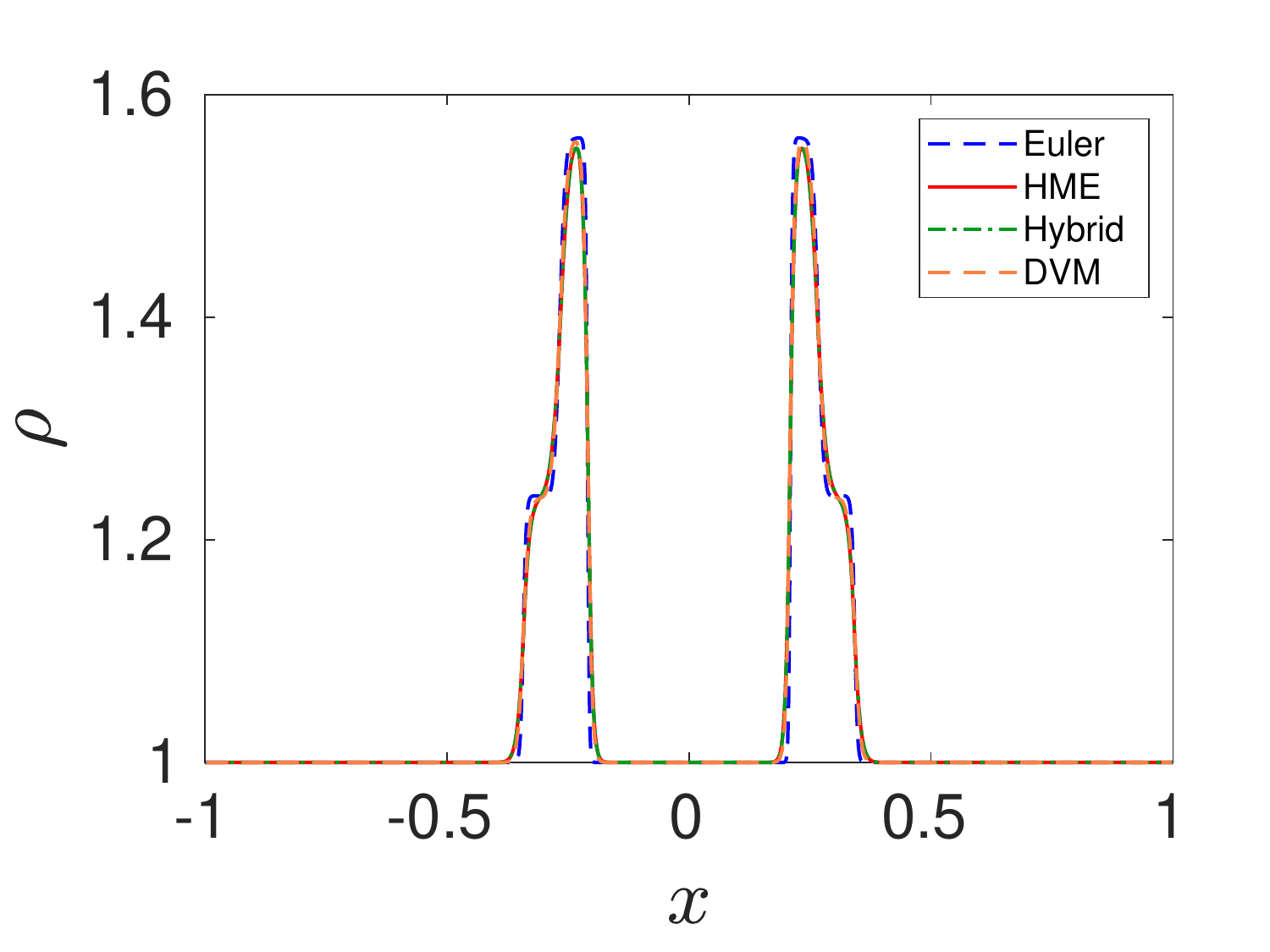}}
\hfill
\subfloat[$\rho, t = 0.15$ ]{
    \label{fig:rho_test4_Kn0p001_t0p15}
  \includegraphics[width=0.3\textwidth]{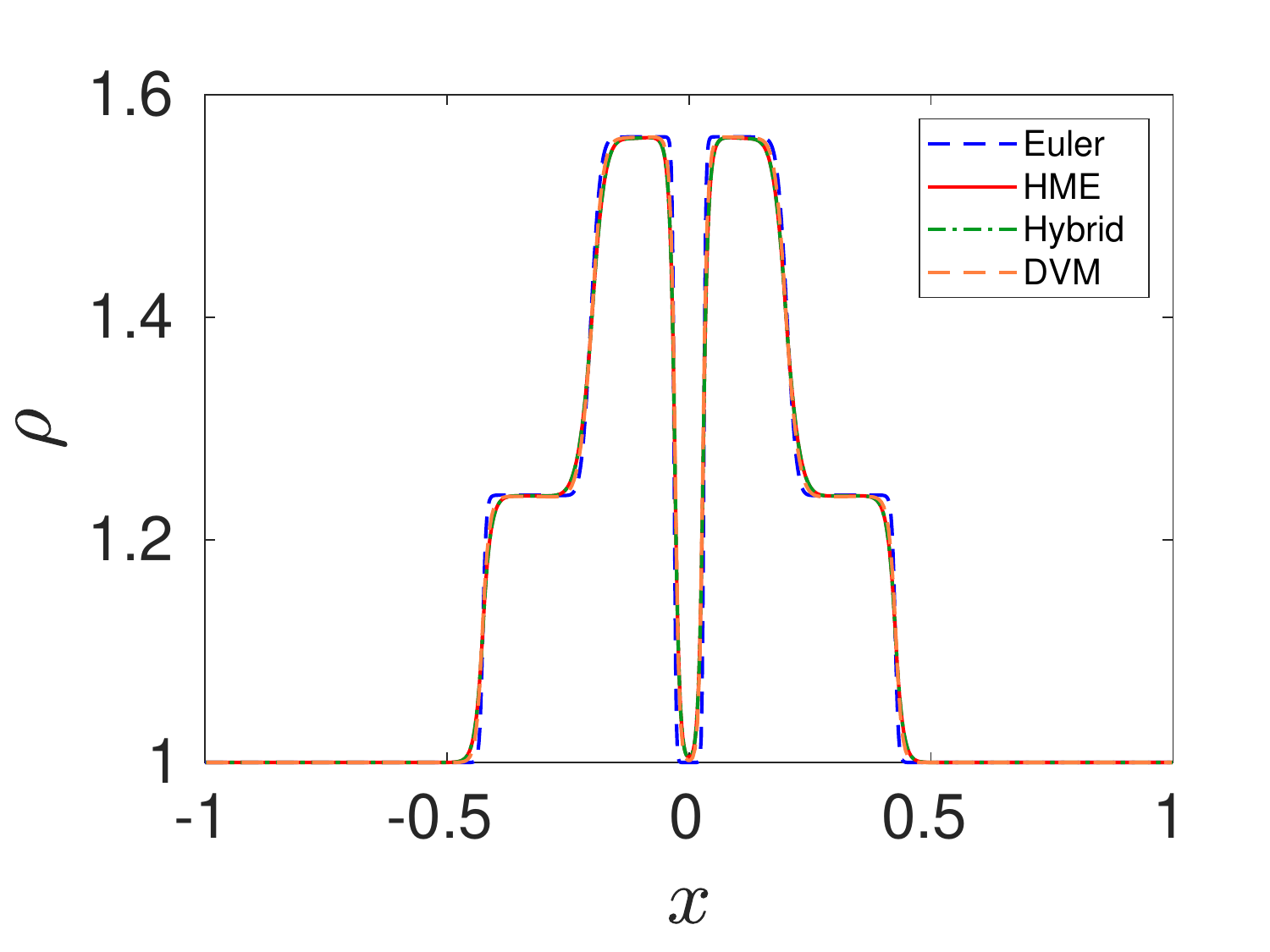}}
\hfill
  \subfloat[$\rho, t= 0 .35$ ]{
    \label{fig:rho_test4_Kn0p001_t0p35}
    \includegraphics[width=0.3\textwidth]{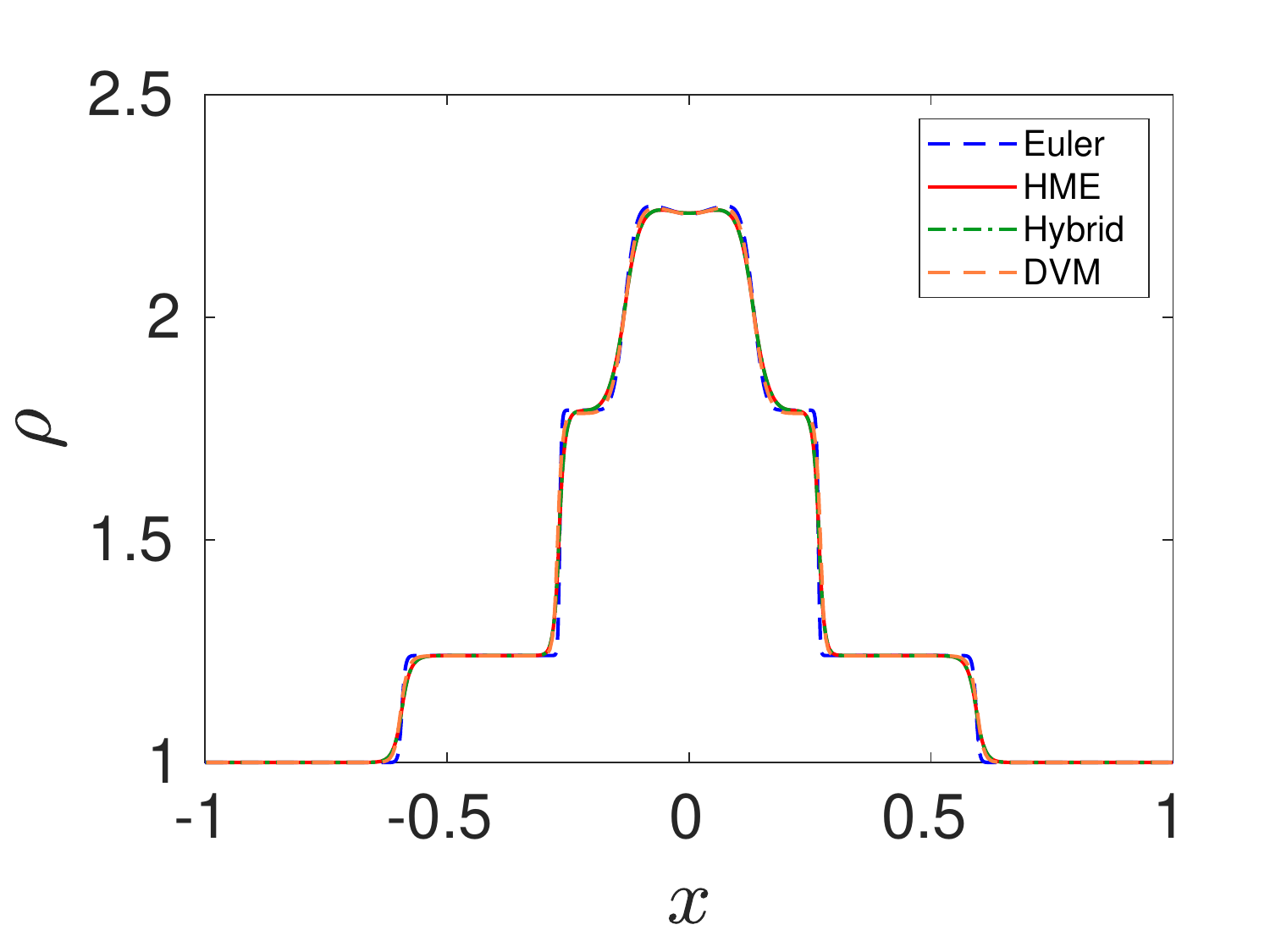}} \\
  \subfloat[$u_x, t = 0.05$]{\label{fig:u_test4_Kn0p001_t0p05}
    \includegraphics[width=0.3\textwidth]{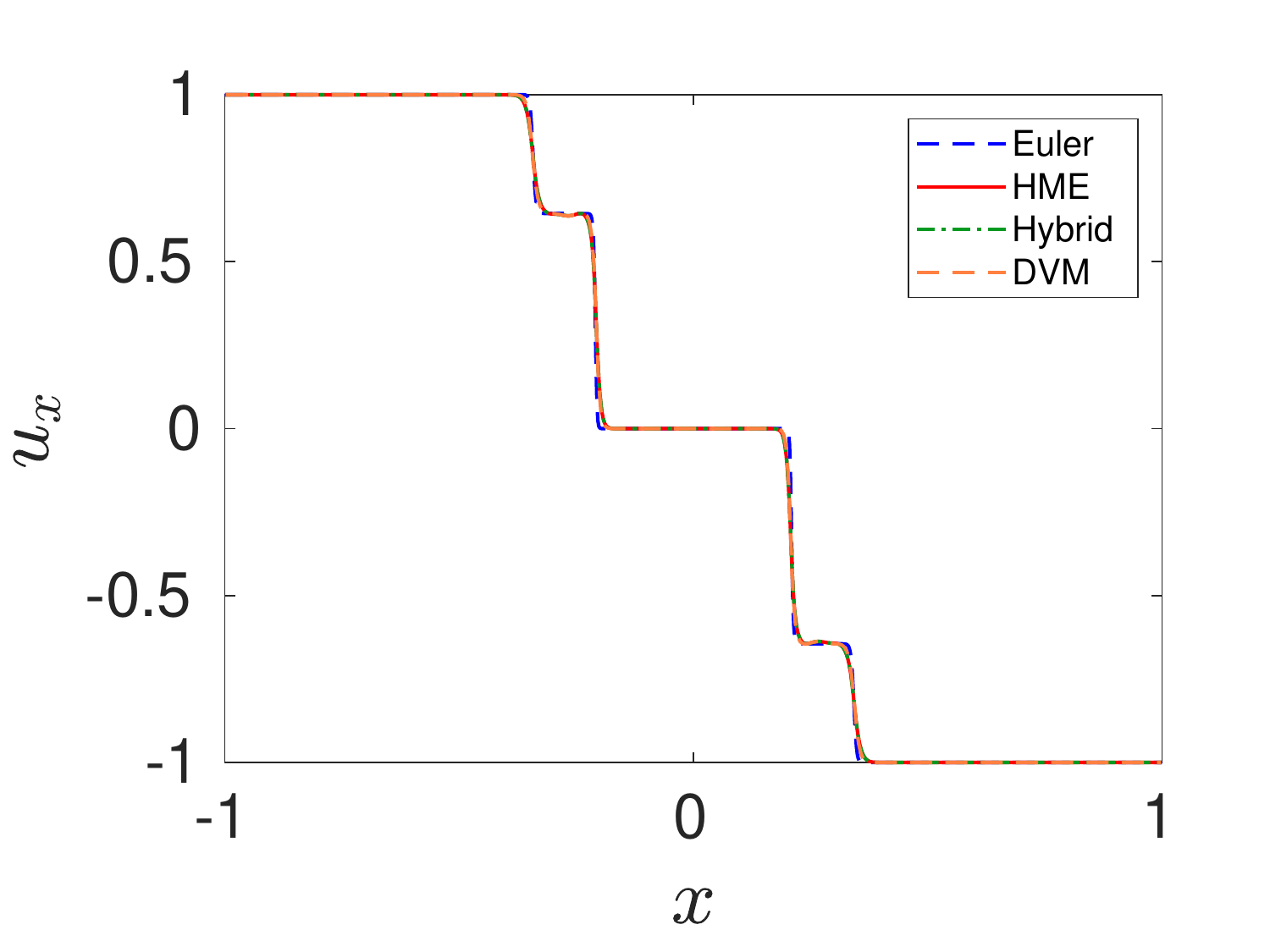}}
  \hfill \subfloat[$u_x, t = 0.15$]{\label{fig:u_test4_Kn0p001_t0p15}
    \includegraphics[width=0.3\textwidth]{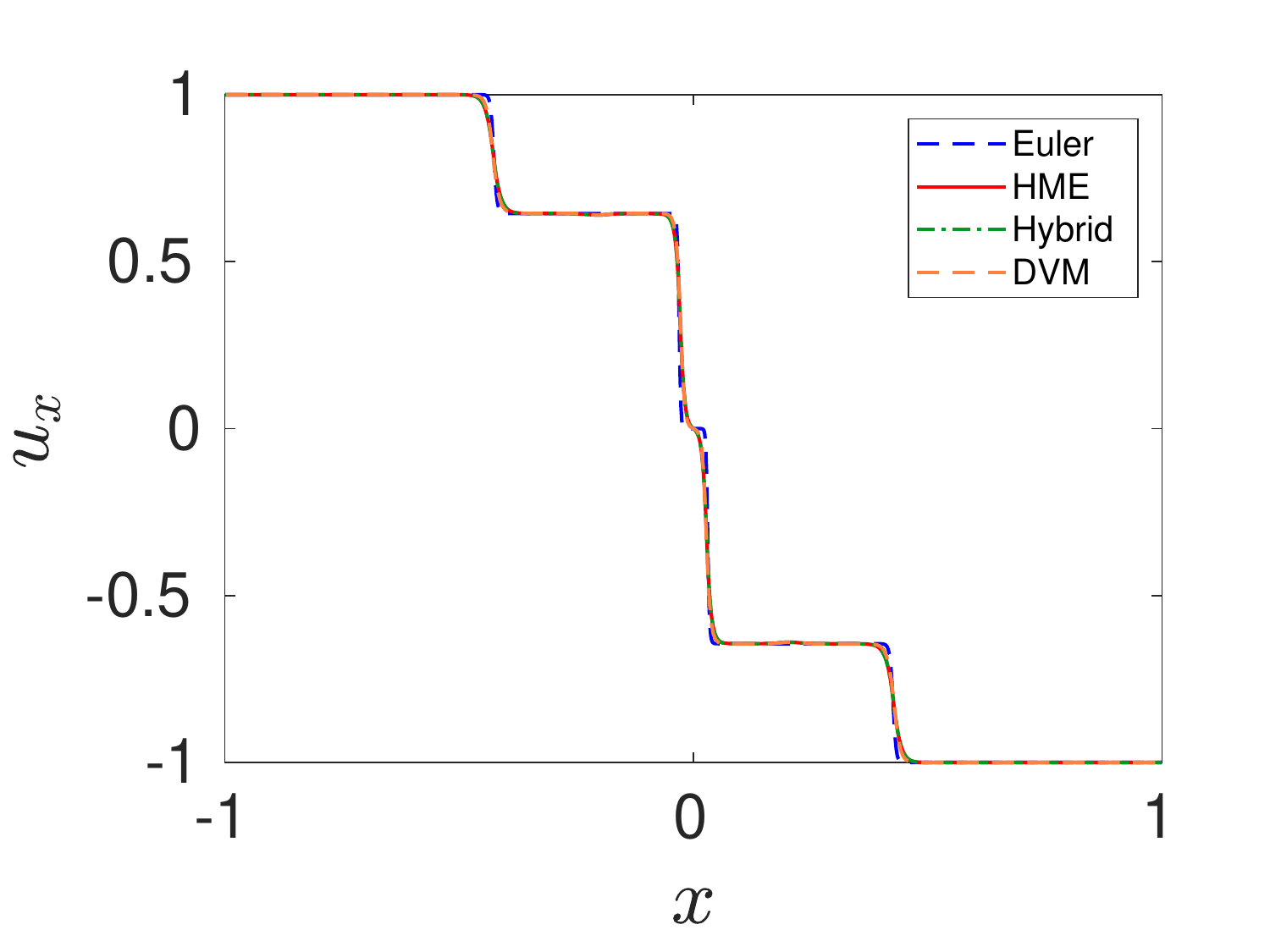}}
  \hfill \subfloat[$u_x, t = 0.35$]{\label{fig:u_test4_Kn0p001_t0p35}
    \includegraphics[width=0.3\textwidth]{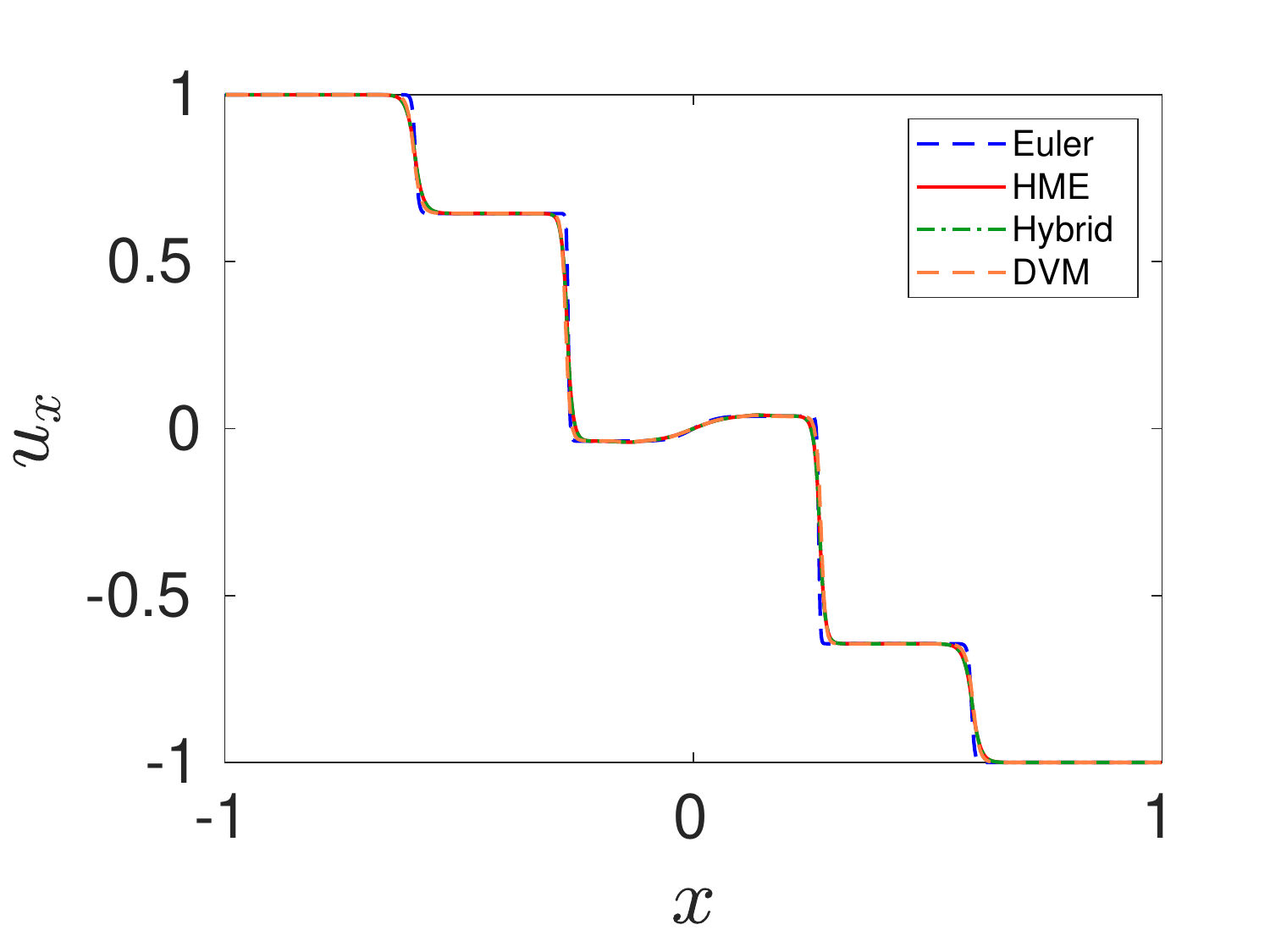}}
  \\
  \subfloat[$\theta, t = 0.05$]{\label{fig:theta_test4_Kn0p001_t0p05}
    \includegraphics[width=0.3\textwidth]{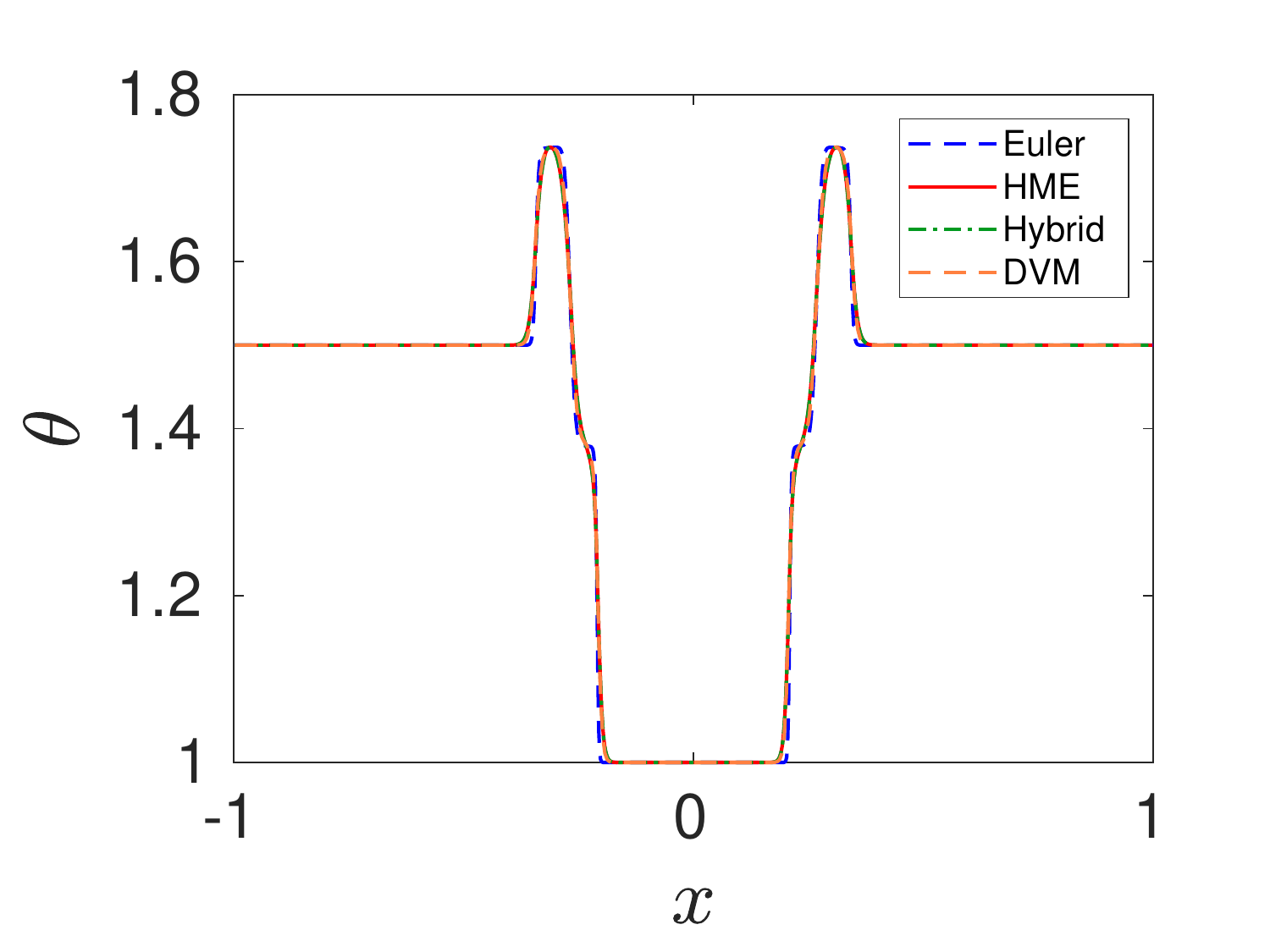}}
  \hfill
  \subfloat[$\theta, t = 0.15$]{\label{fig:theta_test4_Kn0p001_t0p15}
    \includegraphics[width=0.3\textwidth]{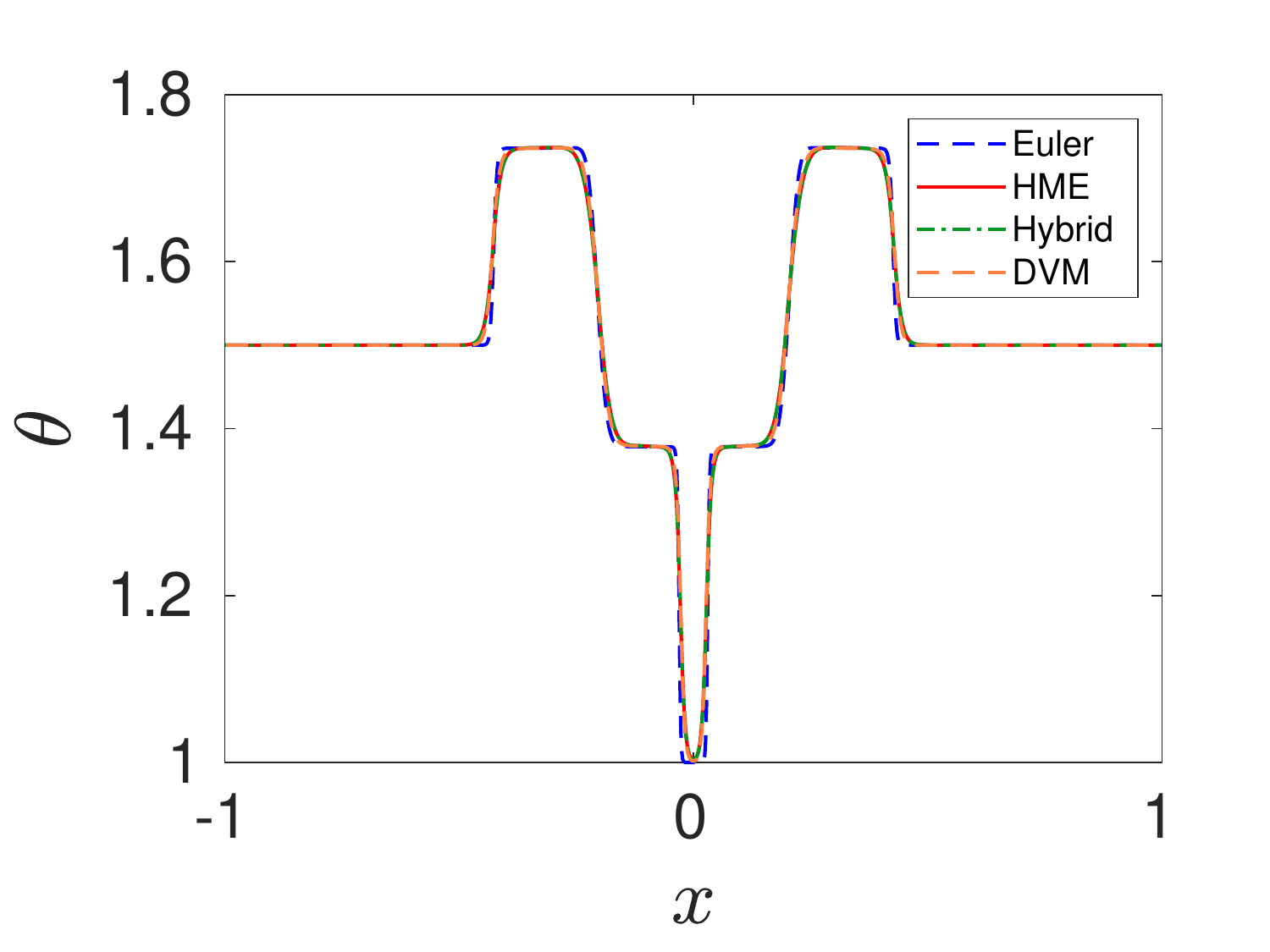}}
  \hfill
  \subfloat[$\theta, t = 0.35$]{\label{fig:theta_test4_Kn0p001_t0p35}
    \includegraphics[width=0.3\textwidth]{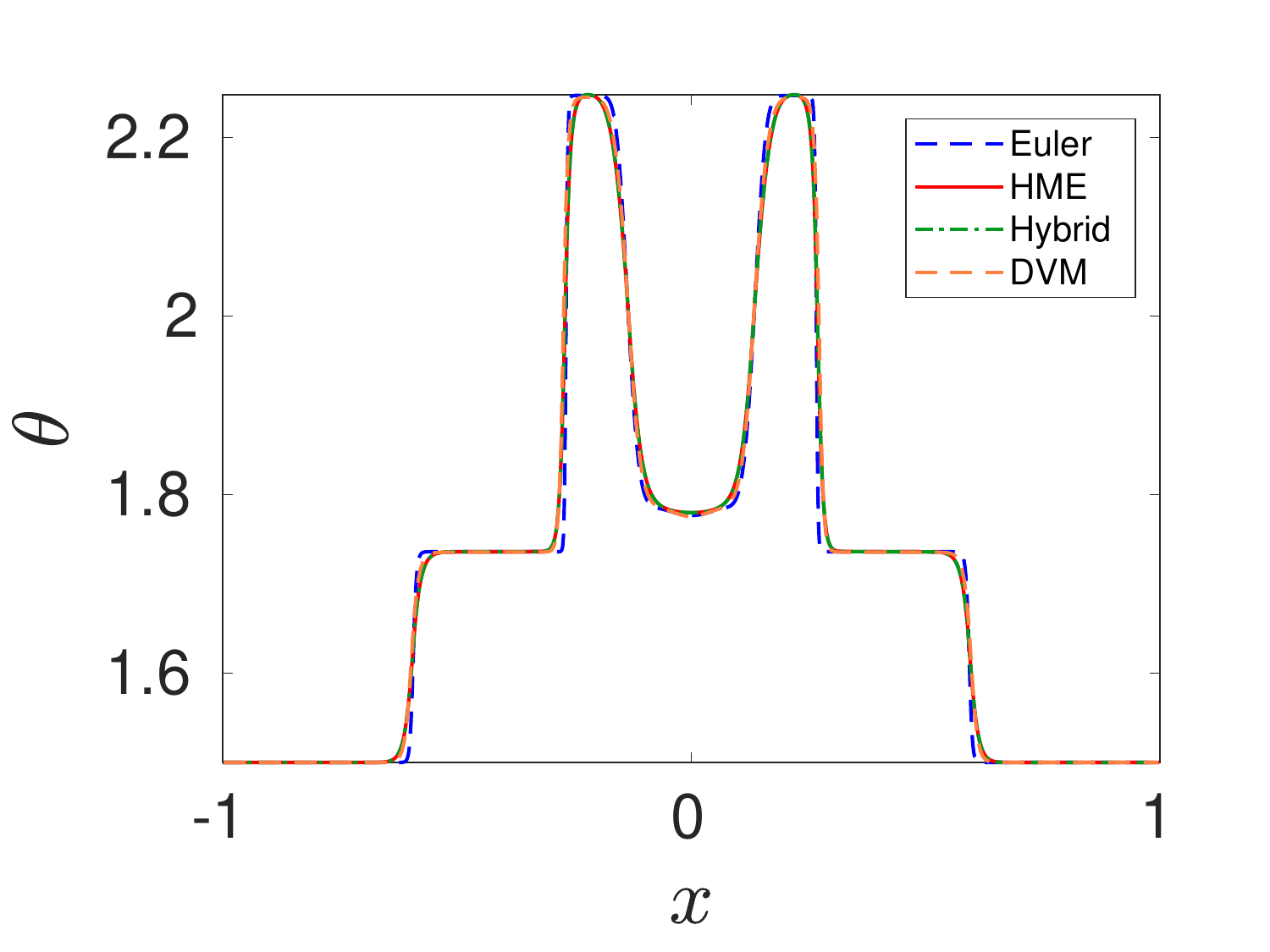}}
  \\
  \subfloat[$q_x, t = 0.05$]{\label{fig:q_test4_Kn0p001_t0p05}
    \includegraphics[width=0.3\textwidth]{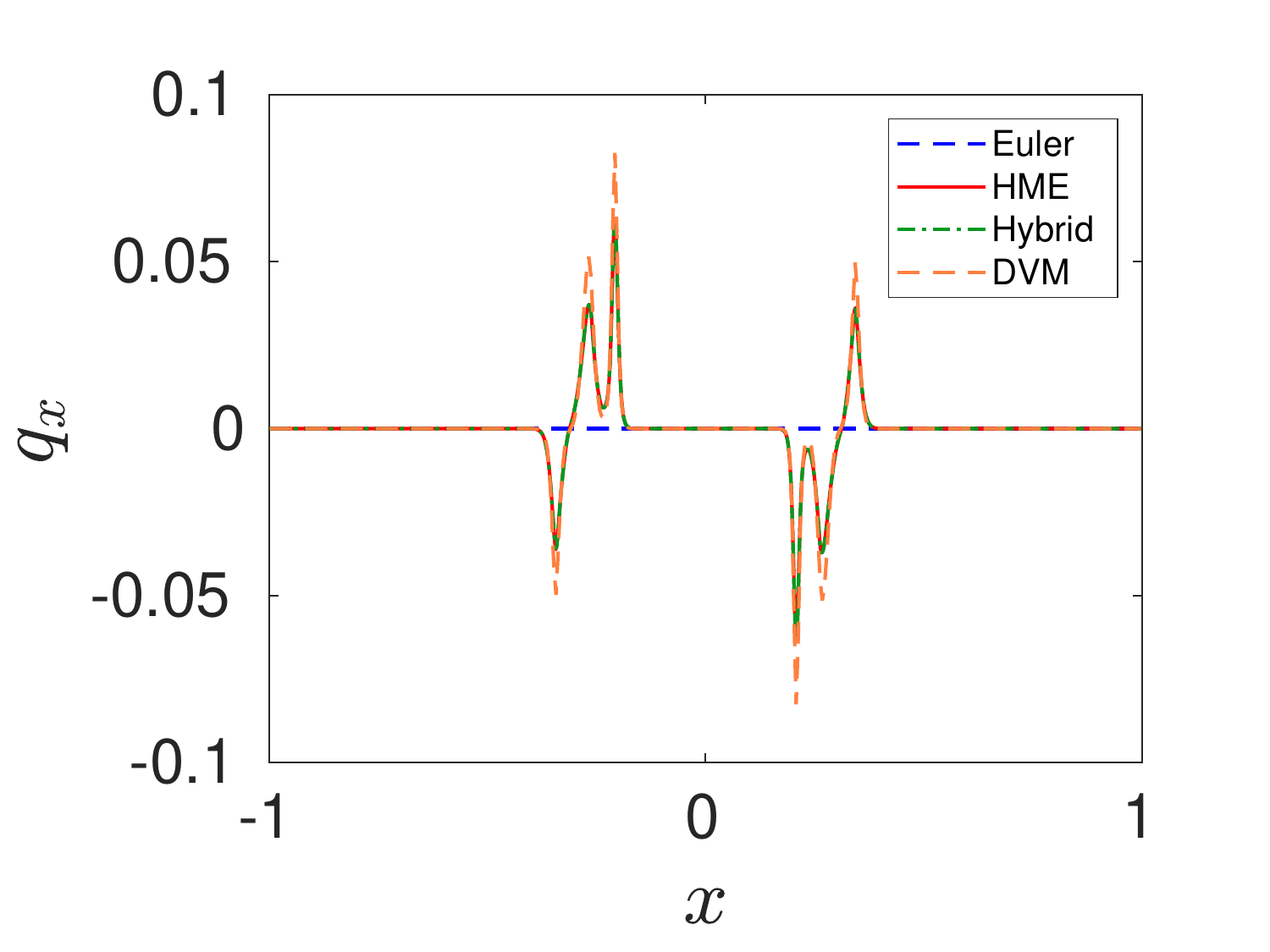}}
  \hfill \subfloat[$q_x, t= 0.15$]{\label{fig:q_test4_Kn0p001_t0p15}
    \includegraphics[width=0.3\textwidth]{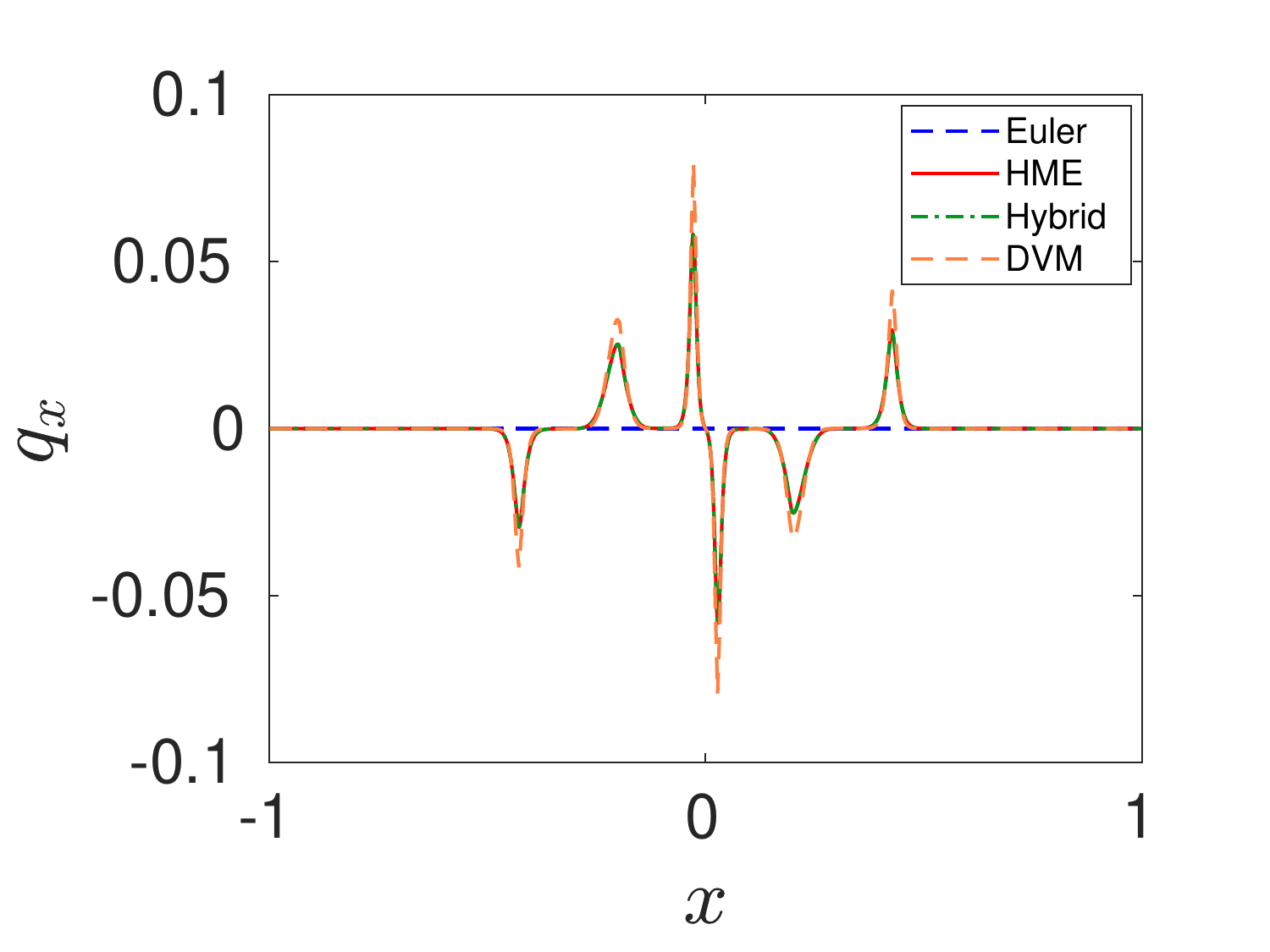}}
  \hfill \subfloat[$q_x, t = 0.35$]{\label{fig:q_test4_Kn0p001_t0p35}
    \includegraphics[width=0.3\textwidth]{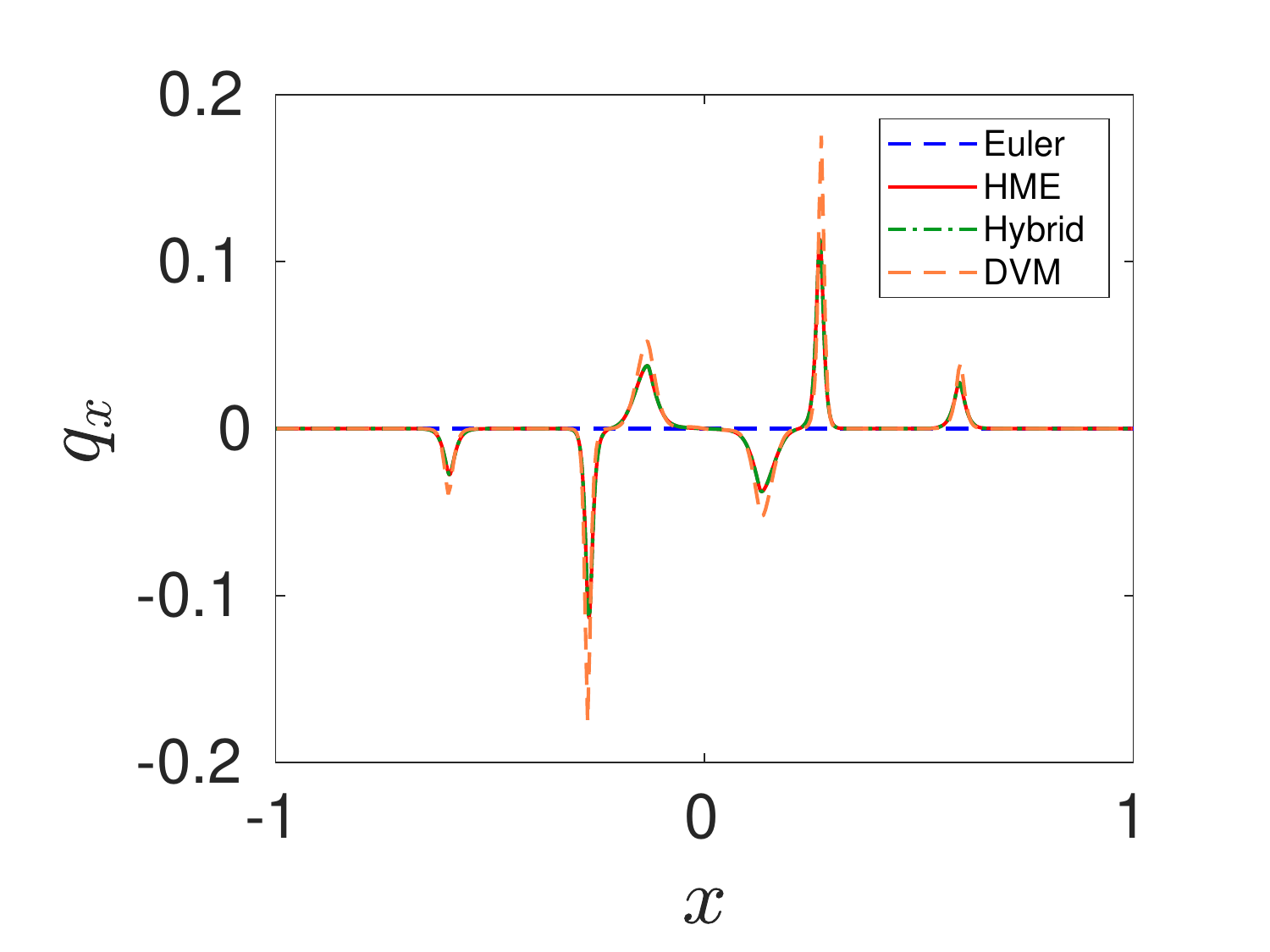}}
\caption{Comparisons of solutions for $\rho$, $u_x$, $\theta$ and
  $q_x$ as functions of $x$ for $\Kn = 0.001$ at time $t = 0.05, 0.15$
  and $0.35$ in Sec \ref{sec:test4}.}
\label{fig:test4_Kn0p001}
\end{figure}

\begin{figure}[htbp]
  \centering
  \subfloat[$\rho, t= 0.05$ ]{
    \label{fig:rho_test4_Kn0p01_t0p05}
  \includegraphics[width=0.3\textwidth]{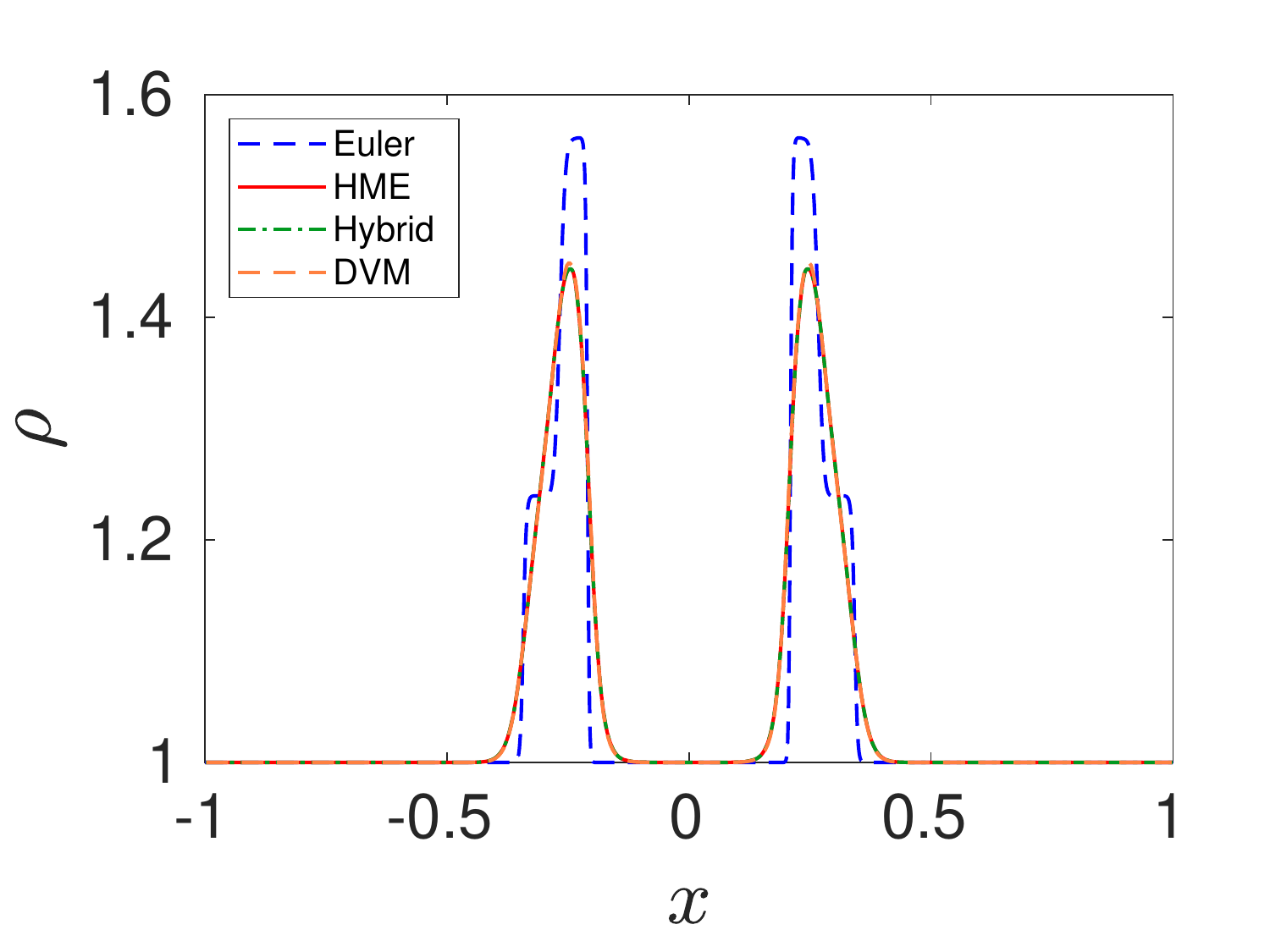}}
\hfill
\subfloat[$\rho, t = 0.15$ ]{
    \label{fig:rho_test4_Kn0p01_t0p15}
  \includegraphics[width=0.3\textwidth]{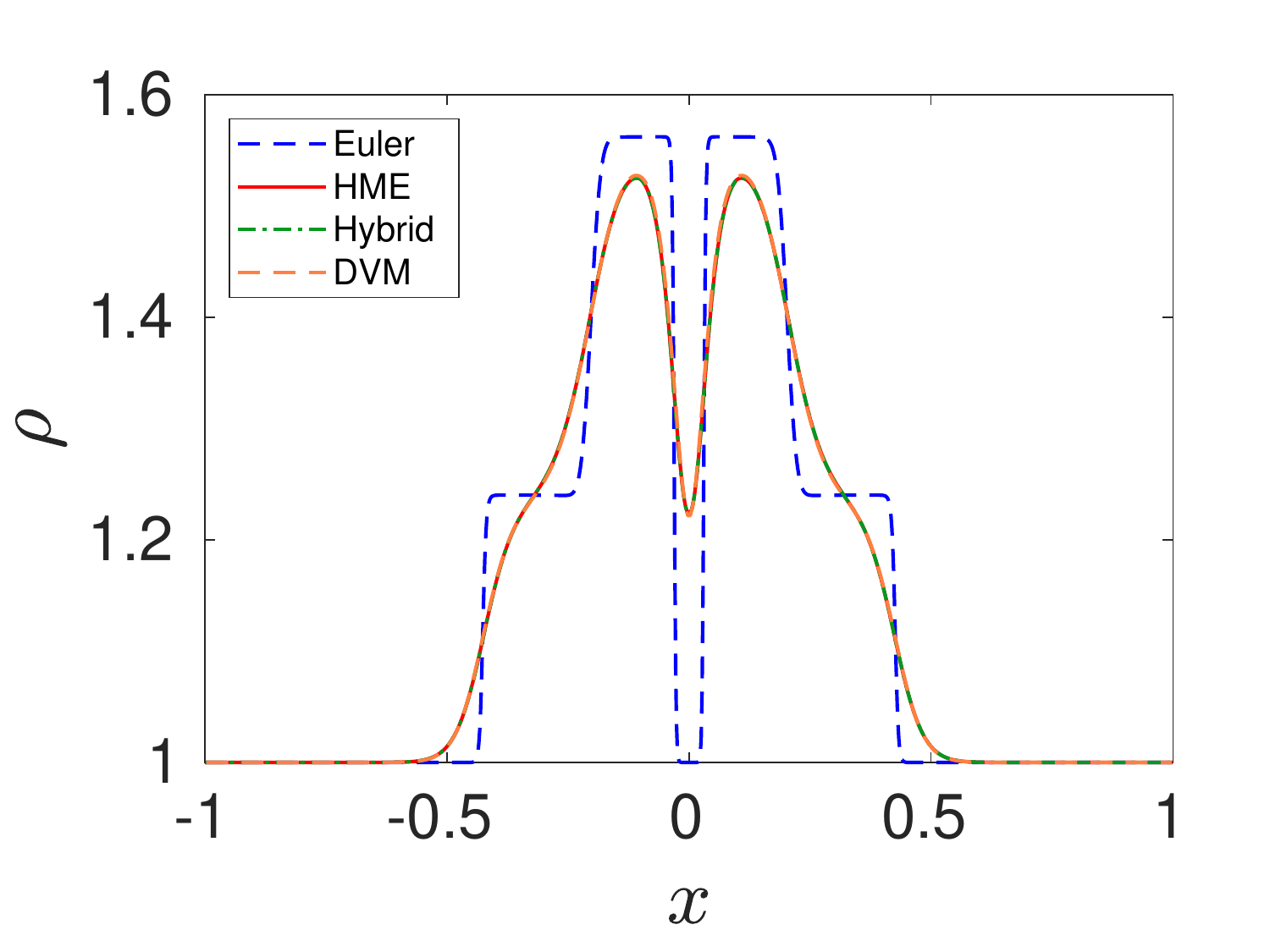}}
\hfill
  \subfloat[$\rho, t= 0 .35$ ]{
    \label{fig:rho_test4_Kn0p01_t0p35}
    \includegraphics[width=0.3\textwidth]{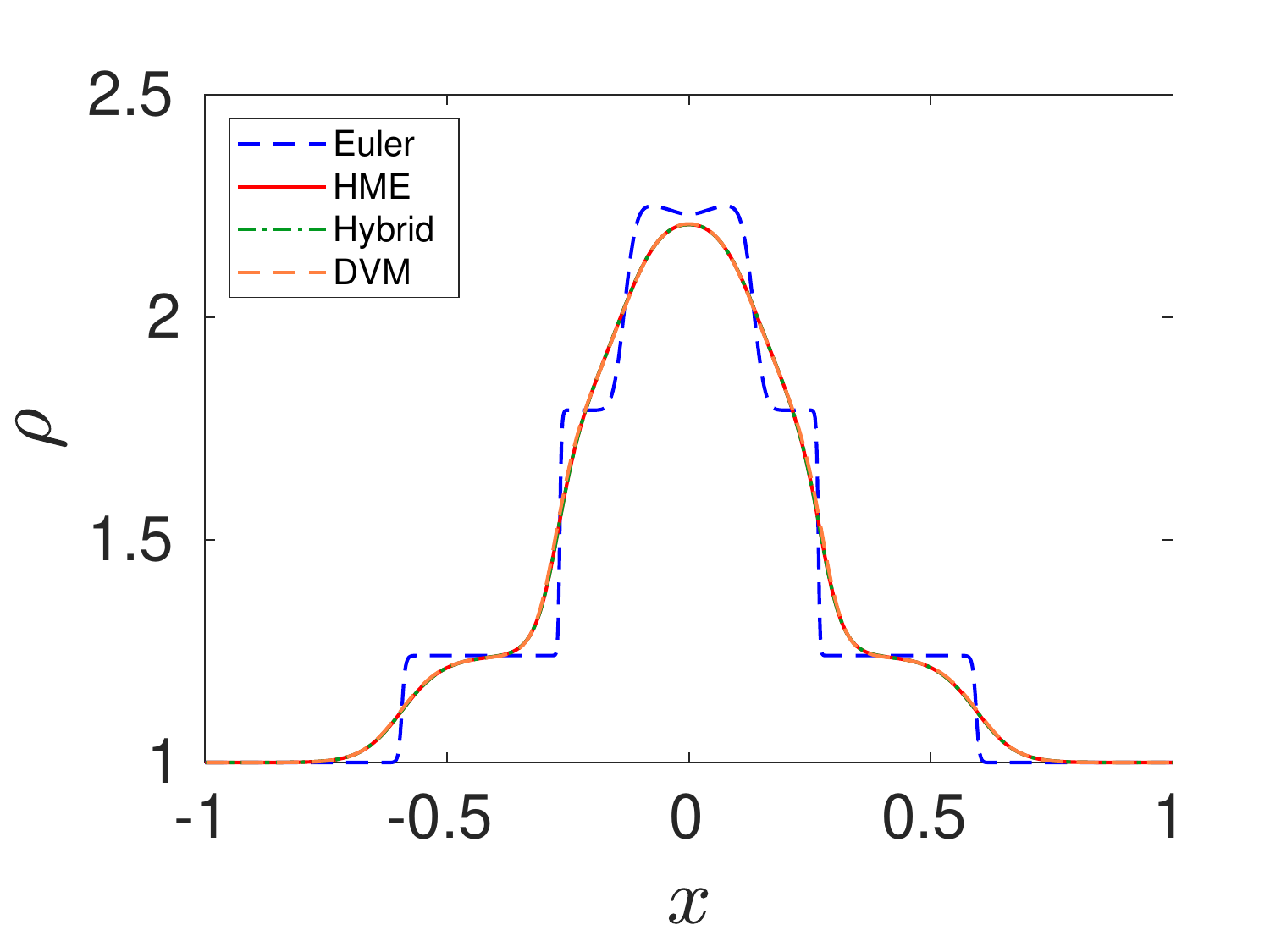}} \\
  \subfloat[$u_x, t = 0.05$]{\label{fig:u_test4_Kn0p01_t0p05}
    \includegraphics[width=0.3\textwidth]{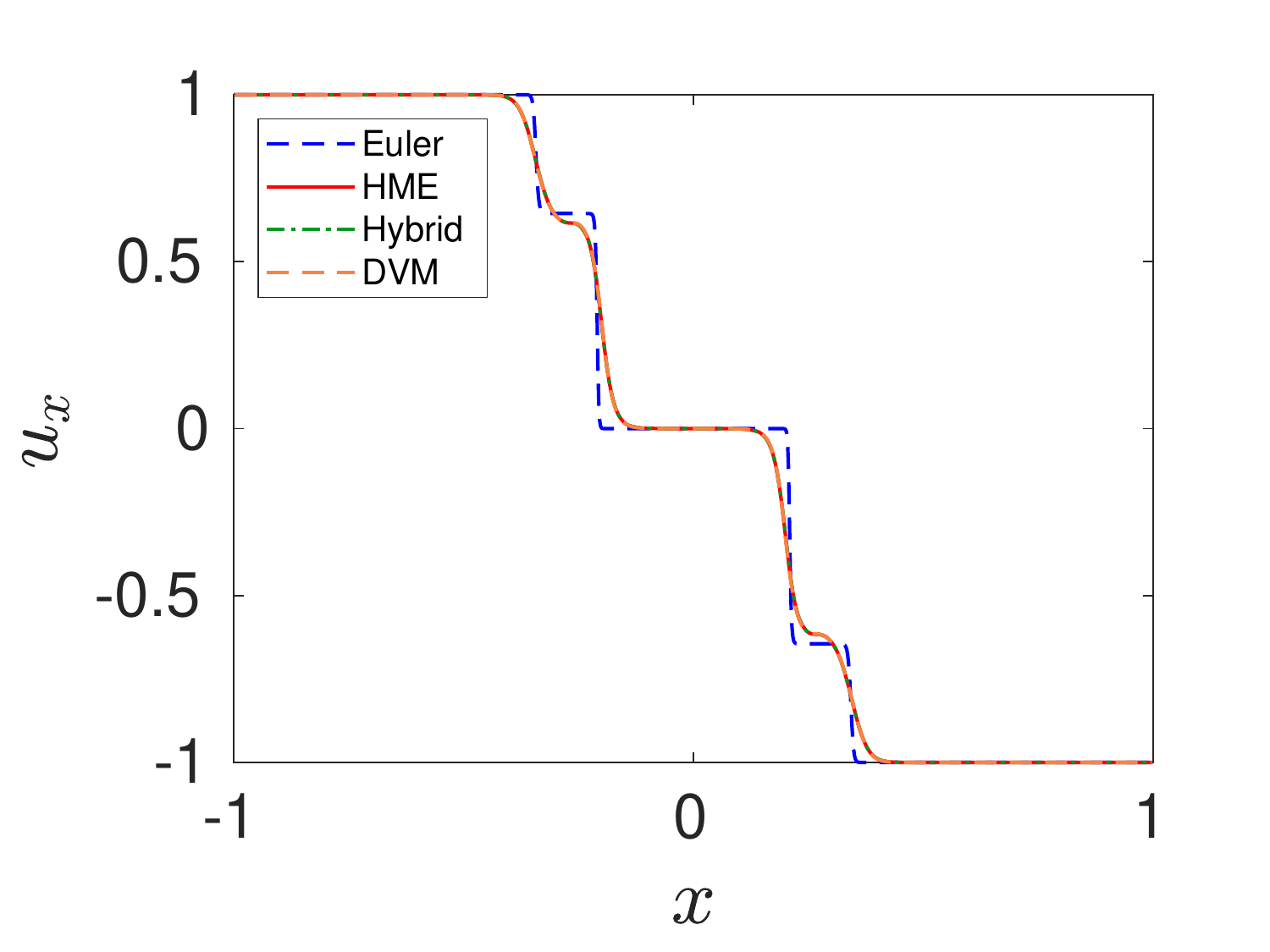}}
  \hfill \subfloat[$u_x, t = 0.15$]{\label{fig:u_test4_Kn0p01_t0p15}
    \includegraphics[width=0.3\textwidth]{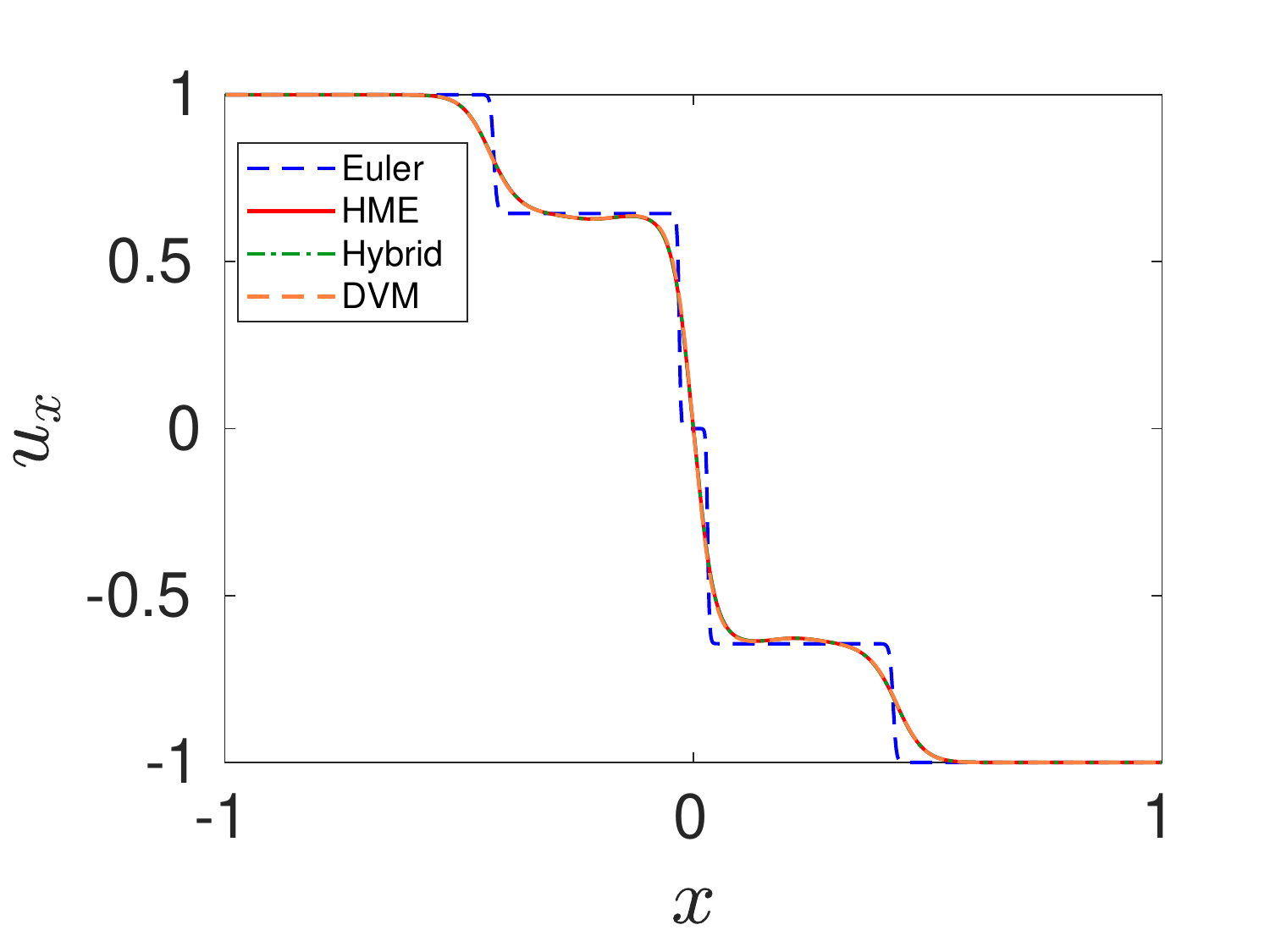}}
  \hfill \subfloat[$u_x, t = 0.35$]{\label{fig:u_test4_Kn0p01_t0p35}
    \includegraphics[width=0.3\textwidth]{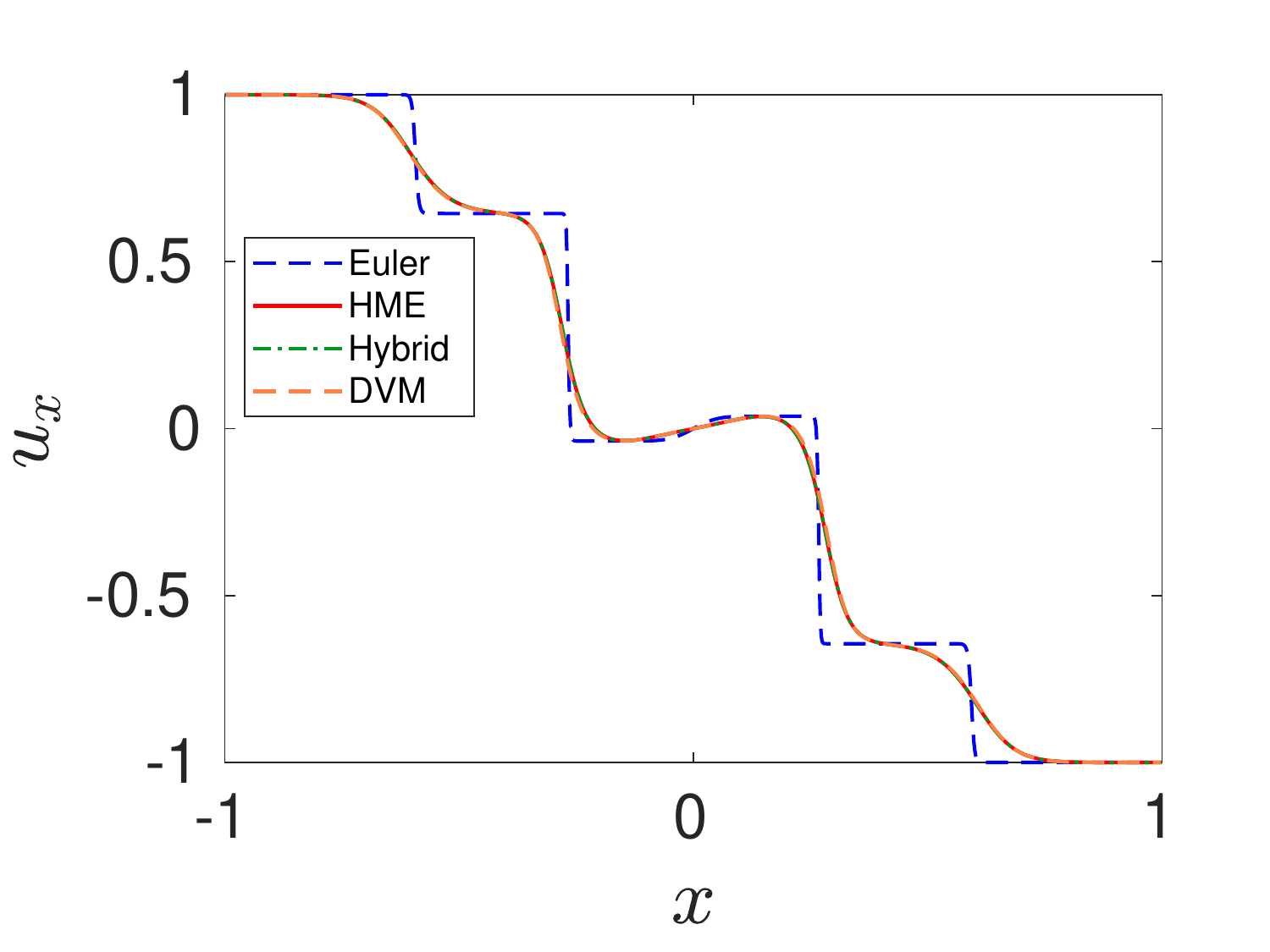}}
  \\
  \subfloat[$\theta, t = 0.05$]{\label{fig:theta_test4_Kn0p01_t0p05}
    \includegraphics[width=0.3\textwidth]{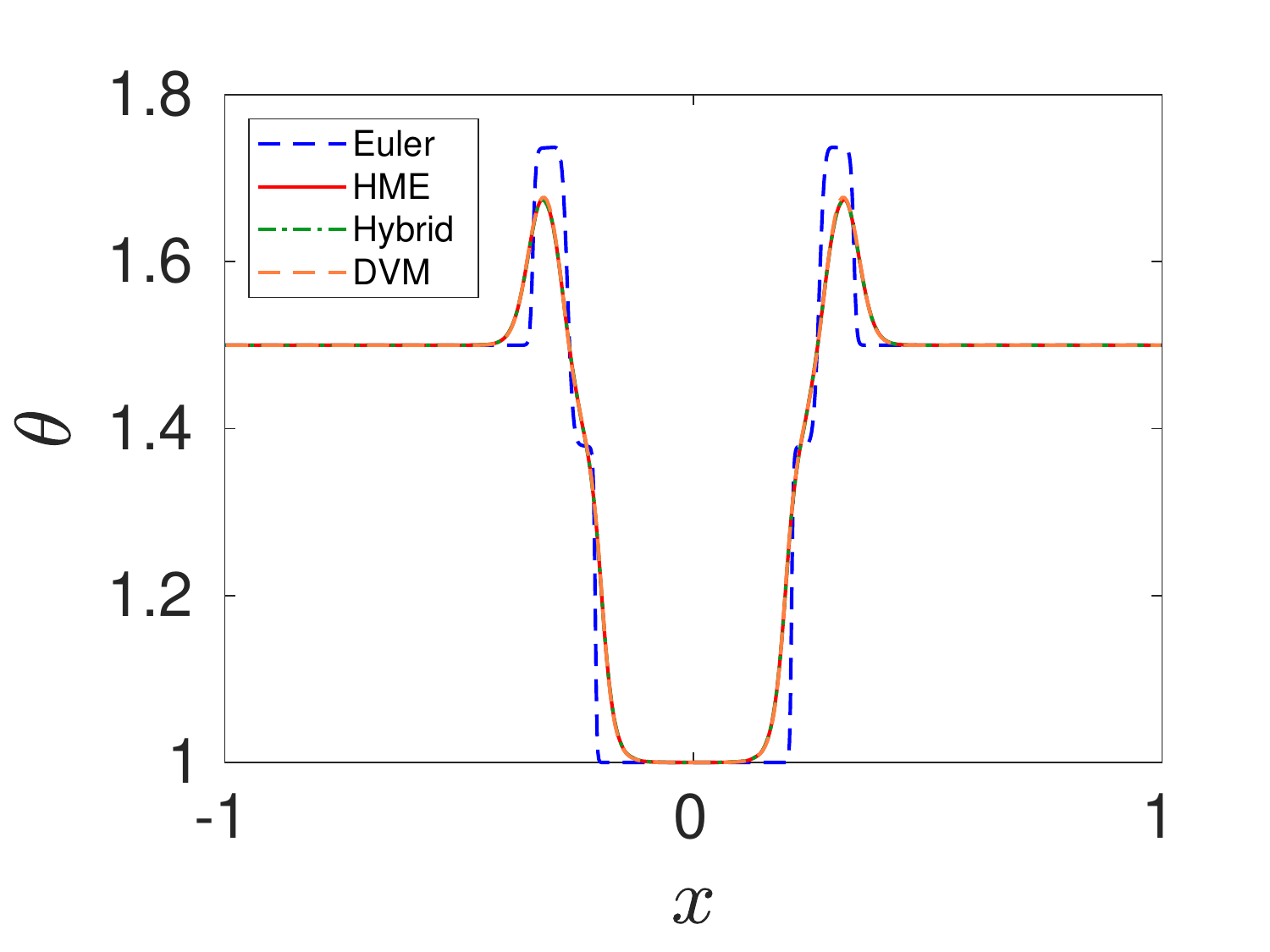}}
  \hfill
  \subfloat[$\theta, t = 0.15$]{\label{fig:theta_test4_Kn0p01_t0p15}
    \includegraphics[width=0.3\textwidth]{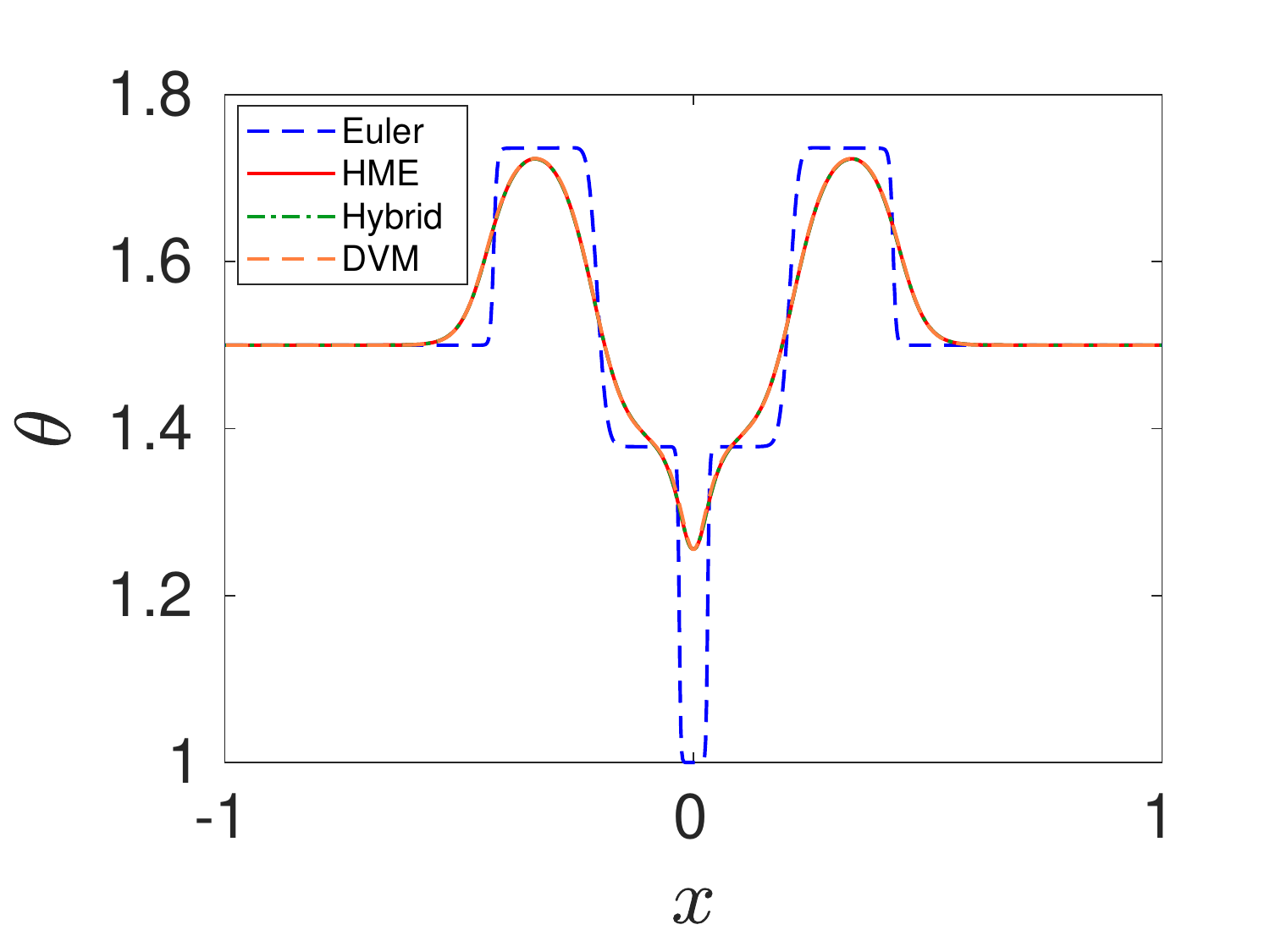}}
  \hfill
  \subfloat[$\theta, t = 0.35$]{\label{fig:theta_test4_Kn0p01_t0p35}
    \includegraphics[width=0.3\textwidth]{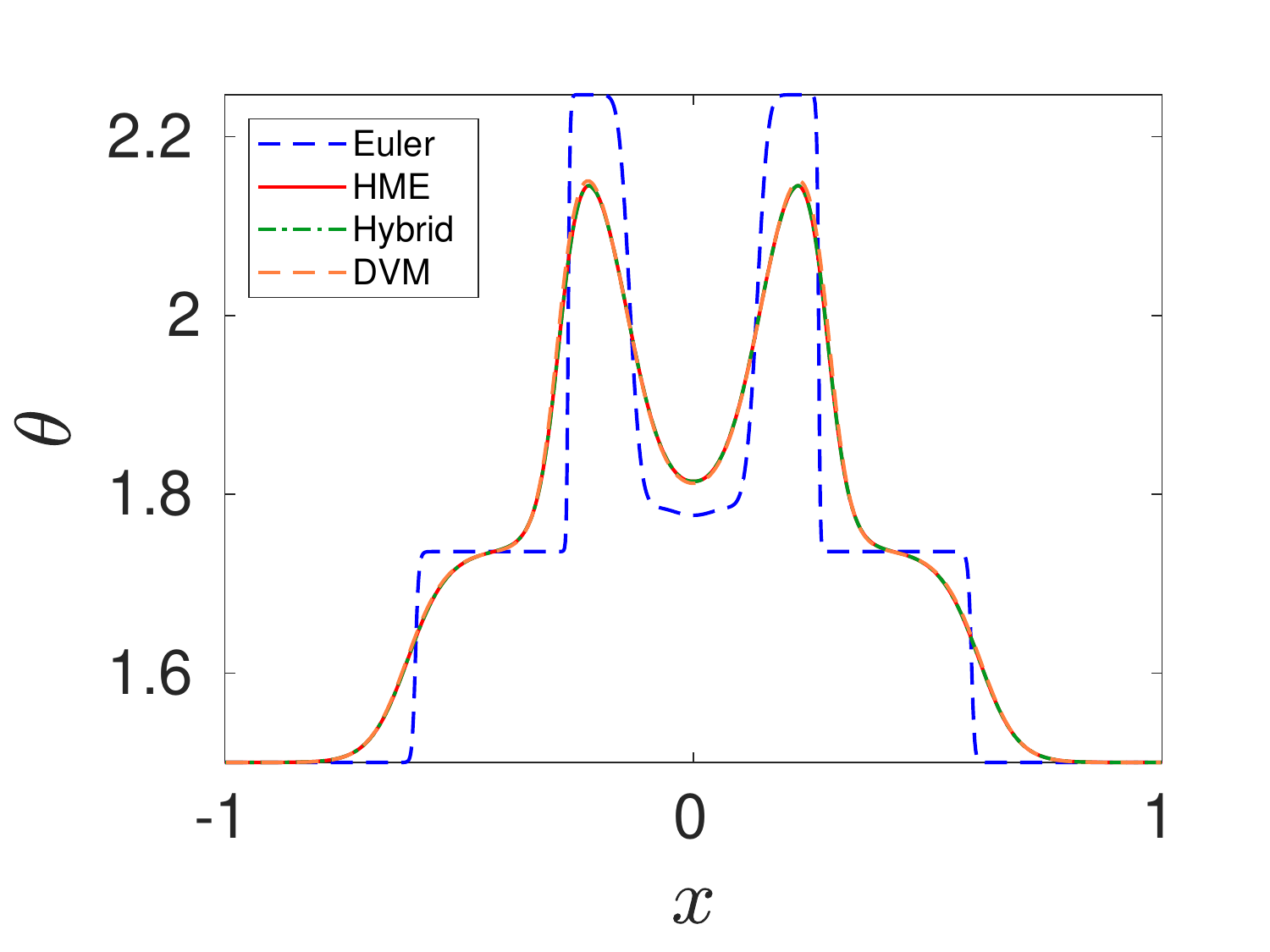}}
  \\
  \subfloat[$q_x, t = 0.05$]{\label{fig:q_test4_Kn0p01_t0p05}
    \includegraphics[width=0.3\textwidth]{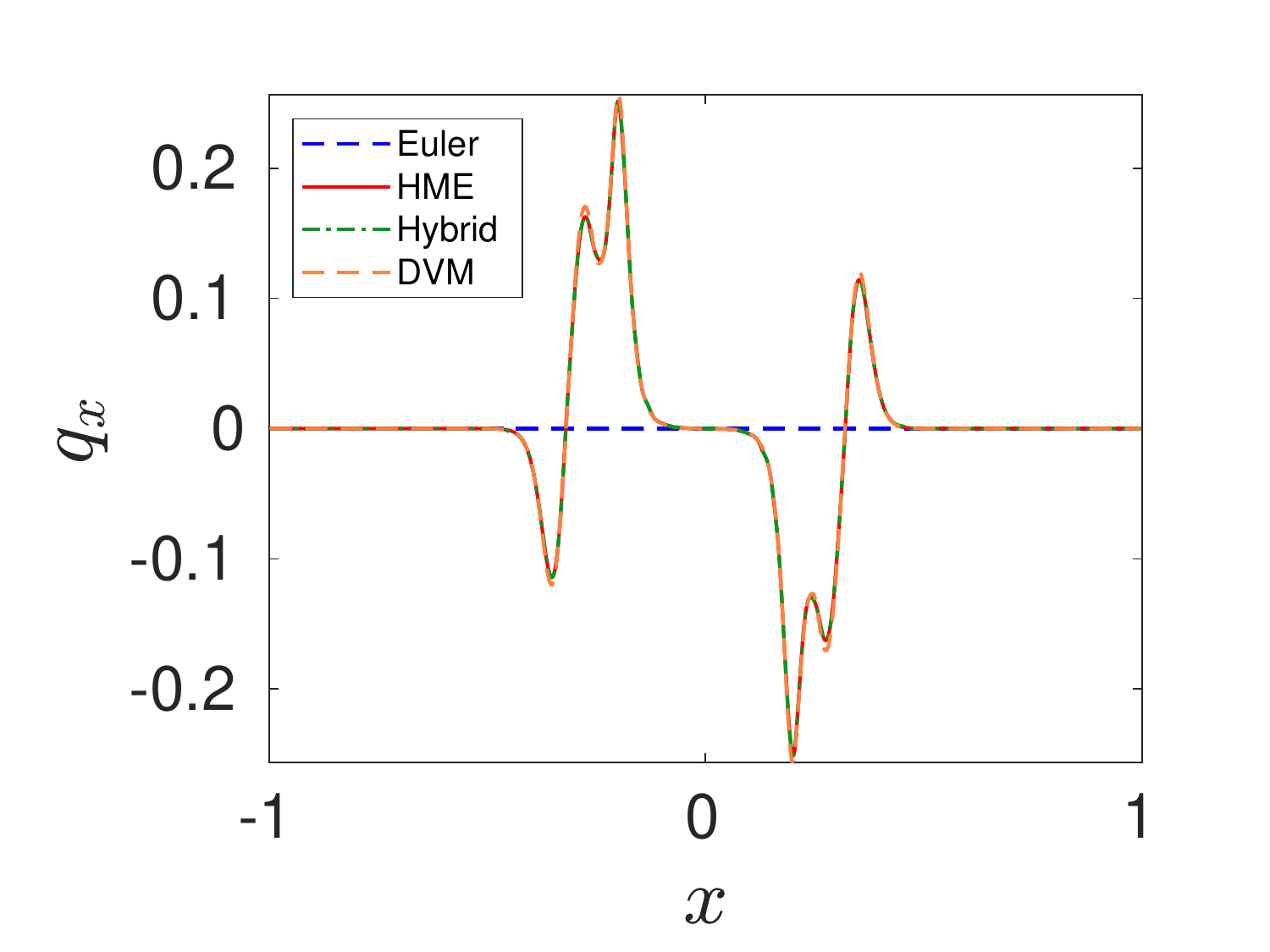}}
  \hfill \subfloat[$q_x, t= 0.15$]{\label{fig:q_test4_Kn0p01_t0p15}
    \includegraphics[width=0.3\textwidth]{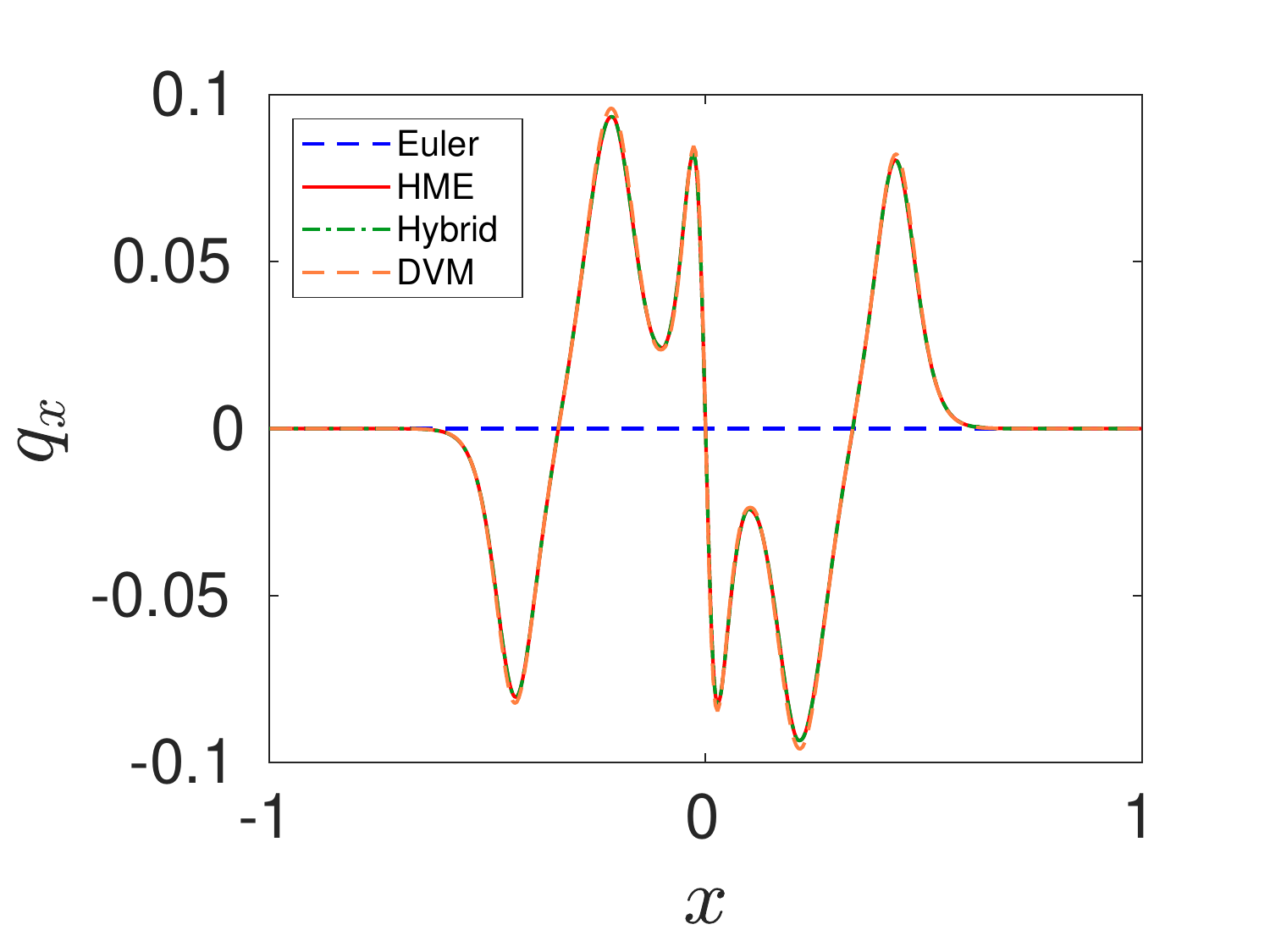}}
  \hfill \subfloat[$q_x, t = 0.35$]{\label{fig:q_test4_Kn0p01_t0p35}
    \includegraphics[width=0.3\textwidth]{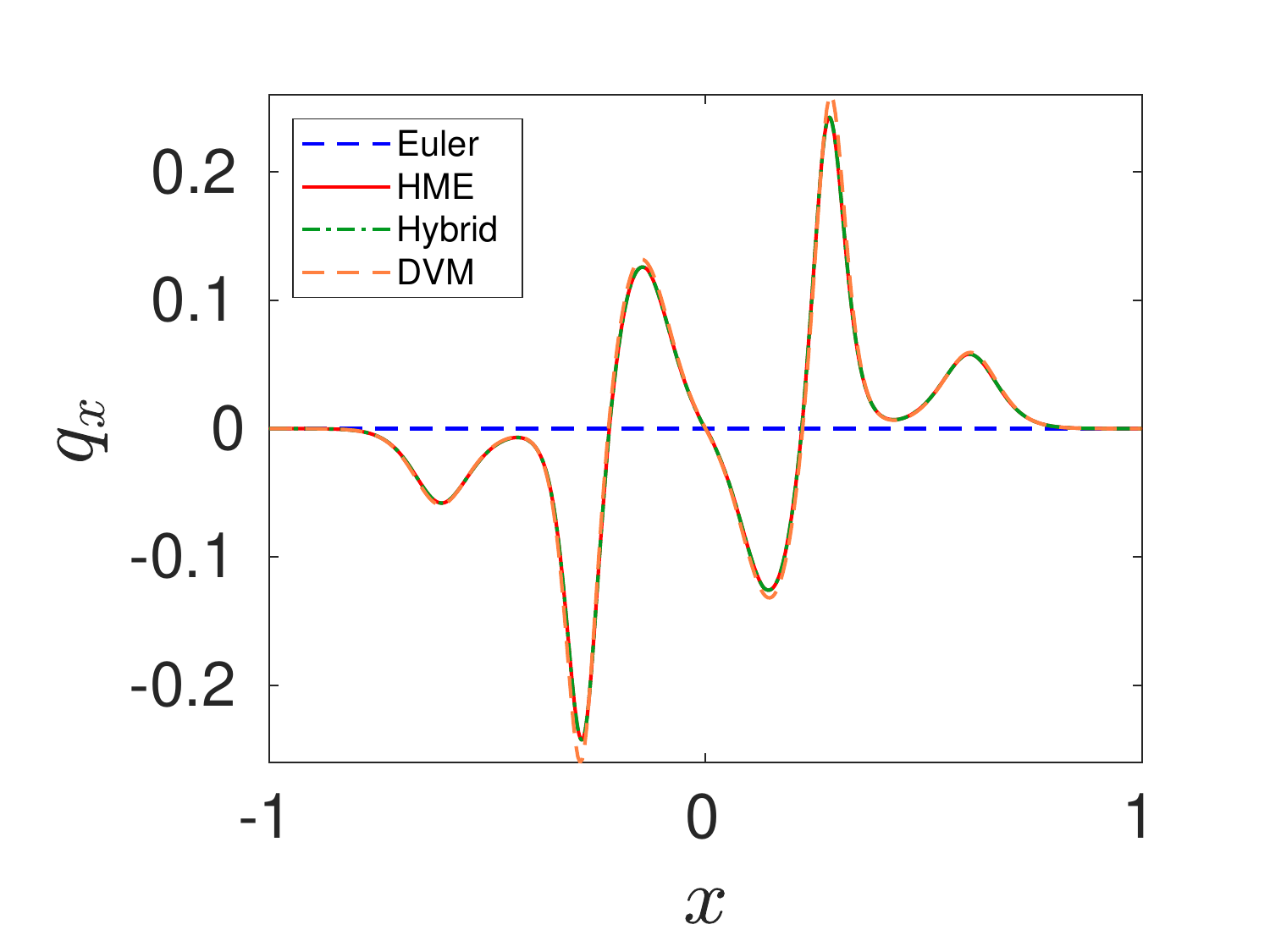}}
\caption{Comparisons of solutions for $\rho$, $u_x$, $\theta$ and
  $q_x$ as functions of $x$ for $\Kn = 0.01$ at time $t = 0.05, 0.15$
  and $0.35$ in Sec \ref{sec:test4}.}
\label{fig:test4_Kn0p01}
\end{figure}

\begin{figure}[htbp]
  \centering
  \subfloat[$\Kn = 0.001$ ]{
    \label{fig:order_test4_0p001}
    \includegraphics[trim=100 225 100
    250,clip,width=0.48\textwidth]{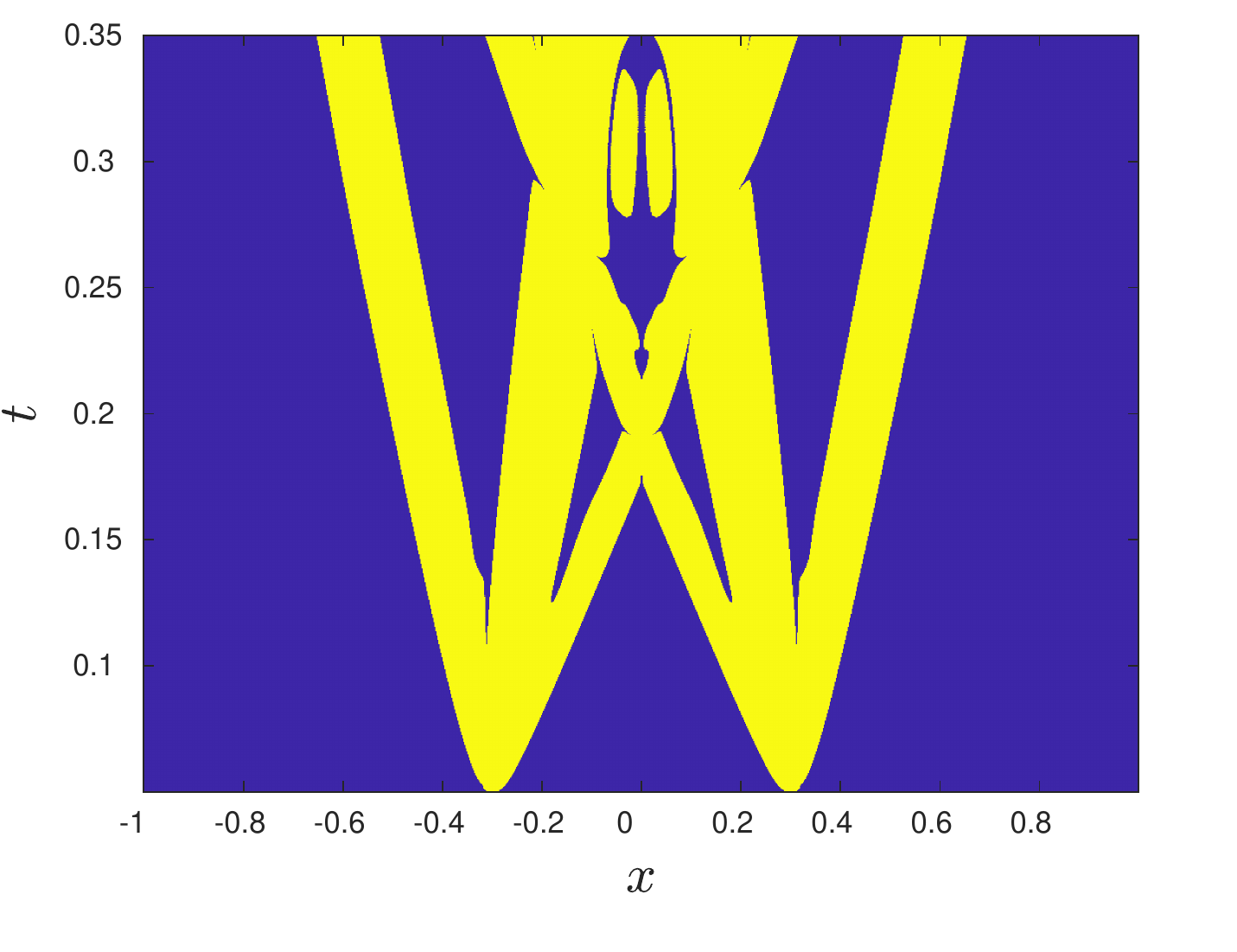}}
  \hfill \subfloat[$\Kn = 0.01$]{\label{fig:order_test4_0p01}
    \includegraphics[trim=100 225 100
    250,clip,width=0.48\textwidth]{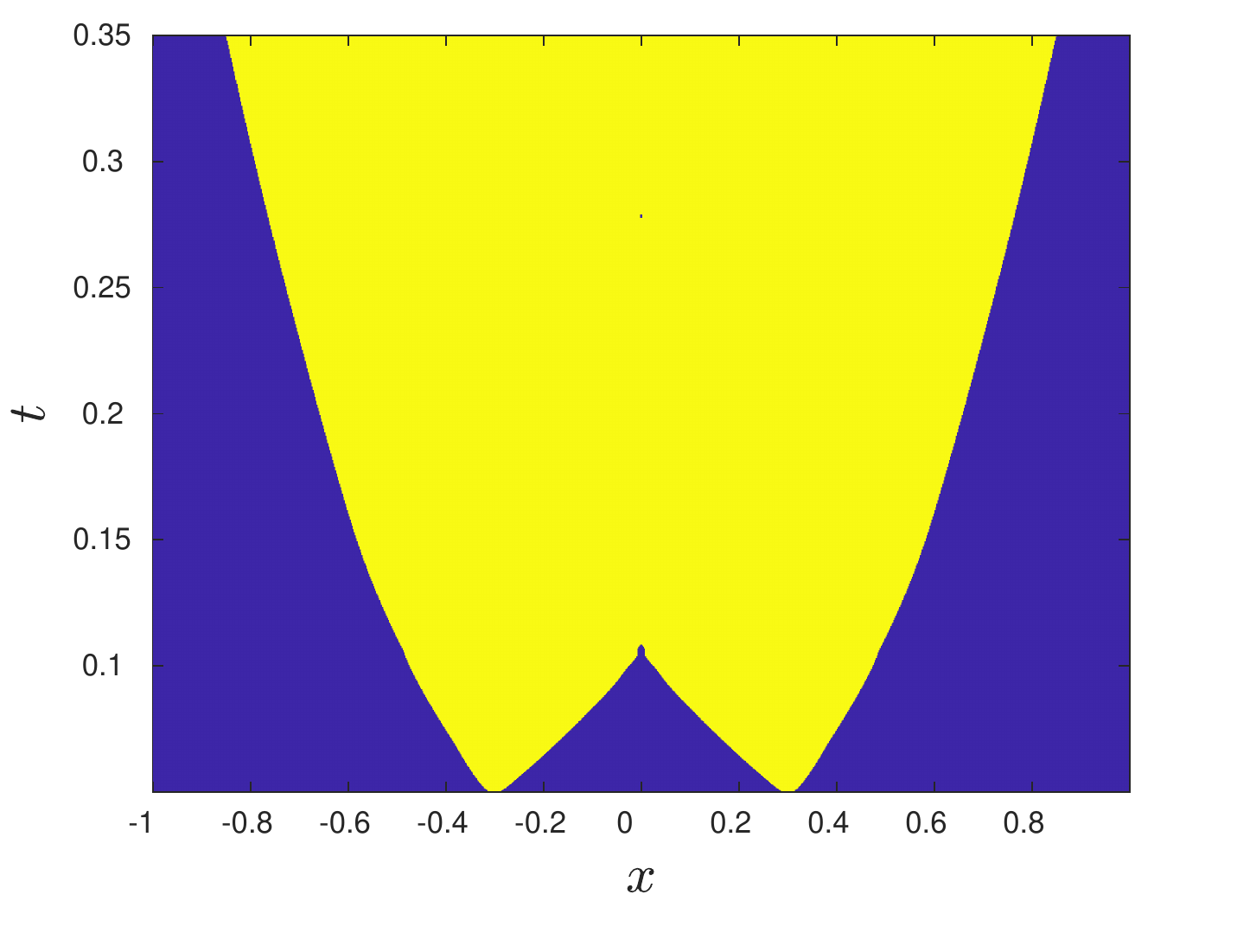}}
  \caption{The maximum order used in all spatial grids at each time
    step for differnt Knudsen numbers in Sec \ref{sec:test4}.}
\label{fig:test4_order}
\end{figure}

\Cref{fig:test4_Kn0p001} and \ref{fig:test4_Kn0p01} show the behavior
of $\rho$, $u_x$, $\theta$ and $q_x$ at time $t = 0.05$, $0.15$ and
$0.35$ for the Knudsen number $\Kn = 0.001$ and $0.01$.  For all these
cases, we see good agreement between the hybrid solver and HME, as
well as the reference solution. When $\Kn$ is small, we can find that
the numerical solutions of HME and hybrid agree well with that of
Euler equations. In \Cref{fig:test4_order}, the time evolution of the
maximum order actually used by the hybrid method is again plotted. We
can find that for the kinetic region in the middle of the spatial
space, the expansion order is $40$, which is the yellow region. For
the fluid regime, the expansion order is $4$, which is the blue
region. Note that when $\Kn = 0.001$, the spatial position of the
yellow region changes as time evolves. This is probably because the
relaxation time depends on $\Kn / \rho$, and as shown in
\Cref{fig:test4_Kn0p001}, the position of the peak values of $\rho$
changes in space with time evolution. This shows our hybrid method could detect
the fluid and kinetic regimes nicely. Moreover, when $\Kn = 0.001$, the
blue region with expansion order $4$ is much larger than that of $\Kn
= 0.01$, which means that more area belongs to the fluid regime when
Knudsen number is smaller. \Cref{tab:timecompare-test4} compares the
computation cost of the hybrid method and HME, where we could see a
significant reduction in computation cost. Moreover, when $\Kn =
0.001$, the hybrid method is more efficient than the case $\Kn =
0.01$, which is also consistent with the knowledge perception and the
variation of the expansion order as time going.

\begin{table}[htbp]
  \centering
  \def\arraystrech{1.5}
  \footnotesize
    \label{tab:timecompare-test4}
    \setlength{\tabcolsep}{5mm}{
      \begin{tabular}{c|ccc|ccc}
        $\Kn$ & \multicolumn{3}{c}{$0.001$} &
                                              \multicolumn{3}{c}{$0.01$}
        \\
        \toprule 
              &    $t_{end}$  & HME (s) & Hybrid (s) & $t_{end}$  &  HME  & Hybrid (s) \\
        \hline 
              &   $0.05$      &   $61$  & $35$ & $0.05$ & $59$ & $35$   \\
        \hline 
              &   $0.05$   & $175$ &  $108$   & $0.15$ &  $181$ &  $125$ \\
        \hline 
              &$0.35$  & $403$ &  $264$   &   $0.35$ &  $422$ &   $336$  \\
      \end{tabular}}
    \caption{Comparison of computational time of the hybrid method and
      HME at different time for different $\Kn$ in Sec
      \ref{sec:test4}.}
\end{table}


%% file: article_conclusion.tex
\section{Conclusion}
\label{sec:conclusion}
This work aims at a uniform hybrid moment method for the multi-scale
kinetic problems. The method is proposed in the framework of the
regularized moment method. Instead of Euler or Navier-Stokes
equations, the fourth order moment system is utilized as the governing
equations in the fluid region, in which case only one set of the
numerical scheme is needed for both fluid and kinetic regions.
Moreover, with the fourth order moment systems, the maximum entropy
method is applied here to derive the high order expansion coefficients
of the distribution function in the interface area from the fluid
regime to the kinetic regime. Several numerical experiments show that
this hybrid numerical algorithm can capture the movement of the
particles in different regions accurately and efficiently.

The method will be further validated in the numerical tests for 
more complex kinetic models, which is one of the ongoing work.  
The extension of the method for transferring information from kinetic 
to fluid regime to higher dimensional problems will also be studied in 
future work.

\section*{Acknowledgements}
We thank Prof. Ruo Li from Peking University for his valuable
suggestions to this research project.  This research of Weiming Li is
supported in part by Science Challenge Project (NO. TZ2016002). The
research of Peng Song is supported in part by the Natural Science
Foundation of China (91630310) and the CAEP foundation
(No.CX20200026). And that of Y. Wang is supported in part by Science
Challenge Project (NO. TZ2016002) and the Natural Science Foundation
of China (U1930402).

%% file: article_appendix.tex
\appendix
\section{Numerical scheme for the regularized moment method}
\label{app:numerical}
In this section, we will introduce the numerical scheme of the
dimensional reduced regularized moment method. We refer readers
\cite{Qiao} for more details.  Let
$\bomega = (\bomega^{(g)}, \bomega^{(h)})$ be the variables of the
reduced system, where $\bomega^{(f)}, f = g, h$ is the corresponding
variable $\bomega$ of the distribution function $f$ in
\eqref{eq:ms_quasi}. The reduced moment system is written as
\begin{equation}
  \label{eq:reduce_ms}
  \pd{\bomega}{t} + \sum_{j=1}^2{\bf B_j} \pd{\bomega}{x} ={\bf Q}\bomega.
\end{equation}
Splitting method is adopted to solve \eqref{eq:reduce_ms}, which is
split into the convection part and the collision part
\begin{itemize}
\item convection part:
  \begin{equation}
    \label{eq:reduce_con}
    \pd{\bomega}{t} + \sum_{j=1}^2{\bf B_j} \pd{\bomega}{x} =0.    
  \end{equation}
\item collision part:
  \begin{equation}
    \label{eq:reduce_col}
    \pd{\bomega}{t} =   {\bf Q}\bomega.
  \end{equation}
\end{itemize}
The detailed algorithm is as below,
\begin{enumerate}
\item Let $n = 0$, and give the initial value of $\bomega_i^n$.
\item Calculate the time step length $\Delta t_n$ by CFL condition.
\item Solve the convection part using the finite volume method with
  the HLL flux, and denote the result by $\bomega_{i}^{n, \ast}$.  The
  detailed distribution function in the hybrid method is given in
  algorithm \ref{alg:hybrid}.
\item Update the collision step using $\bomega_i^{n,\ast}$ as the
  initial condition.
\item Let $t \leftarrow t + \Delta t_n$ and $n \leftarrow n+1$;
  then go to Step 2. 
\end{enumerate}
The numerical scheme for the convection step and the collision is
listed below in detail.

\paragraph{Convection step}
For the convection part, the moment equation \eqref{eq:reduce_con} is
reformulated as
\begin{equation}
  \label{eq:reduce_con_1}
  \pd{\bomega}{t} + \pd{{\bf F}(\bomega)}{x} +{\bf R}(\bomega)  \pd{\bomega}{x}
  = 0. 
\end{equation}
The finite volume method is utilized here. Precisely
\begin{equation}
  \label{eq:flux}
  \bomega_{i}^{n, \ast} = \bomega_i^n - \frac{\Delta t_n}{\Delta x}
  \left(\hat{\bF}_{i+1/2}^n  - \hat{\bF}_{i - 1/2}^n\right)  - \frac{\Delta t_n}{\Delta
    x}\left(\hat{\bR}_{i+1/2}^{n-} - \hat{\bR}_{i-1/2}^{n+}\right),
\end{equation}
where $\hat{\bF}_{i+1/2}^{n}$ is the HLL numerical flux defined by
\begin{equation}
  \label{eq:hll}
  \hat{\bF}_{i+1/2}^{n} = \left\{
    \begin{array}{ll}
      {\bF}(\bomega_i^n),& \lambda_{i+1/2}^{L,n} \geqslant 0, \\[2mm]
      \frac{\lambda_{i+1/2}^{R,n} {\bF}(\bomega_{i}^n) -
      \lambda_{i+1/2}^{L,n} {\bF}(\bomega_{i+1}^n)}{\lambda_{i+1/2}^{R,n} - \lambda_{i+1/2}^{L,n}} +
      \frac{\lambda_{i+1/2}^{L,n} \lambda_{i+1/2}^{R,n} (\bomega_{i+1}^n  -
      \bomega_{i}^n)}{\lambda_{i+1/2}^{R,n} - \lambda_{i+1/2}^{L,n}}, &
                                                                \lambda_{i+1/2}^{L,n}
                                                                < 0 <
                                                                \lambda_{i+1/2}^{R,n},
      \\[2mm]
      \bF(\bomega_{i+1}^n), & \lambda_{i+1/2}^{R,n} \leqslant 0.
                              
    \end{array}
  \right.
\end{equation}
The numerical flux for the nonconservation part
$\hat{\bR}_{i+1/2}^{n\pm}$ is defined by
\begin{equation}
  \label{eq:non_flux}
  \begin{array}{c}
    \hat{\bR}_{i+1/2}^{n-} = \left\{
    \begin{array}{ll}
      0, & \lambda_{i+1/2}^{L,n} \geqslant 0, \\
      -\frac{\lambda_{i+1/2}^{L,n}}{\lambda_{i+1/2}^{R,n} -
      \lambda_{i+1/2}^{L,n}} \bg_{i+1/2}^n ,
         & \lambda_{i+1/2}^{L,n}  < 0  <  \lambda_{i+1/2}^{R,n},  \\
      \bg_{i+1/2}^n, & \lambda_{i+1/2}^{R,n} \leqslant 0, 
    \end{array}
                       \right.    \\ \\[2mm]
    \hat{\bR}_{i+1/2}^{n+} = \left\{
    \begin{array}{ll}
      -\bg_{i+1/2}^n, & \lambda_{i+1/2}^{L,n} \geqslant 0, \\
      -\frac{\lambda_{i+1/2}^{R,n}}{\lambda_{i+1/2}^{R,n} -
      \lambda_{i+1/2}^{L,n}} \bg_{i+1/2}^n ,
                      & \lambda_{i+1/2}^{L,n}  <  0 <  \lambda_{i+1/2}^{R,n},  \\
      0, & \lambda_{i+1/2}^{R,n} \leqslant 0,
    \end{array}
           \right.
  \end{array}
\end{equation}
with
\begin{equation}
  \label{eq:bg}
  \bg_{i+1/2}^n = \int_0^1 \bR(\Phi(s; \bq_i^n; \bq_{i+1}^n))
  \pd{\Phi}{s}(s; \bq_i^n; \bq_{i+1}^n) \dd s,
\end{equation}
where $\Phi(s; \cdot, \cdot)$ is a path to connect the two states. We
refer \cite{dal1995definition} for more details. The characteristic
speeds $\lambda_{i+1/2}^R$ and $\lambda_{i+1/2}^L$ are
\begin{equation}
  \label{eq:characteristic_speed}
  \begin{gathered}
    \lambda_{i+1/2}^{R,n} = \max\left(u_{1,i}^n +
      C_{M+1}\sqrt{\theta_i^n}, u_{1,j}^n
      + C_{M+1} \sqrt{\theta_{i+1}^n}\right), \\
    \lambda_{i+1/2}^{L,n} = \max\left(u_{1,i}^n -
      C_{M+1}\sqrt{\theta_i^n}, u_{1,j}^n - C_{M+1}
      \sqrt{\theta_{i+1}^n}\right),
  \end{gathered}
\end{equation}
where $C_{M+1}$ is the maximal roots of the $M+1$ degree Hermite
polynomial. The time step length is also decided by the
characteristic velocity as
\begin{equation}
  \label{eq:CFL}
  \frac{ \Delta t^n  \max\limits_i \left\{\left|\lambda_{i+1/2}^{R,n}\right|,
      \left|\lambda_{i+1/2}^{L,n}\right|\right\} }{\Delta x} \leqslant {\rm CFL}. 
\end{equation}

\paragraph{Collision step}
For the BGK and Shakhov model, the collision part can be solved
exactly in the reduced regularized moment method. For neatness, the
superscript $n$, $n+1$ and the subscripts $i$ are omitted.  The
solutions are as below:
\begin{itemize}
\item For the BGK model:
  \begin{equation}
    \label{eq:BGK_col}
    \begin{aligned}
      g_{\alpha} &= g_{\alpha}^{\ast} \exp\left(-\frac{\Delta t}{\tau}\right),
      \quad 2 \leqslant |\alpha| \leqslant M, \\
      h_{0}& = g_0^{\ast}\theta^{\ast} \left( 1  -
        \exp\left(\frac{\Delta t}{\tau}\right) \right) +  h_{0}^{\ast} \exp\left(-\frac{\Delta
          t}{\tau}\right), \\
      h_{\alpha}& = h_{\alpha}^{\ast} \exp\left(-\frac{\Delta
          t}{\tau}\right), \quad 1
      \leqslant |\alpha| \leqslant M-2.
    \end{aligned}
  \end{equation}
\item For the Shakhov model:
  \begin{equation}
    \label{eq:Shakhov_col}
    \begin{aligned}
      g_{3e_1} & = g_{3e_1}^{\ast}\exp\left(-\frac{\Delta
          t}{\tau}\right) +
      \frac{1}{5}q_i^{\ast}\left(\exp\left(-\frac{ \Pr \Delta
            t}{\tau}\right) -\exp\left(-\frac{\Delta
            t}{\tau}\right) \right), \\
      g_{\alpha} &= g_{\alpha}^{\ast} \exp\left(-\frac{\Delta
          t}{\tau}\right), \quad |\alpha| = 2 ~{\rm or}~ 4 \leqslant
      |\alpha| \leqslant
      M, \\
      h_{0}& = g_0^{\ast}\theta^{\ast} \left( 1 -
        \exp\left(\frac{\Delta t}{\tau}\right) \right) +
      h_{0}^{\ast} \exp\left(-\frac{\Delta
          t}{\tau}\right), \\
      h_{e_1} & = h_{e_1}^{\ast}\exp\left(-\frac{\Delta
          t}{\tau}\right) +
      \frac{1}{5}q_i^{\ast}\left(\exp\left(-\frac{ \Pr \Delta
            t}{\tau}\right) -\exp\left(-\frac{\Delta t}{\tau}\right)
      \right), \\
      h_{3e_1} & = h_{3e_1}^{\ast}\exp\left(-\frac{\Delta
          t}{\tau}\right) +
      \frac{\theta^{\ast}}{5}q_i^{\ast}\left(\exp\left(-\frac{ \Pr \Delta
            t}{\tau}\right) -\exp\left(-\frac{\Delta t}{\tau}\right)
      \right), \\
      h_{\alpha}& = h_{\alpha}^{\ast} \exp\left(-\frac{\Delta
          t}{\tau}\right), \quad |\alpha| = 2 ~{\rm or}~ 4 \leqslant
      |\alpha| \leqslant  M-2.
    \end{aligned}
  \end{equation}
\end{itemize}

\section{Algorithm for the maximum entropy closure}
\label{app:maximum_entropy_solver}
This section introduces the numerical scheme for solving the specific
maximum entropy distribution function needed in our hybrid moment
method. More details of this algorithm are the subject of a subsequent
article.

With the dimension reduction method in the microscopic velocity space,
the fourth order moment in 1D microscopic velocity space is utilized
in the fluid regime. Consider the maximum entropy moment problem
\begin{equation}\label{eq:maximum_entropy_4order}
  \hat{f} = \text{argmin} \int_{\bbR} f \log f -f\dd v, \quad s.t. \langle \hat{f}(v)
  \rangle_k = \mu_k, \quad k = 0, \cdots, 4.
\end{equation}
We follow the same approach as \cite{Abramov} to reformulate
\Cref{eq:maximum_entropy_4order} into the unconstrained form
\begin{equation}
  \mathcal{L}(\bslambda) = \int_{\bbR} \exp(\sum\limits_{i=0}^4
  \lambda_i v^i) - \sum\limits_{i=0}^4 \lambda_i \mu_i,
\end{equation}
which we solve using the Newton algorithm. The structure of the Newton
algorithm we employ is as follows:
\begin{enumerate}
\item Provide the starting point $\bslambda_0$ by a combination of
  analysis and interpolation from pre-computed data.  Compute the
  corresponding gradient $(\nabla \mathcal{L})_0$ and Hessian matrix
  $H_0$.
\item At iteration $m$, perform the following steps:
  \begin{enumerate}
  \item Solve for the direction of descent by 
    \[
      H_m \bd_m = -(\nabla \mathcal{L})_m;
    \]
  \item Find the next iteration point $\bslambda_{m+1}$ as
    \[
      \bslambda_{m+1} = \bslambda_m + \bd_m;
    \]
  \item At the new iteration point $\bslambda_{m+1}$, compute the
    gradient $(\nabla \mathcal{L})_{m+1}$ and the Hessian
    $H_{m+1}$.
  \end{enumerate}
\end{enumerate}

Direct computation shows that 
\begin{equation}
  (\nabla \mathcal{L})_{m,j} = \int_{\bbR} v^j
  \exp(\sum\limits_{i=0}^4 \lambda_i v^i) - \mu_j, \quad
  H_{m,jk} = \int_{\bbR} v^{i+j} \exp(\sum\limits_{i=0}^4 \lambda_i
  v^i) \dd v.
\end{equation}
Therefore most of the computation effort in solving the optimization
problem lies in computing the moments from the current iteration of
the Lagrange multipliers. In our computation, we utilize the fact that
the maximum entropy distribution function for fourth order system
either contains one peak or two peaks, therefore we employ a cutoff
function to approximate sufficiently accurately the integration on
$\bbR$ by integration on one or two finite domains. A standard
Gauss-Chebyshev quadrature is then employed for computation of the
integration on the finite domains.

%% file: article.bbl
\begin{thebibliography}{10}

\bibitem{Abramov}
R.~Abramov et~al.
\newblock The multidimensional maximum entropy moment problem: A review of
  numerical methods.
\newblock {\em Commun. Math. Sci.}, 8(2):377--392, 2010.

\bibitem{ALAIA20125217}
A.~Alessandro and P.~Gabriella.
\newblock A hybrid method for hydrodynamic-kinetic flow-{P}art {II}-{C}oupling
  of hydrodynamic and kinetic models.
\newblock {\em J. Comput. Phys.}, 231(16):5217 -- 5242, 2012.

\bibitem{Bird}
G.~Bird.
\newblock {\em Molecular Gas Dynamics and the Direct Simulation of Gas Flows}.
\newblock Oxford: Clarendon Press, 1994.

\bibitem{Caflisch2016}
E.~Caflisch, G.~Dimarco, and L.~Pareschi.
\newblock An hybrid method for the {B}oltzmann equation.
\newblock {\em AIP Conference Proceedings}, 1786(1):180001, 2016.

\bibitem{Fan_new}
Z.~Cai, Y.~Fan, and R.~Li.
\newblock Globally hyperbolic regularization of {G}rad's moment system.
\newblock {\em Comm. Pure Appl. Math.}, 67(3):464--518, 2014.

\bibitem{Qiao}
Z.~Cai, Y.~Fan, R.~Li, and Z.~Qiao.
\newblock Dimension-reduced hyperbolic moment method for the {B}oltzmann
  equation with {BGK}-type collision.
\newblock {\em Commun. Comput. Phys.}, 15(5):1368--1406, 2014.

\bibitem{NRxx}
Z.~Cai and R.~Li.
\newblock Numerical regularized moment method of arbitrary order for
  {B}oltzmann-{BGK} equation.
\newblock {\em SIAM J. Sci. Comput.}, 32(5):2875--2907, 2010.

\bibitem{Cai2018}
Z.~Cai and M.~Torrilhon.
\newblock Numerical simulation of microflows using moment methods with
  linearized collision operator.
\newblock {\em J. Sci. Comput.}, 74(1):336--374, 2018.

\bibitem{dal1995definition}
G.~Dal~Maso, P.~Lefloch, and F.~Murat.
\newblock Definition and weak stability of nonconservative products.
\newblock {\em J. Math. Pure. Appl.}, 74(6):483--548, 1995.

\bibitem{Degond2011}
P.~Degond, G.~Dimarco, and L.~Pareschi.
\newblock The moment-guided {M}onte {C}arlo method.
\newblock {\em Int. J. Numer. Meth. Fl.}, 67(2):189--213, 2011.

\bibitem{Dimarco2007}
G.~Dimarco and L.~Pareschi.
\newblock Hybrid multiscale methods ii. kinetic equations.
\newblock {\em SIAM Multiscale Model Sim.}, 6(4):1169--1197, 2008.

\bibitem{Dreyer}
W.~Dreyer.
\newblock Maximisation of the entropy in non-equilibrium.
\newblock {\em J. Phys. A: Math. Gen.}, 20(18):6505--6517, 1987.

\bibitem{Jin2010}
F.~Filbet and S.~Jin.
\newblock A class of asymptotic preserving schemes for kinetic equations and
  related problems with stiff sources.
\newblock {\em J. Comput. Phys.}, 229:7625--7648, 2010.

\bibitem{Filbet2015}
F.~Filbet and T.~Rey.
\newblock A hierarchy of hybrid numerical methods for multiscale kinetic
  equations.
\newblock {\em SIAM J. Sci. Comput.}, 37(3):A1218--A1247, 2015.

\bibitem{FILBET2018841}
F.~Filbet and T.~Xiong.
\newblock A hybrid discontinuous {G}alerkin scheme for multi-scale kinetic
  equations.
\newblock {\em J. Comput. Phy.}, 372:841 -- 863, 2018.

\bibitem{Hu2016}
I.~Gamba, J.~Haack, C.~Hauck, and J.~Hu.
\newblock {A fast spectral method for the {B}oltzmann collision operator with
  general collision kernels}.
\newblock {\em SIAM J. Sci. Comput.}, 39(14):B658--B674, 2017.

\bibitem{Gamba}
I.~Gamba, J.~Haack, and J.~Hu.
\newblock A fast conservative spectral solver for the nonlinear {B}oltzmann
  collision operator.
\newblock In J.~Fan, editor, {\em Proceedings of the 29th International
  Symposium on Rarefied Gas Dynamics}, volume 1628, pages 1003--1008, 2014.

\bibitem{Gamba2018}
I.~Gamba and S.~Rjasanow.
\newblock {G}alerkin-{P}etrov approach for the {B}oltzmann equation.
\newblock {\em J. Comput. Phys.}, 366:341--365, 2018.

\bibitem{goldstein1989}
D.~Goldstein, B.~Sturtevant, and J.~E. Broadwell.
\newblock Investigations of the motion of discrete-velocity gases.
\newblock {\em Progress in Astronautics and Aeronautics}, 117:100--117, 1989.

\bibitem{Grad}
H.~Grad.
\newblock On the kinetic theory of rarefied gases.
\newblock {\em Comm. Pure Appl. Math.}, 2(4):331--407, 1949.

\bibitem{Hauck}
C.~Hauck.
\newblock High-order entropy-based closures for linear transport in slab
  geometry.
\newblock {\em Commun. Math. Sci.}, 9(1):187--205, 2011.

\bibitem{Hu2014Simulation}
Z.~Hu, R.~Li, T.~Lu, Y.~Wang, and W.~Yao.
\newblock Simulation of an $n^{+}\hbox {-}n\hbox {-}n^{+}$ diode by using
  globally-hyperbolically-closed high-order moment models.
\newblock {\em J. Sci. Comput.}, 59(3):761--774, 2014.

\bibitem{Jaynes}
E.~Jaynes.
\newblock Information theory and statistical mechanics.
\newblock {\em Phys. Rev.}, 106(4):620, 1957.

\bibitem{KOLOBOV2007}
V.~Kolobov, R.~Arslanbekov, V.~Aristov, A.~Frolova, and S.~Zabelok.
\newblock Unified solver for rarefied and continuum flows with adaptive mesh
  and algorithm refinement.
\newblock {\em J. Comput. Phys.}, 223(2):589 -- 608, 2007.

\bibitem{Levermore}
C.~Levermore.
\newblock Moment closure hierarchies for kinetic theories.
\newblock {\em J. Stat. Phys.}, 83(5--6):1021--1065, 1996.

\bibitem{Levermore1998}
C.~Levermore, J.~Morokoff, and B.~Nadiga.
\newblock Moment realizability and the validity of the {N}avier–{S}tokes
  equations for rarefied gas dynamics.
\newblock {\em Phys. Fluids}, 10(12):3214--3226, 1998.

\bibitem{Groth}
J.~McDonald and C.~Groth.
\newblock Towards realizable hyperbolic moment closures for viscous
  heat-conducting gas flows based on a maximum-entropy distribution.
\newblock {\em Continuum Mech. Thermodyn.}, 25(5):573--603, 2013.

\bibitem{Mcdonald2013}
J.~McDonald and C.~Groth.
\newblock Towards realizable hyperbolic moment closures for viscous
  heat-conducting gas flows based on a maximum-entropy distribution.
\newblock {\em Continuum Mech. Therm.}, 25(5):573--603, 2013.

\bibitem{Mouhot}
C.~Mouhot and L.~Pareschi.
\newblock Fast algorithms for computing the {B}oltzmann collision operator.
\newblock {\em Math. Comp.}, 75(256):1833--1852, 2006.

\bibitem{Muller1993}
I.~M{\"u}ller and T.~Ruggeri.
\newblock {\em Extended Thermodynamics}, volume~37 of {\em Springer tracts in
  natural philosophy}.
\newblock Springer-Verlag, New York, 1993.

\bibitem{Panferov2002}
A.~Panferov and A.~Heintz.
\newblock {A new consistent discrete-velocity model for the {B}oltzmann
  equation}.
\newblock {\em Math. Method Appl. Sci.}, 25(7):571--593, 2002.

\bibitem{pareschi1996Spectral}
L.~Pareschi and B.~Perthame.
\newblock A fourier spectral method for homogeneous {B}oltzmann equations.
\newblock {\em Transport Theor. Stat.}, 25(3-5):369--382, 1996.

\bibitem{Schaerer}
R.~Schaerer, P.~Bansal, and M.~Torrilhon.
\newblock Efficient algorithms and implementations of entropy-based moment
  closures for rarefied gases.
\newblock {\em J. Comput. Phys.}, 340:138--159, 2017.

\bibitem{Struchtrup2005}
H.~Struchtrup.
\newblock Derivation of 13 moment equations for rarefied gas flow to second
  order accuracy for arbitrary interaction potentials.
\newblock {\em SIAM Multiscale Model. Simul.}, 3(1):221--243, 2005.

\bibitem{Struchtrup2008.1}
H.~Struchtrup and M.~Torrilhon.
\newblock Higher-order effects in rarefied channel flows.
\newblock {\em Phys. Rev. E}, 78:046301, Oct 2008.

\bibitem{TIWARI1998}
S.~Tiwari.
\newblock Coupling of the {B}oltzmann and {E}uler equations with automatic
  domain decomposition.
\newblock {\em J. Comput. Phys.}, 144(2):710 -- 726, 1998.

\bibitem{Torrilhon2006}
M.~Torrilhon.
\newblock Two dimensional bulk microflow simulations based on regularized
  {G}rad's 13-moment equations.
\newblock {\em SIAM Multiscale. Model. Simul.}, 5(3):695--728, 2006.

\bibitem{QuadraticCol}
Y.~Wang and Z.~Cai.
\newblock Approximation of the {B}oltzmann collision operator based on
  {H}ermite spectral method.
\newblock {\em J. Comput. Phys.}, 397(15), 2019.

\bibitem{xiongqiu}
T.~Xiong and J.~Qiu.
\newblock A hierarchical uniformly high order {DG}-{IMEX} scheme for the 1{D}
  {BGK} equation.
\newblock {\em J. Comput. Phy.}, 336:164--191, 2017.

\end{thebibliography}
